\documentclass[aos,preprint]{imsart}

\usepackage{fullpage}
\usepackage{latexsym,amsmath,amssymb,amsthm,epic,eepic,multirow}
\usepackage{natbib}

\usepackage{graphicx}
\usepackage{multirow}
\usepackage{upgreek}
\usepackage{subfigure}
\usepackage{natbib}
\usepackage{algorithm}
\usepackage[normalem]{ulem}
\usepackage{algorithmic}
\usepackage{enumitem}
\usepackage{comment}
\usepackage{mathrsfs}
\usepackage[dvipsnames]{xcolor}
\usepackage{dsfont}

\usepackage{float}
\usepackage{hyperref}
   \hypersetup{ colorlinks = true, linkcolor = blue, citecolor=blue }
\usepackage{xr-hyper}

\newcommand{\R}{\mathbb{R}}

\newcommand{\eps}{\epsilon}
\newcommand{\veps}{\varepsilon}

\newcommand{\Cov}{\mbox{Cov}}

\newcommand{\Var}{\mbox{Var}}

\newcommand{\bX}{\mathbf{X}}

\newcommand{\bU}{\mathbf{U}}
\newcommand{\bV}{\mathbf{V}}

\newcommand{\bel}{\begin{eqnarray}\label}
\newcommand{\eel}{\end{eqnarray}}
\newcommand{\bes}{\begin{eqnarray*}}
\newcommand{\ees}{\end{eqnarray*}}
\newcommand{\bei}{\begin{itemize}}
\newcommand{\eei}{\end{itemize}}

\def\Cov{\hbox{\rm Cov}}
\def\A{{\mathbb{A}}}
\def\E{{\mathbb{E}}}
\def\P{{\mathbb{P}}}
\def\M{{\mathcal{M}}}
\def\veps{\varepsilon}

\def\etabar{{\overline \eta}}

\def\sigmaline{\underline \sigma}
\def\Bbar{{\overline B}}

\def\Xtil{{\widetilde X}}

\def\mutil{{\widetilde \mu}}
\def\nutil{{\widetilde \nu}}

\def\Ttil{{\widetilde T}}

\def\tbX{{\widetilde \bX}}
\def\tbU{{\widetilde \bU}}

\def\pa{\partial }
\def\Rem{\hbox{\rm Rem}}
\def\diag{\hbox{\rm diag}}

\def\Xbar{{\overline X}}

\def\fbar{{\bar{f}}}

\def\scrA{{\mathscr A}}

\def\Omegatil{{\widetilde \Omega}}

\def\etatil{{\tilde \eta}}
\def\atil{{\tilde a}}

\theoremstyle{plain}
\newtheorem{theorem}{Theorem}
\newtheorem{proposition}{Proposition}
\newtheorem{lemma}{Lemma}
\newtheorem{corollary}{Corollary}

\theoremstyle{definition}

\theoremstyle{remark}
\newtheorem{remark}{Remark}

\usepackage{color}

\def\M{\mathfrak{M}}
\def\eqeq{\,\equiv\,}
\def\KS{\eta_n^*}
\def\linesigma{{\overline\sigma}} 

\setattribute{journal}{name}{}

\begin{document}

\begin{frontmatter}
\title{Beyond Gaussian Approximation: \\
Bootstrap for Maxima of\\ Sums of Independent Random Vectors}

\runtitle{Bootstrap for Maxima of Sums of Independent Random Vectors}

\begin{aug}
\author{\fnms{Hang} \snm{Deng}\ead[label=e1]{hdeng@stat.rutgers.edu}},
\and
\author{\fnms{Cun-Hui} \snm{Zhang}\thanksref{t1}\ead[label=e2]{czhang@stat.rutgers.edu}}

\runauthor{Hang Deng \& Cun-Hui Zhang}

\thankstext{t1}{Partially supported by NSF grants DMS-1513378, IIS-1407939, DMS-1721495, IIS-1741390 and CCF-1934924.}

\affiliation{Department of Statistics, Rutger University}

\end{aug}

\maketitle

\begin{abstract}
The Bonferroni adjustment, or the union bound, is commonly used to study rate optimality properties of statistical methods in high-dimensional problems. However, in practice, the Bonferroni adjustment is overly conservative. The extreme value theory has been proven to provide more accurate multiplicity adjustments in a number of settings, but only on ad hoc basis. Recently, Gaussian approximation has been used to justify bootstrap adjustments in large scale simultaneous inference in some general settings when $n \gg (\log p)^7$, where $p$ is the multiplicity of the inference problem and $n$ is the sample size. The thrust of this theory is the validity of the Gaussian approximation for maxima of sums of independent random vectors in high-dimension. In this paper, we reduce the sample size requirement to $n \gg (\log p)^5$ for the consistency of the empirical bootstrap and the multiplier/wild bootstrap in the Kolmogorov-Smirnov distance, possibly in the regime where the Gaussian approximation is not available. New comparison and anti-concentration theorems, which are of considerable interest in and of themselves, are developed as existing ones interweaved with Gaussian approximation are no longer applicable or strong enough to produce desired results. 
\end{abstract}

\end{frontmatter}

{\bf 1. Introduction.}
Let $\bX=(X_1,\ldots, X_n)^T \in \R^{n\times p}$ be a random matrix with
independent rows $X_i = (X_{i,1},\ldots,X_{i,p})^T \in \R^p$, $i=1,\ldots n$, where $p \eqeq p_n$ 
is allowed to depend on $n$. Let
\bes
\Xbar_n =\frac{1}{n}\sum_{i=1}^{n}X_i=(\Xbar_{n,1},\ldots,\Xbar_{n,p})^T.
\ees
We are interested in the consistency of the bootstrap for the maxima
\begin{align}\label{T_n}
\quad T_n = \max_{1\le j\le p}\sqrt{n}\Big(\Xbar_{n,j} - \E\Xbar_{n,j}\Big)
\end{align}
in the case of large $p$, including exponential growth of $p$ at certain rate as $n\to\infty$.

The consistency of the bootstrap for the maxima $T_n$ can be directly used to construct simultaneous
confidence intervals in the many means problem, but the spectrum of its application is much broader.
Examples include sure screening \citep{fanlv07}, removing spurious correlation \citep{fan2016guarding},
testing the equality of two matrices \citep{cai2013two,chang2016comparing},
detecting ridges and estimating level sets \citep{chen2015asymptotic,chen2016density}, and many more.
It can be also used in time series settings \citep{zhang2017gaussian} and high-dimensional regression \citep{zhangzhang11, belloni2014inference,bellonietal15,ZhangCheng2016,dezeure2016high}.
In such modern applications, $p = p_n$ is not fixed and can be much larger than $n$.

In closely related settings, \cite{gine1990} proved the consistency of bootstrap for Donsker classes
of functions, 
{\cite{nagaev1976estimate}, \cite{senatov1980several}, \cite{sazonov1981normal},
\cite{gotze1991rate} and \cite{bentkus1986dependence,bentkus2003dependence}}
for convex sets when $n\ge p^{7/2}$, and \cite{zhilova2016non} for Euclidean balls.
The set $\{T_n\le t\}$ is convex but we are interested in potentially much larger $p$.

More recently, in a groundbreaking paper, \cite{chernozhukov2013} used
Gaussian approximation to prove the consistency of the bootstrap
with a convergence rate of $((\log p)^7/n)^{1/8}$ under certain
moment and tail probability conditions on $\{X_{i,j}\}$.
This convergence rate was improved upon in \cite{chernozhukov2017central} to $((\log p)^7/n)^{1/6}$,
with extensions to the uniform consistency for 
$\P\{\sqrt{n}(\Xbar_n - \E\Xbar_n)\in A\}$ in certain classes of
hyper-rectangular and sparse convex sets $A\subseteq \R^p$.

In this paper, we improve the convergence rate to $((\log p)^5/n)^{1/6}$
for the multiplier/wild bootstrap with third moment match \citep{liu1988bootstrap,mammenwild1993}
and the empirical bootstrap \citep{efron1979} of $T_n$,
so that the sample size requirement is
reduced from $n\gg (\log p)^7$ to $n\gg (\log p)^5$.
We establish this sharper rate by exploiting the fact that under suitable conditions,
the average third moment tensor of $X_i$ is well approximated by its bootstrapped version,
\begin{align}\label{average-moment}
n^{-1}\sum_{i=1}^n \E^*\big(X_i^* - \E^* X_i^*\big)^{\otimes 3}
\approx n^{-1}\sum_{i=1}^n \E\big(X_i - \E X_i\big)^{\otimes 3},
\end{align}
in the supreme norm.
Here and in the sequel,
$\xi^{\otimes m}=(\xi_{i_1}\cdots \xi_{i_m})_{p\times \cdots \times p}$ denotes the
$m$ dimensional tensor/array generated by vector $\xi \in \R^p$.
The benefit of the third and higher moment approximation in bootstrap is well understood in the case of fixed $p$
\citep{singh1981asymptotic, hall1988theoretical, mammenwild1993,shao2012jackknife}.
However, the classical higher order results on bootstrap
were established based on the Edgeworth expansion associated with the central limit theorem,
while we are interested in
high-dimensional regimes in which the consistency of the Gaussian approximation is in question to begin with.
Moreover, as existing approaches of studying the bootstrap in high-dimension
are very much interweaved with the approximation of the average second moment
or the more restrictive approximation of the moments of individual vectors
\begin{align}\label{individual-moment}
\E^*\big(X_i^* - \E^* X_i^*\big)^{\otimes m}
\approx \E\big(X_i - \E X_i\big)^{\otimes m},\ m = 2,3,\ \forall\ i \le n,
\end{align}
our analysis requires new comparison and anti-concentration theorems.
These new comparison and anti-concentration theorems, also proved in this paper, are of
considerable interest in their own right.

The difference between the existing and our analytical approaches can be briefly explained as follows.
The first issue is the comparison between the expectation of smooth functions of the maxima and its
bootstrapped version.
The comparison theorems in \cite{chernozhukov2013,chernozhukov2017central} were derived with a combination of
the \cite{slepian1962one} smart path interpolation and the \cite{stein1981estimation} leave-one-out method.
As this Slepian-Stein approach does not take advantage of the bootstrap approximation of the third moment,
we opt for the Lindeberg approach \citep{lindeberg1922neue,chatterjee2006generalization}.
In fact, the original Lindeberg method was briefly considered in \cite{chernozhukov2013} without an expansion for the third or higher moment match.
As a direct application of the original Lindeberg method requires the more restrictive condition
(\ref{individual-moment}), we develop a coherent Lindeberg interpolation to prove comparison
theorems based on (\ref{average-moment}). This coherent Lindeberg approach and the
resulting comparison theorems are new to the best of our knowledge.
The second issue is the anti-concentration of the maxima, or an upper bound for
the modulus of continuity for the distribution of the maxima, without a valid Gaussian approximation.
We resolve this issue by applying the new comparison theorem to a mixed multiplier bootstrap
with a Gaussian component and a perfect match in the first three moments,
so that the anti-concentration of the Gaussian maxima can be utilized through the mixture.
This solution to the anti-concentration problem is again new to the best of our knowledge. 
For the anti-concentration of the maximum of 
Gaussian vector 
$(\xi_1, \ldots, \xi_p)^T $ with marginal distributions $\xi_j \sim N(\mu_j, \sigma_j^2), 1 \le j \le p$, 
we sharpen the existing upper bound for the density of the maximum from 
$C(2+\sqrt{2\log p})/\sigma_{(1)}$ [based on \cite{klivans2008learning}]
to the potentially much smaller 
$(2+\sqrt{2\log p})/{\overline \sigma}$, where 
\begin{align}\label{sigmas}
{\overline \sigma} = \min_{1\le j \le p} 
\frac{2 + \sqrt{2 \log p}}{1/\sigma_{(1)} +  (1 + \sqrt{2\log j})/\sigma_{(j)}} 
\end{align}
and $\sigma_{(j)}^2$ is the $j$-th smallest average variance 
among $\big\{\sigma_k^2 = n^{-1}\sum_{i=1}^n \Var(X_{i,k}), 1\le k \le p\big\}$. 
Moreover, our anti-concentration bound is sharp up to explicit constants when $\xi_j$ 
are correlated and/or non-central.  
As more weights are given to the smaller $1/\sigma_{(j)}$ in the denominator in \eqref{sigmas}, 
$\sigma_{(1)}\le \linesigma\le \sigma_{(p)}$. 

We organize the paper as follows.  In Section 2, we state our bootstrap consistency theorems
and discuss their implications and applications.
In Section 3, we present new comparison theorems based on the coherent Lindeberg interpolation.
In Section 4, we provide new anti-concentration theorems based mixtures with Gaussian components.
In Section 5, we present some simulation results. The full proofs of all theorems, propositions and lemmas in this paper are relegated to the Supplement Material.

We use the following notation.
We assume $n \to \infty$ and $p=p_n$
to allow $p \to \infty$ as $n \to \infty$.  
We assume $p >1$ for notational simplicity; our analysis remain true for $p=1$ if we replace $\log p$  with $1 \vee (\log p)$. To shorten mathematical expressions, we write moments as tensors as in
(\ref{average-moment}) and (\ref{individual-moment}).
We also write partial derivative operators as tensors
$(\pa/\pa x)^{\otimes m}=\big((\pa/\pa x_{i_1})\cdots (\pa/\pa x_{i_m})\big)_{p\times \cdots \times p}$
for $x=(x_1,\ldots,x_p)^T$,
so that ${f}^{(m)} = (\pa/\pa x)^{\otimes m}f(x)$ is a tensor for functions $f(x)$ of input $x\in\R^p$,
and for two m-th order tensors $f$ and $g$ in $\R ^{p\times \cdots \times p}$,
the vectorized inner product is denoted by
\bes
\big\langle f,g \big\rangle = \sum_{j_1 =1}^p \cdots \sum_{j_m=1}^p  f_{j_1,\ldots, j_m} g_{j_1,\ldots, j_m}
\ees
and $|f|\le |g|$ means $|f_{j_1,\ldots, j_m}|\le |g_{j_1,\ldots, j_m}|$ for all indices ${j_1,\ldots, j_m}$. 
We denote by $\|\cdot\|_{q}$ the $\ell_q$ norm for vectors, 
$\|\cdot\|_{L_q}=\|\cdot\|_{L_q(\P)}$ the $L_q(\P)$ norm 
for random variables under probability $\P$, 
and $\|\cdot\|_{\max}$ the $\ell_{\infty}$ norm for matrices and tensors after vectorization.

We define quantities $M_n$, $\M_m$, $\M_{m,1}$ and $\M_{m,2}$
as follows for the average centered moments of $X_{ij}$ 
under different ways of maximization: 
The maximum average centered moments and the average moments of the maximum are respectively 
\begin{align}\label{M_m}
&& M_m^m = \max_{1\le j \le p} \frac{1}{n}\sum_{i=1}^n \E|X_{i,j} {- \E X_{i,j}}|^m, \quad
\M_m^m =  \frac{1}{n} \sum_{i=1}^n \E \max_{1\le j \le p} |X_{i,j} - \E X_{i,j}|^m,
\end{align}
and the average of the maximum moment
and the expected maximum average power are respectively 
\begin{align}\label{M_m-1}
\M_{m,1}^m = {\frac{1}{n}\sum_{i=1}^n \max_{1\le j \le p} \E\big|X_{i,j} - \E X_{i,j}\big|^m},\quad \M_{m,2}^m = \E \max_{1\le j \le p} \frac{1}{n}\sum_{i=1}^n |X_{i,j} {- \E X_{i,j}}|^m.
\end{align}
Clearly, $M_m\le \M_{m,j}\le \M_m$, $j=1,2$. 

In what follows, we denote by $C_0$ a numerical constant and $C_{\rm index}$ a constant depending
on the ``index'' only. For example, $C_{a,b,c}$ is a constant depending on $(a,b,c)$ only.
To avoid cumbersome calculation of explicit expressions of these constants,
they will be allowed to take different values from one appearance to the next in the proofs. 
Finally, we denote by $\Phi(\cdot)$ the standard normal cumulative distribution function and
$\Phi^{-1}(\cdot)$ the corresponding quantile function.

\medskip
{\bf 2. Consistency of bootstrap.}
Let $T_n$ be the maximum of normalized sum of $n$ independent random vectors
$X_i \in \R^p$ as defined in (\ref{T_n}).
In this section, we present our main theorems on the consistency of bootstrap in
approximating the distribution of $T_n$.
We consider this consistency in two somewhat different perspectives.
In simultaneous inference about the average mean $\E \sum_{i=1}^n X_{i,j}/n$,
we are interested in the performance of the bootstrapped quantile
\bes
t_\alpha^* = \inf\Big[t: \P^*\big\{ T_n^* > t\big\} \le \alpha\Big]
\ees
at a pre-specified significance level $\alpha$,
where $T_n^*$ is the bootstrapped version of $T_n$ and $\P^*$ is the conditional expectation given the
original data.
As an approximation of the $1-\alpha$ quantile of $T_n$, the performance of such $t_\alpha^*$ is
measured by
\bes
\Big| \P\{T_n > t_\alpha^*\} - \alpha\Big|.
\ees
On the other hand, if we are interested in recovering the entire distribution function of $T_n$,
it is natural to consider the Kolmogorov-Smirnov distance
\bes
\KS(T_n,T_n^*) = \sup_{t}\Big|\P\{T_n \le t\} - \P^*\{T_n^* \le t\} \Big|.
\ees
We shall consider Efron's (1979) empirical bootstrap
and the wild bootstrap in separate subsections.

It seems possible to extend our ideas and analysis to more general settings, 
for example the bootstrap schemes in \citep{hall1999intentionally} and \citep{praestgaard1993exchangeably} and the consistency in rectangular sets \citep{chernozhukov2017central}. 
However, we would not pursue these extensions here as they would make the paper 
more technical. 

\medskip
{\bf 2.1. Empirical bootstrap.}
In the empirical bootstrap, we generate i.i.d.\,vectors $X_1^*,\ldots,X_n^*$ from the empirical distribution
of the centered data points $X_1-\Xbar,\ldots,X_n-\Xbar$ from the original sample:
Under the conditional probability $\P^*$ given the original data $\bX = (X_1,\ldots,X_n)^T$,
\begin{align}\label{empirical-bootstrap}
\P^*\Big\{X_i^* = X_k - \Xbar\Big\} = n^{-1}\#\{j: 1\le j\le n: X_j=X_k\},
\ k=1,\ldots, n,\ i=1,\ldots,n,
\end{align}
where 
$\Xbar = \sum_{i=1}^n X_i/n$ is the sample mean.
The bootstrapped version of $T_n$ is defined as
\begin{align}\label{T_n^*}
T_n^* = \max_{1 \le j \le p} \frac{1}{\sqrt{n}}\sum_{i=1}^n X^*_{i,j}.
\end{align}
We state our main theorem on the consistency of empirical bootstrap as follows.

\begin{theorem}\label{cb-empi}
(Empirical Bootstrap)
Let $\bX=(X_1,\ldots, X_n)^T \in \R^{n\times p}$ be a random matrix with independent rows $X_i \in \R^p$,
$X_i^*$ the empirical bootstrapped $X_i$ as in (\ref{empirical-bootstrap}),
and $T_n$ and $T_n^*$ as in (\ref{T_n}) and (\ref{T_n^*}) respectively. Let $M_4$ and $\M_4$ be as in \eqref{M_m} , and $\linesigma$ be as in \eqref{sigmas}. Define
\begin{align}\label{th-empi-new-0}
\gamma^*_{{{\delta}}, M_0} = \bigg(\frac{(\log p)^2 (\log(np/{{\delta}}))^3}{n} \frac{M_0^4}{\linesigma^4} \bigg)^{1/6}. 
\end{align}
Then, with $M \ge M_4$ satisfying 
\begin{align}\label{th-empi-new-1}
\P\Big\{\|\bX  - \E \bX\|_{\max} > \frac{n^{1/3} \linesigma^{1/3}  M^{2/3}}{(\log p)^{1/6}(\log (4np/{{\delta}}))^{1/2}} \Big\}\le \frac{1}{2} \min\Big\{{{\delta}},  \gamma^*_{{{\delta}}, M} \Big\},
\end{align}
there exists a numerical constant $C_0$ such that the Kolmogorov-Smirnov distance between the distributions of $T_n$ and $T_n^*$
is bounded by
\begin{align}\label{th-empi-new-2}
\sup_{t \in \R} \Big|\P\{T_n \le t\} - \P^*\{T_n^* \le t\}\Big| \le C_0 \min \bigg\{ \gamma^*_{{{\delta}}, M}, \, \, \gamma^*_{{{\delta}}, \M_4} \Big[1 \vee \big(\gamma^*_{{{\delta}}, \M_4} /{{\delta}} \big)^{1/5} \Big]  \bigg\} 
\end{align}
with at least probability $1-{{\delta}}$. 
Moreover, with $M \ge M_4$ satisfying \eqref{th-empi-new-1} for ${{\delta}} = 1$,
\bel{th-empi-new-3}
\Big| \P\{T_n\le t^*_\alpha \} - (1-\alpha) \Big| \le C_0 \min \Big\{ \gamma^*_{1, M}, \, \, \gamma^*_{1, \M_4} \Big\}.
\eel
\end{theorem}

Note that the tail probability condition \eqref{th-empi-new-1} is needed only when the first component on the right-hand side of \eqref{th-empi-new-2} and \eqref{th-empi-new-3} is smaller.
Theorem \ref{cb-empi} asserts that
under the fourth moment and tail probability conditions,
Efron's empirical bootstrap provides a consistent estimate
of the distribution of $T_n$ when
\bes
n \gg (\log p)^5.
\ees
This should be compared with the existing results on the Gaussian wild bootstrap 
and empirical bootstrap where
\bes
n \gg (\log p)^7
\ees
is required \citep{chernozhukov2013,chernozhukov2017central}. 
In practice, the significance of the difference between $(\log p)^5$ and $(\log p)^7$ 
would depend on applications even if we ignore the constant factors involved in different theorems. 
If the above conditions are viewed as sample size requirements, 
it would be fair to say that the difference could be quite significant, i.e. a $(\log p)^2$ fold increase 
in $n$, when data are not dirt cheap. More important, our results prove theoretical advantages of bootstrap schemes with third moment match in high-dimension, compared with methods based on Gaussian approximation, as supported by our simulation results in Section 5 for moderately large $p$. 
Moreover, as we show in Corollary \ref{eg-1} below, our theory either requires just the fourth moment $M_4$ or provides the rate $\gamma_n^*\asymp ((B_n/\linesigma)^2(\log(np))^5/n)^{1/2}$ where $B_n$ is the maximum Orlicz norm of $X_{ij}$. 

\medskip
{\bf 2.2. Wild bootstrap.}
In wild bootstrap \citep{wu1986jackknife},
we generate
\begin{align}\label{wild}
X_i^* = W_i\big(X_i - \Xbar\big),
\end{align}
where $\Xbar = \sum_{i=1}^n X_i/n$ is the sample mean,
$W_1,\ldots,W_n$ are i.i.d.\,variables with
\begin{align}\label{second-moment-match}
\E\,W_i = 0,\quad \E\,W_i^2=1,
\end{align}
and the sequence $\{W_i\}$ is independent of the original data $\bX = (X_1,\ldots,X_n)^T$.

This general formulation of the wild bootstrap allows broad choices of the multiplier $W_i$
among them the Gaussian $W_i\sim N(0,1)$ and Rademacher $\P\{W_i=\pm 1\}=1/2$
are the most obvious. \cite{liu1988bootstrap} suggested the use of multipliers satisfying
\bel{third-moment-match}
\E\,W_i = 0,\quad \E\,W_i^2=1,\quad \E W_i^3=1,
\eel
to allow the third moment match $\E(X_i^*)^{\otimes 3} \approx \E\, X_i^{\otimes 3}$,
and explored the benefits of such schemes.
\cite{mammenwild1993} proposed a specific choice of the multiplier $W_i$
satisfying (\ref{third-moment-match}),
\begin{align}\label{mammen-w}
\P \bigg\{W_i= \frac{1\pm\sqrt{5}}{2} \bigg\}= \frac{\sqrt{5}\mp1}{2\sqrt{5}},
\end{align}
and studied extensively the benefit of the third moment match in wild bootstrap.
We note here that while 
(\ref{third-moment-match}) 
holds for many choices of $W_i$, the Gaussian and Rademacher multipliers do not possess this property.
In the following theorem, we assume the sub-Gaussian condition
\begin{align}\label{subgaussian}
\E\, \exp\Big(tW_1\Big) \le \exp\Big(\tau_0^2 t^2/2\Big),\ \forall\ t\in \R,
\end{align}
in addition to the third moment condition (\ref{third-moment-match}).

\begin{theorem}\label{cb-wild}
{(Wild Bootstrap)}
Let $\bX=(X_1,\ldots, X_n)^T \in \R^{n\times p}$ be a random matrix with independent rows
$X_i \in \R^p$, and $X_i^*$ be generated by the
wild bootstrap as in (\ref{wild}) with multipliers satisfying the moment condition
(\ref{third-moment-match}) and the sub-Gaussian condition (\ref{subgaussian}) with a certain $\tau_0<\infty$.
Let $T_n$ and $T_n^*$ be as in (\ref{T_n}) and (\ref{T_n^*}) respectively. Define
\begin{align}\label{th-wild-new-0}
\gamma^*_{{{\delta}}, M_0} = \bigg(\frac{(\log p)^2 (\log (np)) (\log(np/{{\delta}})^2)}{n} \frac{M_0^4}{\linesigma^4} \bigg)^{1/6}. 
\end{align}
Then, with $M \ge M_4$ satisfying 
\begin{align}\label{th-wild-new-1}
\P\Big\{\|\bX  - \E \bX\|_{\max} > \frac{n^{1/3} \linesigma^{1/3}  M^{2/3}}{(\log p)^{1/6}(\log (np))^{1/3}(\log (4np/{{\delta}}))^{1/6}} \Big\}\le \frac{1}{2} \min\Big\{{{\delta}},  \gamma^*_{{{\delta}}, M} \Big\},
\end{align}
there exists a numerical constant $C_{\tau_0}$ such that the Kolmogorov-Smirnov distance between the distributions of $T_n$ and $T_n^*$ is bounded by
\begin{align}\label{th-wild-new-2}
\sup_{t \in \R} \Big|\P\{T_n \le t\} - \P^*\{T_n^* \le t\}\Big| \le C_{\tau_0} \min \bigg\{ \gamma^*_{{{\delta}}, M}, \, \, \gamma^*_{{{\delta}}, \M_{4,2}} \Big[1 \vee \big(\gamma^*_{{{\delta}}, \M_{4,2}} /{{\delta}} \big)^{1/5} \Big] \bigg\} 
\end{align}
with at least probability $1-{{\delta}}$, where $\M_{4,2} \le \M_4$ by its definition in \eqref{M_m-1}. 
Moreover, with $M \ge M_4$ satisfying \eqref{th-wild-new-1} for ${{\delta}} = 1$,
\begin{align}\label{th-wild-new-3}
\Big| \P\{T_n\le t^*_\alpha \} - (1-\alpha) \Big| \le C_{\tau_0} \min \Big\{ \gamma^*_{1, M}, \, \, \gamma^*_{1, \M_{4,2}} \Big\}.
\end{align}
\end{theorem}

\begin{remark}
A user friendly bound of $\M_{4,2}$
\begin{align}\label{M_{4,2}-bd}
\M_{4,2}^4 \le K \Big(M_4^4 + \frac{\log p}{n} \E \max_{i,j} \big|X_{i,j} - \E X_{i,j}\big|^4\Big)
\end{align}
for some universal constant $K$ can be found in 
Lemma 9 of \cite{chernozhukov2015comparison} and Lemma E.3 of \cite{chernozhukov2017central}.
\end{remark}

Theorem \ref{cb-wild} asserts that with the third moment condition (\ref{third-moment-match})
on the multiplier, 
the conclusions of Theorem \ref{cb-empi} are all valid for the wild bootstrap 
under weaker moment condition. Thus, the discussion below Theorem \ref{cb-empi} 
about its significance also applies to Theorem \ref{cb-wild}. 

While the statements of Theorems \ref{cb-empi} and \ref{cb-wild} are almost identical, 
the smaller quantity $\M_{4,2}$ is used in 
\eqref{th-wild-new-2} and \eqref{th-wild-new-3} in Theorem \ref{cb-wild}, 
compared with the larger $\M_4$ in 
\eqref{th-empi-new-2} and \eqref{th-empi-new-3} in Theorem \ref{cb-empi}. 
Theorem \ref{cb-wild} can be further sharpened if Theorems~\ref{th-comp-wild-a}
and \ref{th-comp-wild-b} in Section 3 are applied in full strength. 

As briefly discussed below Theorem \ref{cb-empi}, a 
key point in our theory is the benefit of the third or higher moment match in both the empirical bootstrap and wild bootstrap.
Efron's empirical bootstrap can always match moments but not exactly,
\bes
\E \Big\{ \frac{1}{n} \sum_{i=1}^n \E X_i^{\otimes m} -\frac{1}{n}  \sum_{i=1}^n \E^* (X_i^*)^{\otimes m}  \Big\} \approx 0 \ \ \ \ m= 1,2,\ldots
\ees
An alternative wild bootstrap scheme, $X_i^*=W_iX_i$, which approximates (\ref{wild})
with negligible difference in our analysis under the assumption of $\E X_i=0$,
matches the moments of $X_i$ perfectly,
\bel{moment-match-m}
\E \Big\{ \E X_i^{\otimes m} -  \E^* (X_i^*)^{\otimes m}  \Big\} = 0,\
 \eel
but only up to a certain order; $m=1,2$ for the Gaussian and Rademacher wild bootstrap,
and $m=1,2, 3$ for Mammen's and other wild bootstrap schemes satisfying (\ref{third-moment-match}).
Thus, compared with the proof of Theorem \ref{cb-wild} which directly applies the exact 
moment match in \eqref{moment-match-m}, 
the proof of Theorem~\ref{cb-empi} requires an additional analysis of the 
the difference in the moments, leading to the stronger condition involving $\M_4$. 

If $X_i\in\R^p$ have symmetric distributions,
condition (\ref{moment-match-m}) holds for all $m$ for
the Rademacher wild bootstrap. In this case, the sample size condition $n \gg (\log p)^4$
is sufficient for the consistency of the bootstrap
under sixth moment and tail probability conditions and an anti-concentration condition.

\begin{theorem}\label{cb-rade}
{(Rademacher wild Bootstrap)} Let $\bX=(X_1,\ldots, X_n)^T \in \R^{n\times p}$ be a random matrix with
independent rows $X_i \in \R^p$. 
Suppose $\E (X_i - \E X_i)^{\otimes m} = 0$ for  $m=3$ and $m=5$. Let $X_i^*$ be generated by the Rademacher wild bootstrap, with
$\P\{W_i=\pm 1\}=1/2$ for the multiplier in (\ref{wild}).
Then, for any given constants $c_0$, $c_1$ and $M\ge M_6$,
\begin{align}\label{cb-rade-1}
& \Big| \P\{T_n\le t^*_\alpha \} - (1-\alpha) \Big|
+ \bigg(\E \sup_{t\in \R}\Big|\P\Big\{T_n < t\Big\} - \P^*\Big\{T_n^* < t \Big\} \Big|^2\bigg)^{1/2} 
\\
\le& \quad C_{c_0,c_1}\bigg(\frac{\log p}{n^{1/4}}\bigg)^{4/7}
+ \sup_{t\in \R} \P\bigg\{ t - c_0\bigg(\frac{\log p}{n^{1/4}}\bigg)^{4/7} \le \sqrt{\log p}\,\frac{T_n}{M} \le t\bigg\} \nonumber
\\
& + \bigg[\E \min\bigg\{4, C_{c_0,c_1}\bigg(\frac{\log p}{n^{1/4}}\bigg)^{32/7}
\max_{1\le j\le p}\sum_{i=1}^n
\frac{(X_{i,j}-\E X_{i,j})^6}{M^6n}I_{\{|X_{i,j}-\E X_{i,j}|>a_n\}} \bigg\}\bigg]^{1/3}, \nonumber
\end{align}
where $a_n = c_1M\sqrt{\log p}\big(n^{1/4}/\log p\big)^{10/7}$ and
$C_{c_0,c_1}$ is a constant depending on $\{c_0,c_1\}$ only.
\end{theorem}

The discussion below Theorem \ref{cb-empi} 
about its significance also applies here, 
although $(\log p)^5$ is further improved to $(\log p)^4$ 
and an anti-concentration condition is required in Theorem~\ref{cb-rade}.
In Section 4, we prove that the anti-concentration condition
\bes
\sup_t \P\bigg\{ t - \eps_n \le \sqrt{\log p}\,\frac{T_n}{M} \le t\bigg\}
=o(1)\quad \forall \eps_n=o(1)
\ees
holds when $\sum_{i=1}^n X_i/\sqrt{n}$ is conditionally a Gaussian vector given a certain sigma field
${\cal A}$, with $\Var(\sum_{i=1}^n X_{i,j}/\sqrt{n}|{\cal A}) = \sigma_j^2$ such that
$\P\big\{ \min_j \sigma_j^2 \ge \sigmaline^2 \big\} \to 1$
for a certain constant $\sigmaline>0$.

The condition $\E (X_i - \E X_i)^{\otimes m} = 0$ holds for the leading odd $m \in \{3,5\}$ 
when $X_i$ are symmetric about its mean, 
i.e., $\P\{X_i - \E X_i\in A\} = \P\{\E X_i - X_i\in A\}$ for all Boreal sets $A\subset \R^p$. 
In practice, such conditions could be imposed by the application itself. 
If the validity of such conditions is uncertain, 
we may also test the moment condition when $X_i$ are i.i.d. 
However, a theoretical analysis of such tests and the validity of \eqref{cb-rade-1} for 
the Rademacher wild bootstrap after such tests is beyond the scope of this paper. 

\medskip
{\bf 2.3. Examples.}
In this subsection, we consider some specific examples in which the moment and tail probability
conditions of our theorems hold.
These examples cover many practical problems and applications as discussed in \cite{chernozhukov2013,chernozhukov2017central}, and 
many publications citing their work \citep{dezeure2016high, ning2017general, zhangd2017gaussian, chen2018gaussian, blanchet2019robust, horowitz2019bootstrap}.
Throughout this subsection, we assume the following,
\begin{enumerate}[label=Cond-{\arabic*}:,leftmargin= 1in  ]
	\item 
	$0 < {\overline \sigma} \le 
	(2+\sqrt{2\log p})\big/\big\{1/\sigma_{(1)} + (1+\sqrt{2\log j})/\sigma_{(j)}\big\},\ 
		\forall\ j=1,\ldots, p$, 
	\item $n^{-1}\sum_{i=1}^n \E |X_{i,j}- \E X_{i,j}|^{4} \le M_4^4,\ \forall\ j=1,\ldots, p$,
\end{enumerate}
where $\sigma_{(1)}\le\cdots\le \sigma_{(p)}$ are the ordered values of 
$\sigma_j =\big({n}^{-1} \sum_{i=1}^n\E (X_{i,j} - \E X_{i,j})^2\big)^{1/2}$. 
Here $\linesigma$ and $M_4$ are allowed to depend on $n$ and to diverge to $0$ or $\infty$,
but they can also be treated as constants for simplicity.
Under the above moment conditions, we consider three examples
specified by certain measure $B_n$ of the tail of $\{|X_{i,j}|\}$,
possibly with unbounded $B_n$.

{\medskip
{\it 2.3.1. Exponential tail.}
Here we impose one additional condition on the tail of $X_{i,j}$
in the form of a uniform bound on their Orlicz norm with respect to $\psi_1(x) = e^x-1$:
\bes
\hbox{(E.1):}&& \big\|X_{i,j}\big\|_{\psi_1}
= \inf\Big\{B: \E\, \psi_1(|X_{i,j} - \E X_{i,j}|/B)\le 1\Big\} \le B_n,
\ \forall\ i,j,\ \hbox{ with }\inf\emptyset = \infty.
\ees

\begin{corollary}\label{eg-1}
Suppose $X_i$ are independent. 
Let $T_n$ and $T_n^*$ be as in (\ref{T_n}) and (\ref{T_n^*})  
respectively
and $B_n$ be as in (E.1). 
\begin{enumerate}[label=({\roman*})]
\item Let $X_i^*$ be generated by the empirical bootstrap as in (\ref{empirical-bootstrap}).
Then, \eqref{th-empi-new-2} and \eqref{th-empi-new-3} hold with
\bes
\gamma^*_{{{\delta}}, M} = \max \bigg\{\bigg(\frac{(\log p)^2(\log(np/{{\delta}})^3}{n} \frac{M_4^4}{\linesigma^4}\bigg)^{1/6},
\bigg(\frac{(\log p)(\log(np/{{\delta}}))^4}{n}\bigg)^{1/2} \frac{B_n}{\linesigma} \bigg\}.
\ees

\item Let $X_i^*$ be generated by the wild bootstrap as in (\ref{wild}). Suppose the multipliers $W_i$ satisfy the moment condition (\ref{third-moment-match}) and the sub-Gaussian condition (\ref{subgaussian}) with a $\tau_0<\infty$. 
Then, \eqref{th-wild-new-2} and \eqref{th-wild-new-3} hold with 
\bes
\gamma^*_{{{\delta}}, M} = \max \bigg\{\bigg(\frac{(\log p)^2(\log(np)(\log(np/{{\delta}})^2}{n} \frac{M_4^4}{\linesigma^4}\bigg)^{1/6},
\bigg(\frac{(\log p)(\log (np)) (\log(np/{{\delta}}))^3}{n}\bigg)^{1/2} \frac{B_n}{\linesigma} \bigg\}.
\ees
\end{enumerate}
\end{corollary}

\begin{remark}
As $x^4\le 5\psi_1(x)$ for $x\ge 0$, we have $M_4^4 \le 5 B_n^4$, but $B_n/M_4$ could be unbounded.
We may compare the above result under (E.1) with \cite{chernozhukov2017central} for the maxima. 
For the empirical bootstrap, Propositions 2.1 and 4.3 of \cite{chernozhukov2017central} yields 
the following Kolmogorov-Smirnov distance bound: 
\bes
\sup_t \Big| \P\{T_n \le t\} - \P^*\{T_n^* \le t\} \Big| 
 \le C_{K}\big\{\Bbar_n^2 (\log (np))^7 /n\big\}^{1/6}
\ees
with probability at least $1- {{\delta}}$ when $\log(1/ {{\delta}}) \le K\log(np)$, where 
$\Bbar_n = \max\{M_4^2/\sigma_{(1)}^2,B_n/\sigma_{(1)}\}$ is a scale free version of 
their constant factor with the $B_n$ in (E.1) and $\sigma_{(1)}=\min_{j\le p}\sigma_j$. 
Corollary~\ref{eg-1}~(i) improves the rate of their upper bound 
by at least a factor of $\log^{1/3}(np)({\overline \sigma}/\sigma_{(1)})^{2/3}$. 
When $M_4/\sigma_{(1)} = O(1)$ and $B_n/\sigma_{(1)} \asymp n^{\kappa_0}$ with 
nontrivial $\kappa_0\in (0,1/2)$, the rate improvement is by at least the following factor of polynomial order, 
\bes
\min\Big\{n^{\kappa_0/3}\log^{1/3}(np), n^{(1-2\kappa_0)/3}(\log(np))^{7/6-5/2}\Big\}. 
\ees 
Similarly, for the Gaussian wild bootstrap, the combination of Proposition 2.1 and Corollary 4.2 of \cite{chernozhukov2017central} yields the following Kolmogorov-Smirnov distance bound: 
\bes
\sup_t \Big| \P\{T_n \le t\} - \P^*\{T_n^* \le t\} \Big| 
  \le  C_0 \big\{B_n^2\log^5(np) \log^2(np/ {{\delta}})/n\big\}^{1/6}
\ees
with probability at least $1- {{\delta}}$. 
With the third moment match in wild bootstrap, 
Corollary \ref{eg-1} (ii) improves upon their rate by at least a factor of 
$\log^{1/3}(np)({\overline \sigma}/\sigma_{(1)})^{2/3}$ 
in general, and by at least 
$\min\big\{n^{\kappa_0/3}$, $n^{(1-2\kappa_0)/3}\big\}$polylog$(np/{{\delta}})$ 
when $M_4/\sigma_{(1)} = O(1)$ and 
$B_n/\sigma_{(1)} \asymp n^{\kappa_0}$ with 
$\kappa_0\in (0,1/2)$.  
\end{remark}

We note that the product of sub-Gaussian variables satisfies the sub-exponential condition (E.1) imposed in Corollary \ref{eg-1}.
For example, for testing the equality of the population covariance
matrices of two samples $\{Y_i\}$ and $\{Z_i\}$ in $\R^d$, we just need to set
\bes
X_i = \hbox{\rm vec}\big(Y_iY_i^T - Z_iZ_i^T\Big)\ \hbox{ with }\ p=d(d+1)/2,
\ees
as in \cite{cai2013two} and \cite{chang2016comparing}.

{\medskip
{\it 2.3.2. Conditionally Gaussian vectors with Gaussian tail.}
Suppose
\bes
&& \hbox{$\sum_{i=1}^n X_i/\sqrt{n}$ is conditionally a Gaussian vector given a certain sigma field }
 {\cal A},
\cr \hbox{(E.2):} && \hbox{$\Big(\sum_{i=1}^n X_{i,j}/\sqrt{n}\Big)\Big|{\cal A} \sim N(\mu_j,\sigma_j^2)$,} 
\cr && \|X_{i,j}\|_{\psi_2} = \inf\Big\{B: \E\, \psi_2(|X_{i,j} - \E X_{i,j}|/B)\le 1\Big\} \le B_n,
\hbox{ with }\psi_2 = \exp(x^2)-1.
 \ees
Under (E.2), Theorem \ref{cb-rade} is applicable, and a corollary of it is stated as follows.

\begin{corollary}\label{eg-2}
Let $\bX=(X_1,\ldots, X_n)^T \in \R^{n\times p}$ be a random matrix with
independent rows $X_i \in \R^p$. 
Suppose $\E (X_i - \E X_i)^{\otimes m} = 0$ for  $m=3$ and $m=5$.
Let $X_i^*$ be generated by the Rademacher wild bootstrap, with
$\P\{W_i=\pm 1\}=1/2$ for the multiplier in (\ref{wild}).
Then, under (E.2), we have
\bes
&& \max \bigg\{\Big| \P\{T_n\le t^*_\alpha \} - (1-\alpha) \Big|, \bigg(\E \sup_{t\in \R}\Big|\P\Big\{T_n < t\Big\} - \P^*\Big\{T_n^* < t \Big\} \Big|^2\bigg)^{1/2} \bigg\}
\cr &\le&  C_0 \bigg[\bigg(\frac{\log p}{n^{1/4}}\bigg)^{4/7} \frac{M_6}{\linesigma}
+ \bigg(\frac{\log p}{n^{1/4}}\bigg)^{2}
\frac{B_n \sqrt{\log(np)}}{\linesigma \sqrt{\log p}}
\bigg], 
\ees
where $\linesigma$ is a constant upper bound 
for the soft minimum of $\{\sigma_1,\ldots,\sigma_p\}$ in (E.2) as in \eqref{sigmas}.
\end{corollary}

\medskip
{\it 2.3.3. Moment conditions.} 
Consider the following conditions on moments of the maxima,
\bes
\hbox{(E.3):}&& 
\M_q^q = \hbox{$n^{-1} \sum_{i=1}^n \E\, \max_{1\le j \le p} |X_{i,j} -\E X_{i,j}|^q$} \le B_n^q,
\\[3pt]
\hbox{(E.4):}&& \M_{4,2}^4 = \hbox{$\E\, \max_{1\le j \le p} \sum_{i=1}^n |X_{i,j}-\E X_{i,j}|^4/n$} \le B_n^4,
\\[3pt]
\hbox{(E.5):}&& \M_{6,2}^6 = \hbox{$\E\, \max_{1\le j \le p} \sum_{i=1}^n |X_{i,j}-\E X_{i,j}|^6/n$} \le B_n^6. 
\qquad\qquad\qquad\qquad
\qquad\qquad
\ees
Theorems \ref{cb-empi}, \ref{cb-wild} and \ref{cb-rade} respectively imply the following corollary.

\begin{corollary}\label{eg-3}
Suppose $X_i$ are independent.
Let $T_n$ and $T_n^*$ be as in (\ref{T_n}) and (\ref{T_n^*}) respectively. 
\begin{enumerate}[label=({\roman*})]
\item 
Let $X_i^*$ be generated by the empirical bootstrap as in (\ref{empirical-bootstrap}).
Then, under (E.3), \eqref{th-empi-new-2} and \eqref{th-empi-new-3} hold with constant $C_q$ and
\bes
\gamma^*_{{{\delta}}, M} = \max \bigg\{ \gamma^*_{{{\delta}}, M_4}, \, \, \gamma_1^*(B_n) \bigg[ 1 \vee \Big(  \frac{\gamma_1^*(B_n)}{{{\delta}}} \Big)^{1/q} \bigg] \bigg\},
\ees
where 
\bes
\gamma_1^*(B_n) &=& \bigg( \frac{(\log p)^{1/2}(\log (np/{{\delta}}))}{n^{1/2-1/q}}\frac{B_n}{\linesigma} \bigg)^{q/(q+1)}.
\ees
Moreover, if (E.3) holds with $q=4$, then $\gamma^*_{{{\delta}}, \M_4} = \gamma^*_{{{\delta}}, B_n}$ in \eqref{th-empi-new-2} and \eqref{th-empi-new-3}.

\item Let $X_i^*$ be generated by the
wild bootstrap as in (\ref{wild}) with multipliers satisfying the moment condition
(\ref{third-moment-match}) and the sub-Gaussian condition (\ref{subgaussian}) 
with a certain $\tau_0<\infty$. 
Then, under (E.3), \eqref{th-wild-new-2} and \eqref{th-wild-new-3} hold  with constant $C_{\tau_0, q} $ and \bes
\gamma^*_{{{\delta}}, M} = \max \bigg\{ \gamma^*_{{{\delta}}, M_4}, \,\, \gamma_2^*(B_n) \bigg[ 1 \vee \Big(  \frac{\gamma_2^*(B_n)}{{{\delta}}} \Big)^{1/q} \bigg] \bigg\},
\ees
where 
\bes
\gamma_2^*(B_n) &=& \bigg( \frac{(\log p)^{1/2}(\log (np))^{1/2}(\log (np/{{\delta}}))^{1/2}}{n^{1/2-1/q}}\frac{B_n}{\linesigma} \bigg)^{q/(q+1)}.
\ees
However, under (E.4), \eqref{th-wild-new-2} and \eqref{th-wild-new-3} hold with $\gamma^*_{{{\delta}}, \M_{4,2}} = \gamma^*_{{{\delta}}, B_n}$.

\item Suppose that $\bX$ satisfies the conditions of Theorem \ref{cb-rade} 
and (E.5), and that $\sum_{i=1}^nX_i/\sqrt{n}$ satisfies the conditional Gaussian condition 
in (E.2) with a constant lower bound $\linesigma$ for the soft minimum of 
the conditional standard deviation as in \eqref{sigmas}. 
Let $X_i^*$ be generated by the Rademacher wild bootstrap as in Theorem \ref{cb-rade}. Then,
\bes
&& \Big| \P\{T_n\le t^*_\alpha \} - (1-\alpha) \Big|
+ \bigg(\E \sup_{t\in \R}\Big|\P\Big\{T_n < t\Big\} - \P^*\Big\{T_n^* < t \Big\} \Big|^2\bigg)^{1/2}
\cr &\le& C_0 \bigg[\bigg(\frac{\log p}{n^{1/4}}\bigg)^{4/7}
\frac{B_n\sqrt{\log(np)}}{\linesigma \sqrt{\log p}}+ \bigg(\frac{\log p}{n^{1/4}}\bigg)^{32/21}\bigg].
\ees
\end{enumerate}
\end{corollary}

\begin{remark} 
We compare the above result under (E.3) with \cite{chernozhukov2017central} for the maxima. 
For the empirical bootstrap, 
Corollary \ref{eg-3} (i) implies with at least probability $1-{{\delta}}$, 
the Kolmogorov-Smirnov distance in \eqref{th-empi-new-2}
is bounded by
\begin{align}\label{eg3-explicit-1}
C_q \max\bigg\{\bigg(\frac{(\log p)^2 (\log(np))^3}{n} \frac{M_4^4}{\linesigma^4} \bigg)^{1/6}, \bigg( \frac{(\log p)(\log (np))^2}{n^{1-2/q}}\frac{B_n^2}{\linesigma^2} \bigg)^{\frac{q}{2(q+1)}} \bigg\}
\end{align}
when ${{\delta}}$ is greater than the second component, and
\begin{align}\label{eg3-explicit-2}
C_q \max\bigg\{\bigg(\frac{(\log p)^2 (\log(np))^3}{n} \frac{M_4^4}{\linesigma^4} \bigg)^{1/6}, \bigg( \frac{(\log p)(\log (np))^2}{n^{1-2/q}{{\delta}}^{2/q}}\frac{B_n^2}{\linesigma^2} \bigg)^{1/2} \bigg\}
\end{align}
when ${{\delta}}$ is smaller. Note that $\log(np/{{\delta}}) \asymp \log(np)$ as otherwise ${{\delta}}$ is extremely small so that the second bound is effective but also trivial due to small 
$n^{1-2/q}{{\delta}}^{2/q}$. For the third-moment match wild bootstrap, 
(ii) yields a slightly better result but the above bounds in 
\eqref{eg3-explicit-1} and \eqref{eg3-explicit-2} also apply. 
In \cite{chernozhukov2017central}, the combination of Propositions 2.1 and 4.3 for the empirical bootstrap and the combination of Proposition 2.1 and Corollary 4.2 for the Gaussian wild bootstrap yield the Kolmogorov-Smirnov distance bound as 
\bes
\sup_t \Big| \P\{T_n \le t\} - \P^*\{T_n^* \le t\} \Big| 
 \le C_{q,K} \max\bigg\{\Big(\frac{\Bbar_n^2(\log (np))^7}
{n} \Big)^{1/6}, \Big(\frac{\Bbar_n^2(\log (np))^3}{n^{1-2/q}{{\delta}}^{2/q}}\Big)^{1/3} \bigg\},
\ees
with at least probability $1- {{\delta}}$, 
where $\Bbar_n = \max\{M_4^2/\sigma_{(1)}^2,B_n/\sigma_{(1)}\}$ 
with the $B_n$ in (E.3) and $\sigma_{(1)}=\min_j\sigma_j$. 
It's clear that the first component of the bound in \eqref{eg3-explicit-1} or \eqref{eg3-explicit-2} improves the first rate above by at least a factor of $(\linesigma/\sigma_{(1)}\log(np))^{1/3}$. As $q/(2(q+1)) > 1/3$ for all $q>2$ and the bounds are trivial when $q \le 2$, the second components in \eqref{eg3-explicit-1} and \eqref{eg3-explicit-2} improves the second rate above by at least a factor of
\bes
\bigg( \frac{n^{1-2/q}}{(\log p)(\log (np))^2} \frac{\linesigma^2}{B_n^2} \bigg)^{\frac{q}{2(q+1)} -\frac{1}{3}} \hbox{ for } q>2.
\ees
\end{remark}

In linear regression, we observe $y_i = Z_i^T\beta + \veps_i$.
Suppose the design vectors are deterministic and normalized to
$\sum_{i=1}^n Z_{i,j}^2=n$. Suppose we want to control the spurious correlation
in sure screening based on $\sum_{i=1}^n y_iZ_i/\sqrt{n}$ as in \cite{fanlv07} and \cite{fan2016guarding}.
Let $X_i = y_iZ_i$. We have $X_i - \E X_i = \veps_i Z_i$ and
\bes
T_n = \bigg\| \sum_{i=1}^n y_iZ_i/\sqrt{n} - \E\sum_{i=1}^n y_iZ_i/\sqrt{n}\bigg\|_\infty.
\ees
Suppose $\E \veps_i =0$ and $\E\veps_i^2=\sigma^2$.
For $1\le q\le \infty$ define
\bes
\bigg(\frac{1}{n}\sum_{i=1}^n \E|\veps_i|^{q}\bigg)^{1/q} \le M_{\veps,q},\qquad
\max_{j\le p}\bigg(\frac{1}{n}\sum_{i=1}^n |Z_{i,j}|^{q}\bigg)^{1/q}  \le M_{Z,q}.
\ees
Then, conditions (E.3) with $q=4$, (E.4) and (E.5) can be fulfilled with
\bes
\M_4 \le M_{\veps,4}M_{Z,\infty},\ \M_{m,2} \le M_{\veps,mq}M_{Z,mq/(1-q)},\ 1\le q\le\infty,
\ees
where $m=4,6$ in (E.4), (E.5) respectively.
\cite{dezeure2016high} studied bootstrap simultaneous inference in high-dimensional linear regression
under the sample size condition $n\ge (\log p)^7+ s^2(\log p)^3$ and the moment
condition $M_{\veps,4} + M_{Z,\infty} =O(1)$.

\medskip
{\bf 2.4. L\'evy-Prokhorov pre-distance and anti-concentration.}
The Kolmogorov-Smirnov distance between two distribution functions
can be bounded from the above by a sum of upper bounds for their L\'evy-Prokhorov distance
and the minimum of their modulus of continuity.
For two random elements $T_n$ and $T_n^*$ living in a common metric space equipped with
a probability measure $\P$, the L\'evy-Prokhorov distance is the smallest $\eps>0$ satisfying
\begin{align}\label{Prokhorov-0}
\max\Big[\P\Big\{T_n \in A \Big\} - \P\Big\{T_n^* \in A(\eps) \Big\},
\P\Big\{T_n^*\in A\Big\} - \P\Big\{T_n \in A(\eps) \Big\}\Big] \le \eps
\end{align}
for all Borel sets $A$, where $A(\eps)=\{y: \min_{x\in A} d(x,y)<\eps\}$
is the $\eps$-neighborhood of $A$.
For comparison of the distributions of two maxima $T_n$ and $T_n^*$ for simultaneous testing, it is typically
sufficient to consider one-sided intervals $A = (\infty,t]$ in (\ref{Prokhorov-0}).
Choosing $A = (\infty,t]$ is also sufficient for studying the Kolmogorov-Smirnov distance
between the distribution functions of $T_n$ and $T_n^*$.
Thus, our analysis focuses on the following quantity
\begin{align}\label{Prokhorov}
\eta_n(\eps) \eqeq \eta_n^{{(\P)}}\big(\eps;T_n,T_n^*\big)
= \sup_{t\in \R}\eta_n^{{(\P)}}\big(\eps,t;T_n,T_n^*\big)
\end{align}
with $\eta_n^{{(\P)}}\big(\eps,t;T_n,T_n^*\big) = \max\big[\P\big\{T_n\le t -\eps  \big\} - \P\big\{T_n^* < t \big\},
\P\big\{T_n^*\le t - \eps\big\} - \P\big\{T_n < t \big\},0\big]$.
As the L\'evy-Prokhorov distance over all one-sided intervals is the smallest $\eps$ satisfying
$\eta_n(\eps) \le \eps$, we refer to the quantity $\eta_n(\eps)$ as
L\'evy-Prokhorov pre-distance for convenience.
It does not define a distance between $T_n$ and $T_n^*$, but
satisfies a ``pseudo-triangular inequality" in the sense of
\begin{align}\label{triangle}
\eta_n^{{(\P)}}\big(\eps;T_n,T_n^*\big)
\le \eta_n^{{(\P)}}\big(\eps_1;T_n,\Ttil_n\big) + \eta_n^{{(\P)}}\big(\eps_2;\Ttil_n,T_n^*\big),\quad
\forall\ \Ttil_n,\, \eps_1+\eps_2<\eps,\, \eps_1\wedge\eps_2>0.
\end{align}

It is straightforward by the triangle inequality that the Kolmogorov-Smirnov distance between the
cumulative distribution functions of $T_n$ and $T_n^*$, equal to $\eta_n(0+)$, is bounded by
\begin{align}\label{KS-bound}
\sup_{t\in \R}\Big|\P\Big\{T_n < t\Big\} - \P\Big\{T_n^* < t \Big\} \Big|
 = \eta_n(0+) 
\le \eta_n(\eps) + \min\Big\{\omega_n(\eps;T_n),\omega_n(\eps;T_n^*)\Big\},\ \forall\,\eps>0, 
\end{align}
where $\omega_n(\eps;T_n) = \omega_n^{{(\P)}}(\eps;T_n) = \sup_{t\in \R} \P\{t-\eps < T_n < t\}$
and $\omega_n(\eps;T_n^*) = \omega_n^{{(\P)}}(\eps;T_n^*)$ is defined in the same way with
$T_n$ replaced by $T_n^*$.
The quantity $\omega_n(\eps;T_n)$, which is also called the L\'evy concentration function,
is the modulus of continuity of the cumulative distribution function of $T_n$.

The L\'evy-Prokhorov pre-distance characterizes the convergence in distribution. 
When $T_n$ has a fixed distribution function $H_0$, $T_n^*$ converges in distribution to $H_0$ if and only if $\eta_n(\eps)\to 0\,\forall\,\eps>0$. On the other hand, $\lim_{\eps\to 0+}\omega_n(\eps;T_n) = 0$ 
if and only if $H_0$ is continuous. Of course, if $T_n^*$ converges in distribution to a continuous $H_0$, then the distribution function of $T_n^*$ converges to $H_0$ in the Kolmogorov-Smirnov distance.  
Moreover, as $\eta_n(\eps)$ is decreasing in $\eps$, the condition 
$\eta_n(\eps)\to 0\, \forall \eps>0$ is necessary for the convergence 
$\eta_n(0+)\to 0$ in the Kolmogorov-Smirnov distance. 

Inequality (\ref{KS-bound}) asserts that the Kolmogorov-Smirnov distance is bounded by a sum of two quantities, the L\'evy-Prokhorov pre-distance which allows a shift $\eps$ in the comparison of two distribution functions and the L\'evy concentration as an upper bound for the error introduced by the shift. 
By allowing a shift, the L\'evy-Prokhorov pre-distance can be further bounded by comparison of 
the expectations of smooth functions of $T_n$ and $T_n^*$ so that the Lindeberg interpolation can 
be applied as discussed in detail in Section 3. 
Upper bounds for the L\'evy concentration, called the anti-concentration inequality, will be discussed 
in Section 4. The role of (\ref{KS-bound}) is to explicitly spell out the roles of the comparison and 
anti-concentration theorems and to facilitate the notation in our analysis. 
We note that $\eta_n(\eps)$ is decreasing but $\min\big\{\omega_n(\eps;T_n),\omega_n(\eps;T_n^*)\big\}$ is increasing in $\eps$. 
In our analysis, we pick an $\eps = 1/b_n$ to balance the rate of the two terms in (\ref{KS-bound}). 
For example, as $\omega_n(1/b_n;T_n)\lesssim b_n^{-1}\sqrt{\log p}$ by 
Theorem \ref{anti-general} in Section 4, $b_n^{-1} \asymp ((\log p)^2/n)^{1/6}$ 
is used to achieve the rate $((\log p)^5/n)^{1/6}$ in Theorems \ref{cb-empi} and \ref{cb-wild}.

In bootstrap, we are interested in approximating the distribution of 
$T_n$ under the marginal probability $\P$
by the distribution of the bootstrap $T^*_n$ under the conditional probability $\P^*$ given 
the original data. 
To streamline the notation, we write this comparison under a common probability measure 
by introducing a copy $T_n^0$ of $T_n$ independent of the original data $\bX$, so that 
$\P\big\{T_n\le t\} = \P\big\{T_n^0\le t|\bX\} = \P^*\big\{T_n^0 \le t\}$. This allows us to write 
\begin{align*}
\eta_n^{{(\P^*)}}\big(\eps,t;T^0_n,T_n^*\big) 
=& \, \max\big[\P^*\big\{T_n^0\le t -\eps  \big\} - \P^*\big\{T_n^* < t \big\},
\P^*\big\{T_n^*\le t - \eps\big\} - \P^*\big\{T_n^0 < t \big\},0\big]
\\
=& \, \max\big[\P\big\{T_n\le t -\eps  \big\} - \P^*\big\{T_n^* < t \big\},
\P^*\big\{T_n^*\le t - \eps\big\} - \P\big\{T_n < t \big\},0\big].
\end{align*}
The following lemma connects the consistency of bootstrap to the tail probability of
the random L\'evy-Prokhorov pre-distance under $\P^*$ 
and L\'evy concentration function $\omega_n(\eps;T_n)$.

\begin{lemma}\label{lm:cp-decomp}
Let $t^*_\alpha$ be the $(1-\alpha)$-quantile of $T_n^*$ under $\P^*$. Then,
for all $\eps_n>0$ and $\eta>0$,
\bes
\Big| \P\{T_n\le t^*_\alpha \} - (1-\alpha) \Big|  
\le \sup_t \P\Big\{ \eta_n^{{(\P^*)}}(\eps_n, t ; T_n^0, T_n^*) > \eta \Big\} +\eta  + \omega_n(\eps_n;T_n), 
\ees
and the Kolmogorov-Smirnov distance between 
$\P\big\{T_n\le t\big\}$ and $\P^*\big\{T_n^* < t \big\}$ is bounded by 
\bes 
\sup_{t\in \R}\Big|\P\Big\{T_n < t\Big\} - \P^*\Big\{T_n^* < t \Big\} \Big| 
\le \eta + \omega_n^{(\P)}(\eps_n;T_n)
\ees
when $\eta_n^*(\eps_n) \le \eta$, 
where $\eta_n^*(\eps) \eqeq \eta_n^{{(\P^*)}}(\eps;T_n^0,T_n^*) = \sup_{t\in \R}\eta_n^{{(\P^*)}}(\eps,t;T_n^0,T_n^*)$.
\end{lemma}

We derive in the next two sections upper bounds for the L\'evy-Prokhorov pre-distances $\eta_n(\eps)$ and $\eta_n^*(\eps)$ and 
the L\'evy concentration function $\omega_n(\eps;T_n)$ respectively.

\medskip
{\bf 3. Comparison theorems.}
Let $h_0$ be a smooth decreasing function taking value 1 in $(-\infty,-1]$ and
0 in $[0,\infty)$.
As we will explicitly explain at the beginning of the proof of Theorem \ref{th-comp-gene},
it follows directly from the definition of
the L\'evy-Prokhorov pre-distance in (\ref{Prokhorov}) that
\bes
\eta_n(1/b_n) \le \sup_{t\in \R} \Big| \E h_t\big(b_nT_n\big) - \E h_t\big(b_nT_n^*\big)\Big|,\ \ \forall\ b_n>0,
\ees
where $h_t(\cdot) = h_0(\cdot - t)$ is the location shift of $h_0$.
In this section we develop comparison theorems which provide expansions and bounds for
\bes
\E f(X_1,\ldots,X_n) - \E^* f(X_1^*,\ldots,X_n^*)
\ees
in terms of average moments of $\{X_i,i\le n\}$ and $\{X_i^*, i\le n\}$.
Here $f(x_1,\ldots,x_n)$ is a smooth function of $n$ vectors $x_i\in\R^p$
and $\E$ and $\E^*$ may represent two arbitrary measures.
The bootstrap is treated as a special case where $\E^*$ is the conditional expectation
given $\bX$ under $\E$.

To make a connection between quantities of the form $\E\, h_t\big(b_nT_n\big)$, which is Lipschitz smooth
in $X_i$ at the best, and $\E f(X_1,\ldots,X_n)$, which is required to be more smooth in our analysis,
we approximate the maximum function
$T_n = \max_j \sum_{i=1}^n X_{i,j}/\sqrt{n}$ of $\{X_i\}$ by the
smooth max function $F_\beta(Z_n)$ as in \cite{chernozhukov2013},
where $Z_n = (X_1+\cdots+X_n)/n^{1/2}$ and
\begin{align}\label{smooth-max}
F_\beta(z)= \frac{1}{\beta} \log \bigg(\sum_{j=1}^p e^{\beta z_j} \bigg),\quad
\forall\ z = (z_1,\ldots,z_p)^T.
\end{align}
For $\beta>0$, the function $F_\beta(z)$ is infinitely differentiable and
\bes
\max(z_1,\ldots,z_p) \le F_\beta(z) \le \max(z_1,\ldots,z_p)  + \beta^{-1}\log p.
\ees
It follows that, cf. Proof of Theorem \ref{th-comp-gene} in the Appendix, for $\beta_n = 2b_n\log p$,
\begin{align}\label{F_beta-approx}
\eta_n(1/b_n) \le \sup_{t\in \R}
\Big| \E h_t\big(2b_nF_{\beta_n}(Z_n)\big) - \E h_t\big(2b_nF_{\beta_n}(Z_n^*)\big)\Big|,
\end{align}
where $Z_n^* = (X_1^*+\cdots+X_n^*)/n^{1/2}$.
In the Appendix, we provide upper bounds for the derivatives of $F_\beta(z)$ and
$f = h\circ(b_nF_\beta)$  via the Faa di Bruno formula.

We shall put $\bX$ and $\bX^*$ in the same probability space to better present our analysis.
For this purpose, we use slightly different notation between
the general and bootstrap cases.
In the general case where both $\E$ and $\E^*$ are treated as deterministic,
the problem does not involve the joint distribution between $\{X_i\}$ and $\{X_i^*\}$.
This allows us to assume without loss of generality that
$(X_i, X_i^*)\in \R^{p\times 2}$, $1\le i\le n$, are independent matrices under $\E$,
so that the problem concerns
\bes
\Delta_n(f) = \E\Big\{ f(X_1,\ldots,X_n) - f(X_1^*,\ldots,X_n^*)\Big\}.
\ees
In the bootstrap case, $\E^*$ is the conditional expectation given $\bX$ and we consider
\bel{Delta^*}
\Delta_n^*(f) = \E^*\Big\{ f(X_1^0,\ldots,X_n^0) - f(X_1^*,\ldots,X_n^*)\Big\}
= \E f(X_1,\ldots,X_n) - \E^* f(X_1^*,\ldots,X_n^*)
\eel
where $\bX^0=(X_1^0,\ldots,X_n^0)^T$ is an independent copy of $\bX$.
As $(X_i^0, X_i^*)$ are still independent random matrices under $\E^*$, we can conveniently write
the mean squared approximation error as
\bes
\E\bigg[\E^*\Big\{ f(X_1^0,\ldots,X_n^0) - f(X_1^*,\ldots,X_n^*)\Big\}\bigg]^2.
\ees
In either cases, we assume throughout this section that $\E X_i = \E^* X_i^* = 0$, so that
the average centered moments are
\begin{align}\label{moments}
\mu^{{(m)}} = \frac{1}{n}\sum_{i=1}^n \E X_i^{\otimes m},\quad
\nu^{{(m)}} = \frac{1}{n}\sum_{i=1}^n \E^*\big(X_i^*\big)^{\otimes m}.
\end{align}

We consider in separate sections the Lindeberg method and comparison bounds
for two general measures, the maxima, the empirical bootstrap, and the wild bootstrap.

\medskip
{\bf 3.1. A coherent Lindeberg interpolation.}
Let $(X_i,X_i^*)\in \R^{p\times 2}$ be independent random matrices under $\E$,
$\bU_{i} = (X_1,\ldots,X_{{i}-1},0,X^*_{{i}+1},\ldots,X^*_n)$, and
$\bV_{i} = (X_1,\ldots,X_{i},X^*_{{i}+1},\ldots,X^*_n)$.
The original Lindeberg (1922) proof of the central limit theorem begins with the decomposition
\bes
\Delta_n(f) = \E\Big\{f(\bV_n) - f(\bV_0)\Big\}
= \sum_{{i}=1}^n \E \Big\{f(\bV_{i}) - f(\bV_{{i}-1})\Big\},
\ees
followed by a Taylor expansion of the increments
$f(\bV_{i}) - f(\bV_{{i}-1})$ at $\bU_{i}$, so that
\begin{align}\label{Lindeberg-1}
\Delta_n(f)
=\sum_{m=1}^{m^*-1}\Delta_{n,m} + \Rem,\quad
\Delta_{n,m} = \frac{1}{m!}\sum_{{i}=1}^n \Big\langle \E {f}^{(m)}_i(\bU_{i}),
\E\,X_{i}^{\otimes m} - \E(X_{i}^*)^{\otimes m}\Big\rangle,
\end{align}
where ${f}^{(m)}_i(x_1,\ldots,x_n) =(\pa/\pa x_{i})^{\otimes m}f(x_1,\ldots,x_n)$.
To prove the central limit theorem, Lindeberg (1922) took $m^*=3$ and Gaussian $X_i^*$
with the same first two moments as $X_i$, so that $\Delta_n(f) =\Rem$.
In this approach, $f(\bV_{i})$ can be viewed as an interpolation between $f(\bV_n) = f(\bX)$
and $f(\bV_0)= f(\bX^*)$.
The ideal has found much broader applications recently;
See for example \cite{chatterjee2006generalization}.
However, the decomposition (\ref{Lindeberg-1}) may not yield the best bounds for
$\Delta_n(f)$ when $\E\,X_{i}^{\otimes m} - \E(X_{i}^*)^{\otimes m}$ are heterogeneous,
for example in the case of the empirical bootstrap with heteroscedastic $X_i$.

We further develop the Lindeberg approach (\ref{Lindeberg-1}) as follows to bound the quantity
$\Delta_n(f)$ in terms of the difference of the average moments of $\{X_i\}$ and $\{X_i^*\}$,
\begin{align}\label{diff-aver}
\frac{1}{n}\sum_{i=1}^n\E\,X_{i}^{\otimes m} - \frac{1}{n}\sum_{i=1}^n\E(X_{i}^*)^{\otimes m},
\end{align}
instead of the difference in the moments of individual $X_i$ and $X_i^*$ as in a direct application
of (\ref{Lindeberg-1}).
This improvement, which can be viewed as a ``coherent" Lindeberg interpolation and
facilitates our analyses of the bootstrap for the maxima of the sums of $X_i$,
is achieved by taking the average of the Lindeberg interpolation
over all permutations of the index $i$.

Consider permutation invariant functions $f(x_1,\ldots,x_n)$ of $x_i\in \R^p, 1\le i\le n$, satisfying
\bes
f(x_1,\ldots,x_n) = f\big(x_{\sigma_1},\ldots,x_{\sigma_n}\big)
\ees
for all permutations $\sigma = (\sigma_1,\ldots,\sigma_n)$ of $\{1,\ldots,n\}$.
While $\Delta_n(f)$ of (\ref{Lindeberg-1}) is invariant in the permutation $\sigma$,
the individuals components $\Delta_{n,m}$ and the remainder term on the right-hand side are not.
Thus, the worst scenario bounds for $|\Delta_{n,m}|$ and $|\Rem|$ may not yield
optimal results compared with the coherent Lindeberg interpolation,
which we formally describe as follows.

Suppose $\E X_i =\E X_i^*=0$.
For permutations $\sigma = (\sigma_1,\ldots,\sigma_n)$ of $\{1,\ldots,n\}$, let
\bes
\bU_{\sigma,{k}} = \big(X_{\sigma_1},\ldots,X_{\sigma_{k-1}},X^*_{\sigma_{k+1}},\ldots,X^*_{\sigma_n}\big).
\ees
As $\Delta_n(f)$ invariant under permutation of the index $i$, 
for each permutation $\sigma$ (\ref{Lindeberg-1}) yields
\bes
\Delta_n(f) = \sum_{m=2}^{m^*-1}\Delta_{n,m,\sigma} + \Rem_\sigma,
\ees
with $\Delta_{n,m,\sigma} = (m!)^{-1}\sum_{k=1}^n \big\langle
\E {f}^{(m)}_{\sigma_k}(\bU_{\sigma,{k}},0),\,
\E\,X_{\sigma_k}^{\otimes m} - \E(X_{\sigma_k}^*)^{\otimes m}\big\rangle$.
This leads to the expansion
\begin{align}\label{Lindeberg-2}
\Delta_n(f) = \E\Big\{f(X_1,\ldots,X_n) - f(X_1^*,\ldots,X_n^*) \Big\}
= \sum_{m=2}^{m^*-1}\A_\sigma\big(\Delta_{n,m,\sigma}\big) + \A_\sigma\big(\Rem_\sigma\big),
\end{align}
where $\A_\sigma$ is the operator of averaging over all permutations $\sigma$ of $\{1,\ldots,n\}$.
The expansion in (\ref{Lindeberg-2}) can be viewed as a coherent version of the original
one in (\ref{Lindeberg-1}) as the fluctuation with respect to the choice of $\sigma$ is
removed by taking average over all permutations.
The following lemma will be used to approximate $\A_\sigma(\Delta_{n,m,\sigma})$ and
$\A_\sigma(\Rem_\sigma)$
by quantities of the same form with the difference of the average moments (\ref{diff-aver})
in place of $\E X_{i}^{\otimes m} - \E(X_i^*)^{\otimes m}$.  Define
\bes
\zeta_{{k},{i}} = \delta_{k}X_{i} + (1-\delta_{k})X_{i}^*,
\ees
where $\{\delta_k\}$ are Bernoulli variables independent of $\{X_i,X_i^*, i\le n\}$ under $\E$
with $\P\{\delta_{k}=1\} = {k}/(n+1)$.
Let $\A_{\sigma,k}$ be the operator of taking the average over all permutations $\sigma$ and
all $k=1,\ldots,n$ and the expectation with respect to $\delta_k$, conditionally on $\{X_i,X_i^*, i\le n\}$,
\begin{align}\label{A_sigma,k}
& \A_{\sigma,k}{h}(\sigma,k,\zeta_{k,\sigma_k},\bX,\bX^*)
\\ 
\nonumber =&\ \frac{1}{n}\sum_{k=1}^n \frac{1}{n!}\sum_\sigma
\bigg\{\frac{k {h}(\sigma,k,X_{\sigma_k},\bX_{(\sigma)},\bX^*_{(\sigma)})}{n+1} +
\frac{(n+1-k) {h}(\sigma,k,X_{\sigma_k}^*,\bX_{(\sigma)},\bX^*_{(\sigma)})}{n+1}\bigg\},
\end{align}
for all Borel functions $h$, where $\bX_{(\sigma)}$ is the permutation over rows of $\bX$.

\begin{lemma}\label{k-nonspecific}
For all permutation invariant functions $f(x_1,\ldots,x_n)$,
\bes
\A_{\sigma,k}\Big(I_{\{\sigma_k=i\}}f(\bU_{\sigma,{k}},\zeta_{{k},{i}})\Big)
\ees
does not depend on ${i}$.
Consequently, for any function $g_i(\cdot,\cdot)$, $1\le i\le n$,
\bes
\A_{\sigma,k}\Big\langle f(\bU_{\sigma,{k}},\zeta_{{k},{\sigma_k}}), g_{\sigma_k}(\bX,\bX^*)\Big\rangle
= \bigg\langle \A_{\sigma,k}\Big(f(\bU_{\sigma,{k}},\zeta_{{k},{\sigma_k}})\Big),
\frac{1}{n}\sum_{i=1}^n g_{i}(\bX,\bX^*)\bigg\rangle.
\ees
\end{lemma}

Consider smooth functions with slightly stronger permutation invariance properties.
Suppose that for certain permutation invariant functions ${f}^{(m,0)}(x_1,\ldots,x_n)$,
\begin{align}\label{cond-perm-1}
{f}^{(m)}_n(x_1,\ldots,x_{n-1},0) = {f}^{(m,0)}(x_1,\ldots,x_{n-1},0),\ m=0,2,\ldots,m^*-1,
\end{align}
where ${f}^{(m)}_n(x_1,\ldots,x_n) = (\pa/\pa x_n)^{\otimes m} f(x_1,\ldots,x_n)$ is
as in (\ref{Lindeberg-1}).
Such ${f}^{(m,0)}$ exist if $f(x_1,\ldots,x_n) = f_0(x_1,\ldots,x_n,0)$
for a permutation invariant $f_0(x_1,\ldots,x_n,x_{n+1})$ involving $n+1$ vectors,
e.g. a function of the sum $x_1+\cdots+x_n$. In this case, we may pick
\bes
{f}^{(m,0)}(x_1,\ldots,x_n)
= (\pa/\pa x_{n+1})^{\otimes m} f_0(x_1,\ldots,x_n,x_{n+1})\big|_{x_{n+1}=0}.
\ees

It follows from (\ref{Lindeberg-2}), Lemma \ref{k-nonspecific} and (\ref{cond-perm-1}) that
\begin{align}\label{approx-Delta_{n,m}}
\A_\sigma\big(\Delta_{n,m,\sigma}\big)
=&\ n\,\A_{\sigma,k}\Big((m!)^{-1}\Big\langle \E {f}^{(m,0)}(\bU_{\sigma,{k}},0),
\E\,X_{\sigma_k}^{\otimes m} - \E(X_{\sigma_k}^*)^{\otimes m}\Big\rangle\Big)
\\
\approx&\nonumber \ n\,\A_{\sigma,k}\Big((m!)^{-1}\Big\langle \E {f}^{(m,0)}(\bU_{\sigma,{k}},\zeta_{k,\sigma_k}),
\E\,X_{\sigma_k}^{\otimes m} - \E(X_{\sigma_k}^*)^{\otimes m}\Big\rangle\Big)
\\ 
\nonumber = & \ \bigg\langle \frac{n}{m!}
\A_{\sigma,k}\Big(\E {f}^{(m,0)}(\bU_{\sigma,{k}},\zeta_{k,\sigma_k})\Big),
\frac{1}{n}\sum_{i=1}^n\E\,X_{i}^{\otimes m} - \frac{1}{n}\sum_{i=1}^n\E(X_{i}^*)^{\otimes m}\bigg\rangle,
\end{align}
so that $\A_\sigma\big(\Delta_{n,m,\sigma}\big)$ is small when the average moments between $\{X_i\}$
and $\{X_i^*\}$ are close to each other.
Interestingly, a combination of Slepian's (1962)
smart path interpolation and Stein's (1981)
leave-one-out method also allows comparison of the average of the second moment,
but not the third moment and beyond.
The Edgeworth expansion, a classical tool for high-order analysis of the bootstrap,
is not available in our analysis as we are interested in the regime where the Gaussian
approximation may fail to begin with.

\medskip
{\bf 3.2. A general comparison theorem.}
In this subsection, we present upper bounds for the absolute value of
$\Delta_n(f)$ in (\ref{Lindeberg-2})
for smooth permutation invariant functions $f(x_1,\ldots,x_n)$ in a general setting, where
$(X_i, X_i^*)\in \R^{p\times 2}$, $1\le i\le n$, are assumed to be independent random
matrices under $\E$. Conditions up to the $m^*$-th moment will be imposed,
e.g. $m^*=4$ in (\ref{Lindeberg-2}).

In addition to invariance condition (\ref{cond-perm-1}),
we assume the following stability condition on derivatives of order $m^*$.
For integers $m_1\ge 2$ and $m_2\ge 0$ with $m_1+m_2\le m^*$, define
\bes
{f}^{(m_1,m_2)}(x_1,\ldots,x_{n-1},x_n)
= \big((\pa/\pa x_n)^{\otimes m_2}\big)\otimes f^{(m_1,0)}(x_1,\ldots,x_{n-1},x_n).
\ees
Here $\big((\pa/\pa x_n)^{\otimes m_2}\big)\otimes f^{(m_1,0)}$,
a product of two tensors, is treated
as an $m=m_1+m_2$ dimensional tensor with elements
$(\pa/\pa x_{n,j_1})\cdots(\pa/\pa x_{n,j_{m_2}})f^{(m_1,0)}_{j_{m_2+1},\ldots,j_{m_2+m_1}}$.
Suppose that for $m_1\ge 2$ and $m_2 = m^*-m_1$, e.g. $(m_1,m_2) = (2,2)$ or $(3,1)$ for $m^*=4$,
\begin{align}\label{cond-1a}
\P\left\{\begin{matrix}
\Big|{f}^{(m_1,m_2)}_{j_1,\ldots,j_{m^*}}(x_1,\ldots,x_{n-1}, t\xi_i)\Big| \le g(\|\xi_i\|/u_n)
\fbar^{{(m^*)}}_{j_1,\ldots,j_{m^*}}(x_1,\ldots,x_{n-1},0),
\cr \Big|{f}^{({m^*})}_{j_1,\ldots,j_{m^*}}(x_1,\ldots,x_{n-1}, t\xi_i)\Big| \le g(\|\xi_i\|/u_n)
\fbar^{{(m^*)}}_{j_1,\ldots,j_{m^*}}(x_1,\ldots,x_{n-1},0)\end{matrix}\right\}=1
\end{align}
for all $0\le t\le 1$ and $1\le i\le n$, where $\xi_i$ is either $X_i$ or $X_i^*$.
Suppose further that for some permutation invariant $f_{\max}^{{(m^*)}}(x_1,\ldots,x_n)$,
\begin{align}\label{cond-1b}
\P\Big\{ \fbar^{{(m^*)}}_{j_1,\ldots,j_{m^*}}(x_1,\ldots,x_{n-1},0) \le g(\|\xi_i\|/u_n)
\Big({f}^{({m^*})}_{\max}(x_1,\ldots,x_{n-1},\xi_i)\Big)_{j_1,\ldots,j_{m^*}}\Big\}=1
\end{align}
for the same $\xi_i$.
Define $G_k = \big(\E \big\{1/g\big(\|X_k\|/u_n\big)\big\}\big)\wedge
\big(\E \big\{1/g\big(\|X_k^*\|/u_n\big)\big\}\big)$ and
\begin{align}\label{mu_max}
\mu^{{(m)}}_{\max} =&\
\bigg(\bigg[\max\bigg\{\sum_{k=1}^n \frac{\E |X_k|^m g\big(\|X_k\|/u_n\big)}{nG_k},
\sum_{k=1}^n \frac{\E |X_k^*|^m g\big(\|X_k^*\|/u_n\big)}{nG_k},
\\ 
\nonumber &\ \qquad\qquad \sum_{k=1}^n \frac{\E |X_k|^m \,\E\, g\big(\|X_k^*\|/u_n\big)}{nG_k},
\sum_{k=1}^n \frac{\E |X_k^*|^m\,\E\, g\big(\|X_k\|/u_n\big)}{nG_k}\bigg\}
\bigg]^{1/m}\bigg)^{\otimes m}.
\end{align}
When $g(t)$ is increasing in $t$ and
$\P\big\{\max_{1\le {i}\le n}\Big(\|X_{i}\|\vee \|X_{i}^*\|\Big)\le c u_n\big\}=1$ for a constant $c$,
\bes
\mu^{{(m)}}_{\max} \le g^2(c)
\bigg(\bigg(\max\bigg\{\sum_{k=1}^n \frac{\E |X_k|^m}{n},
\sum_{k=1}^n \frac{\E |X_k^*|^m}{n}\bigg\}\bigg)^{1/m}\bigg)^{\otimes m}.
\ees
Let $\bU_{\sigma,{k}}$ and $\zeta_{k,i}$ be as in Lemma \ref{k-nonspecific} and define
\bes
F^{{(m)}} = \frac{n}{m!} \A_{\sigma,k}\Big(\E {f}^{(m,0)}(\bU_{\sigma,{k}},\zeta_{k,\sigma_k})\Big)
= \sum_{{k}=1}^n \frac{1}{m!n!}\sum_{{i}=1}^n\sum_{\sigma,\sigma_{k}={i}}
\E {f}^{(m,0)}(\bU_{\sigma,{k}},\zeta_{k,i}),
\ees
where $\A_{\sigma,k}$ is the operator defined in (\ref{A_sigma,k}). Similarly, 
define
\bes
F^{{(m)}}_{\max} = \frac{n}{m!} \A_{\sigma,k}\Big(\E {f}^{(m)}_{\max}(\bU_{\sigma,{k}},\zeta_{k,\sigma_k})\Big)
= \sum_{{k}=1}^n\frac{1}{m!n!}\sum_{{i}=1}^n\sum_{\sigma,\sigma_{k}={i}}
\E {f}^{(m)}_{\max}(\bU_{\sigma,{k}},\zeta_{k,i}).
\ees

\begin{theorem}\label{comparison}
Let $(X_i, X_i^*)\in \R^{p\times 2}$, $1\le i\le n$, 
be independent random matrices under expectation $\E$. 
Let $m^*\in \{3,4\}$. Suppose (\ref{cond-1a}) and (\ref{cond-1b}) hold. Then,
\bes
\E f(X_1,\ldots,X_n) - \E  f(X_1^*,\ldots,X_n^*)
= \sum_{m=2}^{m^*-1}\Big\langle F^{{(m)}},\mu^{{(m)}} - \nu^{{(m)}}\Big\rangle + \Rem,
\ees
where $\mu^{{(m)}} = n^{-1}\sum_{i=1}^n \E X_i^{\otimes m}$
and $\nu^{{(m)}} = n^{-1}\sum_{i=1}^n \E (X_i^*)^{\otimes m}$ as in (\ref{moments}), and
\bes
\Big|\Rem\Big| \le \bigg\{ 2+4\sum_{m=2}^{m^*-1}{m^*\choose m}\bigg\}
\Big\langle F^{{(m^*)}}_{\max},\mu^{{(m^*)}}_{\max}\Big\rangle.
\ees
\end{theorem}

We may apply Theorem \ref{comparison} directly to $\{X_i\}$ and $\{X_i^*\}$ or
their truncated versions as we will show in Theorems \ref{th-comp-gene}
and \ref{th-comp-empi} in the next two subsections.

In Theorem \ref{comparison}, the difference between the left- and right-hand sides of
(\ref{approx-Delta_{n,m}}) is absorbed in the remainder term, which itself is expressed
in terms of the average of moment-like quantities in (\ref{mu_max}), under
conditions (\ref{cond-1a}) and (\ref{cond-1b}).

\medskip
{\bf 3.3. Comparison theorem for the maxima of sums.}
As in (\ref{T_n}) and (\ref{T_n^*}), let
\bes
T_n = \bigg\|\sum_{i=1}^n X_i/\sqrt{n}\bigg\|_\infty,\quad
T_n^* = \bigg\|\sum_{i=1}^n X_i^*/\sqrt{n}\bigg\|_\infty.
\ees
For random matrices $\tbX=(\Xtil_1,\ldots,\Xtil_n)$ and $\tbX^*=(\Xtil_1^*,\ldots,\Xtil_n^*)$
and $b_n>0$, define
\begin{align}\label{Omega_0}
\Omega_0 = \bigg\{\bigg\|\sum_{i=1}^n\frac{X_i-\Xtil_i}{n^{1/2}}\bigg\|_\infty > \frac{1}{4b_n}\Big\},\quad
\Omega_0^* = \bigg\{\bigg\|\sum_{i=1}^n\frac{X_i^*-\Xtil^*_i}{n^{1/2}}\bigg\|_\infty > \frac{1}{4b_n}\Big\}.
\end{align}

\begin{theorem}\label{th-comp-gene}
Let $(X_i, X_i^*)\in \R^{p\times 2}$, $1\le i\le n$, be independent random matrices
under expectation $\E$, $m^*\in \{3,4\}$,  
$\eta_n(\eps)$ be the L\'evy-Prokhorov pre-distance in (\ref{Prokhorov}), 
and $u_n =\sqrt{n}/(2b_n\log p)$.
\\
(i) Let $\mu^{{(m)}}_{\max}$ be given in (\ref{mu_max}) with $g(t) = e^{2m^* t}$. Then,
\begin{align}\label{th-comp-gene-1}
\eta_n(1/b_n) \le
C_{m^*}\Bigg( \sum_{m=2}^{m^*-1} \frac{b_n^m(\log p)^{m-1}}{n^{m/2-1}}\big\|\mu^{{(m)}} - \nu^{{(m)}}\big\|_{\max}
+ \frac{b_n^{m^*}(\log p)^{m^*-1}}{n^{m^*/2-1}}\|\mu_{\max}^{{(m^*)}}\|_{\max}\Bigg)
\end{align}
where $\mu^{{(m)}}$ and $\nu^{{(m)}}$ are as in Theorem \ref{comparison}. \\
(ii) Let $\tbX=(\Xtil_1,\ldots,\Xtil_n)$ and $\tbX^*=(\Xtil_1^*,\ldots,\Xtil_n^*)$.
Suppose that $(\Xtil_i,\Xtil^*_i)$ are independent matrices under $\P$, $\E \Xtil_i=\E\Xtil_i^*=0$,
and $\P\big\{\|\tbX\|_{\max} \vee \|\tbX^*\|_{\max} \le c_1 u_n\big\} =1$ for a constant $c_1$. Then,
\begin{align}\label{th-comp-gene-2}
\eta_n(1/b_n) \le&\
C_{m^*,c_1}\sum_{m=2}^{m^*}\frac{b_n^m(\log p)^{m-1}}{n^{m/2-1}}\big\|\mutil^{{(m)}} - \nutil^{{(m)}}\big\|_{\max}
\\ 
\nonumber &\ + C_{m^*,c_1}\frac{b_n^{m^*}(\log p)^{m^*-1}}{n^{m^*/2-1}}\big\|\mutil^{{(m^*)}}\big\|_{\max}
+ \P\big\{\Omega_0\big\} + \P\big\{\Omega_0^*\big\},
\end{align}
where $\mutil^{{(m)}} = n^{-1}\sum_{i=1}^n \E \Xtil_i^{\otimes m}$, 
$\nutil^{{(m)}} = n^{-1}\sum_{i=1}^n \E (\Xtil_i^*)^{\otimes m}$, 
and $\Omega_0$ and $\Omega_0^*$ are as in \eqref{Omega_0}.
\end{theorem}

We may consider $\tbX =(\Xtil_{i,j})_{n\times p}=(\Xtil_1,\ldots,\Xtil_n)$
as a truncated version of $\bX$ given by
\begin{align}\label{Xtil-a_n}
\Xtil_{i,j} = X_{i,j}I_{\{|X_{i,j}| \le a_n\}} - \E X_{i,j}I_{\{|X_{i,j}|\le a_n\}}.
\end{align}
In this case, the following lemma can be used to bound $\P\{\Omega_0\}$.

\begin{lemma}\label{lm-truncate-new}
Let $M_m$ be as in (\ref{M_m}) with $m>2$,
$\tbX$ as in (\ref{Xtil-a_n}) with $a_n$ satisfying $M_m\big\{n/\log(p/{{\eps_n}})\big\}^{1/m} \le a_n \le \atil_n = \{c_1n^{1/2}/(b_n\log (p/\eps_n))\}$ with $c_1>0$, and $\Omega_0$ as in (\ref{Omega_0}).
Then, for sufficiently large constant $C_{m,c_1}$, it implies by
$b_n^m (\log (p/\eps_n))^{m-1} M_m^m/n^{m/2-1} \le 1/C_{m,c_1}$ that
\begin{align}\label{lm-truncate-new-1}
\P\big\{\Omega_0\big\}
\le  \eps_{n} +\P\big\{\Omegatil_0\big\}
\le \eps_{n} + C_{m,c_1}\frac{b_n^m (\log (p/\eps_n))^{m-1}}{n^{m/2-1}}\M_{m,2}^m,
\end{align}
where
$\Omegatil_0 = \big\{\max_{1\le j\le p}\big|n^{-1/2}\sum_{i=1}^n X_{i,j}I_{\{|X_{i,j}|> \atil_n\}}\big| >1/(8b_n)\big\}$
and $\M_{m,2}$ is as in (\ref{M_m-1}).
\end{lemma}

We note that the upper bound for $a_n$ is no smaller than the lower bound due to the condition 
\bes
b_n^m (\log (p/\eps_n))^{m-1} M_m^m/n^{m/2-1} \le 1/C_{m,c_1}.
\ees

\medskip
{\bf 3.4. Efron's empirical bootstrap.}
We have already obtained upper bounds for the L\'evy-Prokhorov pre-distance (\ref{Prokhorov})
in terms of the average moments of $X_i$ and $X_i^*$ in Theorem \ref{th-comp-gene}.
In bootstrap, the L\'evy-Prokhorov pre-distance is a random variable due to the involvement of $\P^*$,
\begin{align}\label{eta_n-boot}
\eta_n^*(\eps) \eqeq \eta_n^{{(\P^*)}}(\eps;T_n^0,T_n^*) = \sup_{t\in \R}\eta_n^{{(\P^*)}}(\eps,t;T_n^0,T_n^*),
\end{align}
where $\eta_n^{{(\P^*)}}(\eps,t;T_n^0,T_n^*)
= \max\big[\P^*\big\{T_n^0\le t -\eps  \big\} - \P^*\big\{T_n^* < t \big\},
\P^*\big\{T_n^*\le t - \eps\big\} - \P^*\big\{T_n^0 < t \big\},0\big]$ 
as in Lemma \ref{lm:cp-decomp}, 
and $T_n^*$ is the bootstrapped $T_n$.
Recall that $\P^*\{T_n^0\le t\} = \P\{T_n\le t\}$ as $T_n^0$ is an independent copy of $T_n$.
In this subsection, we derive more explicit bounds for $\eta_n^*(\eps)$ in terms of the
average moments of $\{X_i\}$ for Efron's empirical bootstrap.

For the empirical bootstrap, the difference of the average moments between $X_i$ and $X_i^*$ is
\bes
\nu^{{(m)}} - \mu^{{(m)}} &=& \frac{1}{n}\sum_{i=1}^n (X_i - \Xbar)^{\otimes m}
- \frac{1}{n}\sum_{i=1}^n \E X_i^{\otimes m}
\cr &=& \frac{1}{n}\sum_{i=1}^n \Big(X_i^{\otimes m} - \mu^{{(m)}}\Big)
+ \sum_{k=1}^m {m\choose k}\hbox{Sym}
\bigg(\big(-\Xbar\big)^{\otimes k}\sum_{i=1}^n \frac{X_i^{\otimes (m-k)}}{n} \bigg),
\ees
where $\nu^{{(m)}}$ and $\mu^{{(m)}}$ are as in (\ref{moments}) with the assumption $\mu^{{(1)}} =0$
and Sym$(A)$ denotes the symmetrization of tensor $A$ by taking the average over all permutations of
the index of its elements.
It can be seen from the above expression that
the quantities $\|\mu^{{(m)}} - \nu^{{(m)}}\|_{\max}$ in the right-hand side of (\ref{th-comp-gene-1}),
and $\|\mu_{\max}^{{(4)}}\|_{\max}$ as well, can be bounded by empirical process methods.
However, as high moments are involved, some level of truncation may still be
needed to obtain sharp results when $\|\bX\|_{\max}$ is unbounded.
Therefore, a direct application of the error bound (\ref{th-comp-gene-2}) with truncation is natural.
This approach is taken here.

\begin{theorem}\label{th-comp-empi}
Let $X_i\in\R^p$ be independent random vectors and $X_i^*$ generated by the
empirical bootstrap. Let $b_n>0$, $M_4$ be as in (\ref{M_m}),
$\{c_1,c_2\}$ be fixed positive constants,
and $\atil_n = 
{c_1}\sqrt{n}/(b_n\log(p/{{\eps_n}}))$.
Suppose $\log(p/{{\eps_n}})\le {c_2}n$. Then,
\begin{align}\label{th-comp-empi-2}
\eta_n^*(1/b_n)
\le C_{c_1,c_2} b_n^2 (\log(p/{{\eps_n}}) )^{3/2}M_4^2 \big/n^{1/2} + 2\eps_n + \P\Big\{\|\bX\|_{\max}>\atil_n\Big\}
\end{align}
with at least probability $1 - \big( \P \{\|\bX\|_{\max} > \atil_n \} + 2\eps_n\big) $,
and with $\M_4$ as in (\ref{M_m}),
\begin{align}\label{th-comp-empi-3}
\P\bigg\{\eta_n^*(1/b_n)
> C_{c_1,c_2} \Big(\eps_n + b_n^4 (\log (p/\eps_n))^3\M_4^4/(\eps_n n) \Big) \bigg\}
\le \eps_n.
\end{align}
\end{theorem}

\medskip
{\bf 3.5. Wild bootstrap.}
Let $\{W_i\}$ be a sequence of i.i.d.\,variables independent of $\bX$ and satisfying
$\E W_i=0$ and $\E W_i^2=1$.
The wild bootstrap \citep{wu1986jackknife, liu1988bootstrap, mammenwild1993} is defined
in (\ref{wild}). Recall that we assume $\E X_i=0$ without loss of generality in our analysis.
As $\big\|\sum_{i=1}^n W_i\overline{X}/\sqrt{n}\big\|_\infty = O_P(1)\big\|\overline{X}\big\|_\infty$
is typically negligible in the analysis of the maxima of the sum of $X_i^*$
under mild conditions, for simplicity we may study
\begin{align}\label{simple-wild}
X_i^*=W_iX_i.
\end{align}

Suppose the moments of individual $X_i^*$ matches that of $X_i$ under the joint expectation
$\E$,
\begin{align}\label{moment-match}
\E X_i^{\otimes m} = \E(W_iX_i)^{\otimes m},\ m = 1,\ldots,m^*-1,
\end{align}
where $m^*$ represents the highest order of expansion involved in the comparison theorem.
Condition (\ref{moment-match}) holds for $m^*=4$ when $\E W_i^3=1$
\citep{liu1988bootstrap, mammenwild1993},
and all $m^*$ for the Rademacher wild bootstrap when 
$\E X_i^{\otimes m}=0$ for all positive odd $m$ smaller than $m^*$, 
\begin{align}\label{moment-match-2}
\{ \E W_i^3=1\ \hbox{ and }\ m^*=4\}\ \hbox{ or }\ \{ \E W_i^4=1\ \hbox{ and }\ 
\E X_i^{\otimes m}=0\ \forall \hbox{ odd } m \in [1,m^*) \}.
\end{align}
We note that (\ref{moment-match}) always holds for $m^*=3$ due to the default conditions
$\E W_i=0$ and $\E W_i^2=1$.
Under this moment condition and the sub-Gaussian condition (\ref{subgaussian}) on $W_i$,
a modification of the proof of Theorem \ref{th-comp-empi} yields the following result.

\begin{theorem}\label{th-comp-wild-a}
Let $X_i\in\R^p$ be independent random vectors and $X_i^*$ generated by the
wild bootstrap as in (\ref{wild}).
Suppose (\ref{moment-match}) holds with $m^*\in \{3,4\}$ 
and (\ref{subgaussian}) holds with $\tau_0<\infty$.
Let $M_{m^*}$ and $\M_{m^*,2}$ be as in (\ref{M_m}) and (\ref{M_m-1}) respectively. 
Let $b_n>0$, $\eps_n \le {\overline{\eps}_n }$ and $\atil_n = c_1\sqrt{n}/\big\{ (b_n (\log (p/{\overline{\eps}_n }))^{1/2} (\log (p/\eps_n))^{1/2} \big\}$.
Suppose $\log p \le c_2n$ with a constant $c_2>0$ and $M = M_{m^*} \big( n/\log(p /\overline{\eps}_n) \big)^{1/m^*-1/4}$. 
Then, for a sufficiently large constant $C_{m^*,\tau_0,c_1,c_2}$,
\begin{align}\label{th-comp-wild-a-1}
\eta_n^*(1/b_n)
\le&\ 
C_{m^*, \tau_0,c_1,c_2} b_n^2 (\log (p/{\overline{\eps}_n }))^{1/2} (\log (p/\eps_n)) /n^{1/2} M^2
+ \overline{\eps}_n
\\
&\nonumber \qquad \qquad + \, \P\Big\{\max_{1\le j\le p}\Big|\sum_{i=1}^n 
\frac{X_{i,j}I_{\{|X_{i,j}|> \atil_n\}} }{\sqrt{n}} \Big| >1/(8b_n)\Big\}
\end{align}
with at least probability $1 - \big( \P\big\{ C_0\tau_0^2b_n^2\log(p/\overline{\eps}_n)\max_{1\le j\le p}
\sum_{i=1}^n X_{i,j}^2I_{\{|X_{i,j}| > \atil_n\}} > n\big\} + 2\eps_{n} \big)$, and
\begin{align}\label{th-comp-wild-a-2}
&& \P\bigg\{ \eta_n^*(1/b_n)
> C_{m^*,\tau_0,c_1, c_2}' \Big( {\overline{\eps}_n } + \frac{b_n^{m^*}(\log(p/\overline{\eps}_n))^{m^*/2-1}(\log(p/\eps_n))^{m^*/2}}{\eps_n \cdot n^{m^*/2-1}}\M_{m^*,2}^{m^*} \Big) \bigg\} \le \eps_{n}.
\end{align}
\end{theorem}

While (\ref{th-comp-wild-a-1}) is comparable with (\ref{th-comp-empi-2}) in Theorem \ref{th-comp-empi}, (\ref{th-comp-wild-a-2}) requires the weaker
moment $\M_{m^*,2}$ than the $\M_{m^*}$ in (\ref{th-comp-empi-3}).

In the rest of the subsection, we study the implication of a martingale structure in the original
Lindeberg expansion (\ref{Lindeberg-1}) for wild bootstrap.
This would lead to a comparison theory more useful for the high order $m^*> 4$. 
Let
\bes
\bU_{i}^0 = (X_1^0,\ldots,X_{{i}-1}^0,0,X^*_{{i}+1},\ldots,X^*_n),\quad
\bV_{i}^0 = (X_1^0,\ldots,X_{i}^0,X^*_{{i}+1},\ldots,X^*_n),
\ees
where $\bX^0 = (X_1^0,\ldots,X_n^0)^T$ is an independent copy of $\bX$.
Let ${f}^{(m)}_i = (\pa/\pa x_i)^m f$ and $\Delta^*_n(f)$ be as in (\ref{Delta^*}).
The bootstrap version of the Lindeberg expansion (\ref{Lindeberg-1}) is
\begin{align}\label{Lindeberg-4}
\Delta^*_n(f)
= \sum_{m=2}^{m^*-1} \Delta_{n,m}^*
+ \Rem
\end{align}
with $\Delta_{n,m}^*= (m!)^{-1}\sum_{{i}=1}^n \big\langle \E^*{f}^{(m)}_i(\bU_{i}^0),
\E^*(X_{i}^0)^{\otimes m} - \E^*(X_{i}^*)^{\otimes m}\big\rangle$.

Consider the case where $X^*_i$ are defined as in (\ref{simple-wild}).
By (\ref{moment-match}), $\E\big\{\E^*(X_{i}^0)^{\otimes m} - \E^*(X_{i}^*)^{\otimes m}\big\}=0$.
As $\E^*{f}^{(m)}_i(\bU_{i}^0)$ is a function of
$(X_{{i}+1},\ldots,X_n)$, $\Delta_{n,m}^*$
is a sum of martingale differences. This directly leads to the comparison inequalities in
Proposition \ref{prop-wild} below. Consider functions $f$ satisfying
\begin{align}\label{cond-f}
\P\bigg\{\,\begin{matrix}\big|{f}^{(m^*)}_i(x_1,\ldots,x_{{i}-1},t\xi_i,x_{{i}+1},\ldots,x_n)\big|
\qquad\qquad\qquad
\cr \le  g(\|\xi_i\|/u_n)f_{\max}^{{(m^*)}}(x_1,\ldots,x_{{i}-1},0,x_{{i}+1},\ldots, x_n)
\end{matrix} \bigg\}=1,\ 0\le t\le 1,
\end{align}
for $\xi_i =X_i$ or $X_i^*$, with real-valued $g(t)$ and $m^*$-tensor-valued $f_{\max}^{{(m^*)}}$.
Let
\begin{align*}
& s_{n,m,i} = \Big\langle \big(\E^*{f}^{(m)}_i(\bU_{i}^0)\big)^{\otimes 2},
\E \big(X_{i}^{\otimes m} - \E\, X_{i}^{\otimes m}\big)^{\otimes 2}\big(\E\,W_i^m\big)^2\Big\rangle^{1/2},
2\le m < m^*,
\\
& s_{n,m^*,i} = \Big\langle \big(\E^* f_{\max}^{{(m^*)}}(\bU_{i}^0)\big)^{\otimes 2},
\E\big(\E^*g(\|X_{i}^*\|/u_n)|X_{i}^*|^{\otimes m^*}
- \E g(\|X_{i}^*\|/u_n)|X_{i}^*|^{\otimes m^*} \big)^{\otimes 2}\Big\rangle^{1/2},
\\
& {\overline \Rem} = \frac{1}{m^*!}
\sum_{{i}=1}^n \Big\langle \E^* f_{\max}^{{(m^*)}}(\bU_{i}^0),
\E\, g(\|X_{i}^0\|/u_n)|X_{i}^0|^{\otimes m^*} +
\E\, g(\|X_{i}^*\|/u_n)|X_{i}^*|^{\otimes m^*}\Big\rangle.
\end{align*}

\begin{proposition}\label{prop-wild}
Let $X_i$ and $X_i^*$ be as in (\ref{simple-wild}) and
$\Delta_n^*(f)$ as in (\ref{Lindeberg-4}).
Suppose (\ref{moment-match}) and (\ref{cond-f}). Then,
\begin{align}\label{prop-wild-1}
\E\Big|\Delta_n^*(f)\Big| \le&\
\sum_{m=2}^{m^*-1}\frac{1}{m!}\bigg(\sum_{{i}=1}^n \E s_{n,m,i}^2 \bigg)^{1/2}
+ \E\Big({\overline \Rem}\Big),
\\
\nonumber \Big(\E\Big|\Delta_n^*(f)\Big|^2\Big)^{1/2} \le &\
\sum_{m=2}^{m^*}\frac{1}{m!}\bigg(\sum_{{i}=1}^n \E s_{n,m,i}^2 \bigg)^{1/2}
+ \Big(\E\Big({\overline \Rem}\Big)^2\Big)^{1/2}.
\end{align}
\end{proposition}

For Efron's empirical bootstrap,
\begin{align}\label{empi-mome}
\E^*(X_i^*)^{\otimes m} = n^{-1}\sum_{k=1}^n (X_k-\overline{X})^{\otimes m}
\end{align}
involves all data points, so that the martingale argument does not directly apply. An application of the martingale Bernstein inequality \citep{steiger1969best, freedman1975tail}
leads to the following theorem.

\begin{theorem}\label{th-comp-wild-b}
Theorem \ref{th-comp-wild-a} is still valid for general $m^* > 2$ 
when $\eps_{n}$ is defined by
\begin{align}\label{th-comp-wild-b-1}
\eps_{n} = b_n^2(\log p)\{\log(1/\eps_{n})/n\}^{1/2}M^2 + 
\kappa_{n,m^*} \big( \M_{m^*,1} /M_{m^*} \big)^{m^*}
\end{align}
provided that $M \ge \M_{4,1}$ with the $\M_{m,1}$ in (\ref{M_m-1}).
\end{theorem}

Consider $m^*=6$. When $M^6 \asymp\M_{6,1}$
and $M b_n \asymp \sqrt{\log p}$, the second term in (\ref{th-comp-wild-b-1})
is of no greater order than $\{b_n^2(\log p)n^{-1/2}M^2\}^4$,
so that by Theorem~\ref{th-comp-wild-b}
\bes
(\log p)^4/n\to 0 \ \Rightarrow\ \eps_n\to 0.
\ees
In this case, taking $m^*>6$ does not improve
the order of $\eps_{n}$ in Theorem~\ref{th-comp-wild-b}.

Next, we derive upper bounds for
\begin{align}\label{eta_n}
\eta_n^{{(q)}}(\eps) = \sup_{t\in \R}\Big[\E\big\{\eta_n^{{(\P^*)}}(\eps,t;T_n^0,T_n^*)\big\}^q\Big]^{1/q}
\end{align}
with the $\eta_n^{{(\P^*)}}(\eps,t;T_n^0,T_n^*)$ in (\ref {eta_n-boot}).
The quantity $\eta_n^{{(q)}}(\eps)$ can be viewed as a weak L\'evy-Prokhorov pre-distance,
as the supreme is taken outside the expectation.
However, this weak version of the L\'evy-Prokhorov pre-distance is still stronger than the unconditional one.
In fact, we have
\bes
\eta_n(\eps) \eqeq \sup_{t\in\R}\eta_n^{{(\P)}}(\eps,t;T_n,T_n^*)
\le \eta_n^{{(q)}}(\eps) \le \Big\|\eta_n^*(\eps)\Big\|_{L_q(\P)},\quad q\ge 1,
\ees
where $\eta_n^{{(\P)}}(\eps,t;T_n,T_n^*)$ is as in (\ref{Prokhorov}).
See (\ref{eta_n-boot}) and the discussion below (\ref{Prokhorov}).

In addition to the average moments $M_m$ defined in (\ref{M_m}), we use quantities
\begin{align}\label{M_{4,m}}
M_{m,{1}} =&\ \bigg\|\sum_{i=1}^n\frac{\E \exp(2m\|W_iX_{i}\|_\infty/u_{n})|W_iX_{i}|^m
}{ n\E \exp(-2m\|X_{i}\|_\infty/u_n)}\bigg\|_\infty^{1/m},
\\
\nonumber M_{m,{2}} =&\ \bigg\|\frac{1}{n}\sum_{i=1}^n\frac{\E |W_iX_{i}|^{m}}
{\E \exp(-2m\|X_{i}\|_\infty/u_n)}\bigg\|_\infty^{1/m},\quad
\end{align}
to bound the $\eta_n^{{(q)}}(\eps)$ in (\ref{eta_n}).
When $\P\{\|X_i\|_\infty \le a_n\}=1$,
\bes
M_{m,{1}}\le e^{2a_n/u_n} \Big(\E |W_1|^m \E e^{2m|W_1|a_n/u_n}\Big)^{1/m}M_m,\quad
M_{m,{2}} \le e^{2a_n/u_n}\big(\E |W_1|^{m}\big)^{1/m}M_m.
\ees
In any case, controlling $M_{m,{1}}$ requires $W_1$ to have a finite moment
generating function in the interval $[0,2m^*a_n/u_n]$.

\begin{theorem}\label{th-comp-wild}
Let $a_n= {c_1}\sqrt{n}/(b_n\log p)$, $m^*\ge 3$ and $\eta_n^{{(q)}}(\cdot)$ be as in (\ref{eta_n}). \\
(i) Let $X_i^*$ be as in (\ref{simple-wild}). Suppose (\ref{moment-match}) holds. 
Let $b_n > 0$ and $u_n=\sqrt{n}/(2 b_n\log p)$ in (\ref{M_{4,m}}).
Then,
\begin{align}\label{th-comp-wild-1}
\eta_n^{{(1)}}(1/b_n) \le C_{m^*}\bigg(
\sum_{m=2}^{m^*-1}\big|\E W_1^m\big| \frac{b_n^{m}(\log p)^{m-1}}{n^{m/2-1/2}}M_{2m,{2}}^{m}
+ \frac{b_n^{m^*}(\log p)^{m^*-1}}{n^{m^*/2-1}}M_{m^*,{1}}^{m^*} \bigg).
\end{align}
(ii) Let $X_i^*$ be as in (\ref{wild}).
Suppose (\ref{moment-match-2}) and (\ref{subgaussian}) hold. Then, for $1\le q\le 2$,
\begin{align}\label{th-comp-wild-2}
& \eta_n^{{(q)}}(1/b_n)
\\ 
\nonumber \le &\ C_{m^*,\tau_0,{c_1}}\bigg(
\frac{b_n^2\log p}{n^{1/2}}M_{4}^2
+ \kappa_{n,m^*}^{1/q}\bigg)
+\bigg[\E \min\bigg\{2, C_{\tau_0}\frac{b_n^2\log p}{n}
\max_{1\le j\le p}\sum_{i=1}^n X_{i,j}^2I_{\{|X_{i,j}|>a_n\}}\bigg\}\bigg]^{1/q}
\\ 
\nonumber \le &\ C_{m^*,\tau_0,{c_1}}\bigg(
\frac{b_n^2\log p}{n^{1/2}}M_{4}^2
+ \kappa_{n,m^*}^{1/q}\bigg)
\\ 
\nonumber &\ +\bigg[\E \min\bigg\{2,C_{m^*,\tau_0,c_1}\frac{b_n^{m^*}(\log p)^{m^*-1}}{n^{m^*/2}}
\max_{1\le j\le p}\sum_{i=1}^n |X_{i,j}|^{m^*}I_{\{|X_{i,j}|>a_n\}}\bigg\}\bigg]^{1/q},
\end{align}
where $\kappa_{n,m} = b_n^m(\log p)^{m-1} n^{1-m/2} M_m^m$. Moreover,
\begin{align}\label{th-comp-wild-3}
\Big(\E\Big| \eta_n^*(1/b_n)\Big|^q\Big)^{1/q}
\le (1+q)\big\{q^{-1}2^{1/q}\eta_n^{{(q)}}(1/b_n)\big\}^{q/(q+1)}.
\end{align}
\end{theorem}

Compared with the first term on the right-hand side of (\ref{th-comp-empi-2}),
the first term on the right-hand side of (\ref{th-comp-wild-2}) is of smaller order
by at least a factor $\sqrt{\log(p/\kappa_{n,4})}$.

The proof of Theorem \ref{th-comp-wild}, given in the Appendix, involves two issues.
The first one is to relate the maxima $T_n$ in (\ref{T_n}) and $T_n^*$ in (\ref{T_n^*})
to smooth functions $f(x_1,\ldots,x_n)$ in Proposition \ref{prop-wild}.
This is done via the smooth max function in (\ref{smooth-max}) as discussed at the beginning of
this section. The second issue involves heterogeneity among $X_i$.
When $\P\{\|\bX\|_{\max}\le u_n\}=1$, the quantities in (\ref{M_{4,m}}) are bounded
under the condition $M_{m^*}=O(1)$ on the average moments.
However, a direct application of (\ref{prop-wild-1}) requires the stronger condition
\bes
\frac{1}{n}\sum_{i=1}^n \max_{1\le j\le p} \E |X_{i,j}|^{m^*} = O(1)
\ees
as in Theorem \ref{th-comp-wild-b}.
This issue is again resolved through Lemma \ref{k-nonspecific}.

\medskip
{\bf 4. Anti-concentration of the maxima.}
As we have discussed at the end of Section 2, the Kolmogorov-Smirnov distance between two
distribution functions can be bounded from the above by the sum of the L\'evy-Prokhorov pre-distance
and the minimum of the L\'evy concentration of the two distribution functions,
\begin{align}\label{anti-conc-1}
\sup_{t\in \R}\Big|\P\Big\{T_n\le t\Big\} - \P\Big\{T_n^* < t \Big\} \Big|
\le \eta_n(\eps) + \min\Big\{\omega_n(\eps;T_n),\omega_n(\eps;T_n^*)\Big\}
\end{align}
as in (\ref{KS-bound}).
{The above two terms are also required if one wants to use Lemma \ref{lm:cp-decomp}
to derive an upper bound for $\big| \P\{T_n\le t^*_\alpha \} - (1-\alpha) \big|$.}
As upper bounds for the L\'evy-Prokhorov pre-distance $\eta_n(\eps)$ and its bootstrap version
$\eta_n^*(\eps)$ have already been established in
Section 3, the aim of this section is to develop anti-concentration inequalities to bound
the L\'evy concentration function $\omega_n(\eps;T_n)$ from the above.
We note that once a comparison theorem becomes available as an upper bound for $\eta_n(\eps)$,
an anti-concentration inequality for $T_n$ can be established from one for $T_n^*$, as
\begin{align}\label{anti-conc-2}
\omega_n(\eps;T_n)
\le \omega_n(\eps;T_n^*)  +  2\sup_{t\in \R}\Big|\P\Big\{T_n\le t\Big\} - \P\Big\{T_n^* < t \Big\} \Big|
\le 3\omega_n(\eps;T_n^*) + 2\eta_n(\eps)
\end{align}
by the triangle inequality and (\ref{anti-conc-1}), and vice versa.

To study the consistency of the Gaussian wild bootstrap, say $T_n^{*,{\rm Gauss}}$ for the approximation
of the distribution of $T_n$,
the Kolmogorov-Smirnov distance of interest is bounded by
\bes
\sup_{t\in \R}\Big|\P\Big\{T_n\le t\Big\} - \P^*\Big\{T_n^{*,{\rm Gauss}} < t \Big\} \Big|
\le \eta_n^*(\eps) + \min\Big\{\omega_n^{{(\P)}}(\eps;T_n),\omega_n^{{(\P^*)}}(\eps;T_n^{*,{\rm Gauss}})\Big\},
\ees
where $\eta_n^*(\eps)= \eta_n^{{(\P^*)}}(\eps;T_n^0,T_n^{*,{\rm Gauss}})$
and $\omega_n^{{(\P)}}(\eps;T_n)$ are as in (\ref{eta_n-boot}) and (\ref{KS-bound})
respectively. Thus, an anti-concentration inequality for the Gaussian maxima $T_n^{*,{\rm Gauss}}$
under $\P^*$ suffices \citep{chernozhukov2015comparison}.
This approach has been taken in \cite{chernozhukov2013, chernozhukov2017central} among others.
However, the inequality (\ref{anti-conc-2}) with $T_n^* = T_n^{*,{\rm Gauss}}$, which requires a small
L\'evy-Prokhorov pre-distance $\eta_n(\eps) = \eta_n^{{(\P)}}(\eps;T_n,T_n^{*,{\rm Gauss}})$,
is not helpful in our study as we are interested in scenarios
where the Gaussian approximation may not hold.

Our idea is to derive anti-concentration inequalities for the maxima $T_n$ of sums of possibly skewed
independent random vectors through a mixed wild bootstrap which has a Gaussian component
and also provides the third moment match as 
\cite{liu1988bootstrap} and \cite{mammenwild1993} advocated.
Compared with the Gaussian wild bootstrap, such a mixed wild bootstrap enjoys both
the anti-concentration properties of the Gaussian component through conditioning
and sharper approximation of the distribution of $T_n$
through the fourth order comparison theorems developed in Section~3.

The multiplier of the above mixed wild, bootstrap can be defined as
\begin{align}\label{multiplier-mix}
W_i^{**} = a_0\delta_i Z_i + b_0(1-\delta_i)W^0_i,
\end{align}
where $\delta_i,Z_i,W^0_i, i=1,\ldots,n$, are independent random variables,
$\delta_i$ are Bernoulli variables with $\P\{\delta_i=1\} = p_0 = 1-\P\{\delta_i=0\}$,
$Z_i\sim N(0,1)$, and $W^0_i$ can be taken as Mammen's bootstrap multiplier in (\ref{mammen-w}).
In this mixed wild bootstrap, $a_0$, $b_0$ and $p_0$ are positive constants satisfying
\begin{align}\label{multiplier-mix-coef}
0< p_0 <1,\quad \E\big(W_i^{**}\big)^2 = a_0^2p_0+b_0^2(1-p_0)=1,\quad
\E\big(W_i^{**}\big)^3 = b_0^3(1-p_0)=1.
\end{align}
For any $p_0 \in (0,1)$, the values of $a_0$ and $b_0$ are determined by
\bes
b_0 = (1-p_0)^{-1/3},\quad a_0 = \sqrt{p_0^{-1}\big(1- (1-p_0)^{1/3}\big)}.
\ees
For example, $a_0 = 0.6423387$ and $b_0 = 1.259921$ for $p_0=1/2$.

Suppose $\E X_i=0$ as in Section 3.
Given the multiplier (\ref{multiplier-mix}) and the original data $X_i=(X_{i,1},\ldots,X_{i,p})^T, i=1,\ldots,n$, the mixed wild bootstrap for $T_n$ is defined through
\begin{align}\label{mixed-wild-1}
X_i^{**} = W_i^{**}X_i,\quad Z^{**}_n = (Z^{**}_{n,1},\ldots,Z^{**}_{n,p})^T = \frac{1}{\sqrt{n}}\sum_{i=1}^nX_{i}^{**},\
\hbox{ and }\ T_n^{**}  = \max_{1\le j\le p}Z^{**}_{n,j}.
\end{align}
We conveniently avoid the complication of subtracting the sample mean from $X_i$
as the primary purpose of this mixed wild bootstrap is to provide a vehicle to derive
anti-concentration inequalities for the maxima $T_n$ for the original data.
Once an upper bound for $\omega_n^{{(\P)}}(\eps;T_n)$ is established, the consistency of
the bootstrap can be studied through (\ref{anti-conc-1}) and Lemma \ref{lm:cp-decomp}.

Let $\P^{**}$ be the conditional probability given $\{X_i,\delta_i,W^0_i, i=1,\ldots,n\}$.
We find that under $\P^{**}$, $Z^{**}_n$ is a Gaussian vector
with individual mean and standard deviation
\begin{align}\label{conditional}
\mu_j^{**} = \E^{**}\big(Z^{**}_{n,j}\big) = \frac{b_0}{\sqrt{n}}\sum_{i=1}^n (1-\delta_i)W^0_iX_{i,j},\quad
\sigma_j^{**} = \sqrt{\Var^{**}\big(Z^{**}_{n,j}\big)}
= \bigg(\frac{a_0^2}{n}\sum_{i=1}^n \delta_i X_{i,j}^2\bigg)^{1/2}.
\end{align}
Anti-concentration inequalities for $T_n^{**}$ under the marginal probability $\P$ can be derived
from the conditional one under $\P^{**}$ via
\begin{align}\label{un-mix}
\omega_n^{{(\P)}}\big(\eps;T_n^{**}\big) \le \E\Big[\omega_n^{{(\P^{**})}}\big(\eps;T_n^{**}\big)\Big],
\end{align}
where $\omega_n^{{(\P^{**})}}\big(\eps;T_n^{**}\big)$, a function of
the random vector $(\mu_j^{**},\sigma_j^{**}, 1\le j\le p)$, is the L\'evy concentration function of $T_n^{**}$
under the conditional probability $\P^{**}$ as in (\ref{KS-bound}).

In what follows we present anti-concentration inequalities for the maxima of Gaussian vectors,
sums in the mixed wild bootstrap, and sums of general independent vectors with zero mean.

\begin{theorem}\label{anti-concentration}
Let $\xi = (\xi_1,\ldots,\xi_p)^T$ be a multivariate Gaussian vector with
marginal distributions $\xi_j\sim N(\mu_j,\sigma_j^2)$, 
$\sigma_{(1)}\le \cdots\le \sigma_{(p)}$ be the ordered values of $\sigma_1,\ldots,\sigma_p$. Then, for all $x_m\ge 1$, 
\begin{align}\label{lm-anti-2}
\sup_{x} \frac{d}{dx}\P\Big\{\max_{1\le j\le p}\xi_j \le x\Big\}
\le \max_{1\le m\le p}\bigg\{\frac{x_m}{\sigma_{(m)}}
+ \sum_{k=1}^{m-1}\frac{\varphi(x_k-1/x_k)}{\sigma_{(k)}}\bigg\}. 
\end{align}
Consequently, with ${\overline\sigma} 
= (2+\sqrt{2\log p})/\{1/\sigma_{(1)} + \max_{1\le m\le p}(1+\sqrt{2\log m})/\sigma_{(m)}\} 
\ge \sigma_{(1)}$, 
\begin{align}\label{lm-anti-3}
\P\Big\{ a < \max_{1\le j\le p}\xi_j \le a+\eps\Big\} \le 
\frac{\eps}{\overline\sigma}\Big(2 + \sqrt{2\log p} \Big) 
,\quad\forall\ \eps>0,\ a\in\R. 
\end{align}
Given $\{\sigma_j\}$, there exist certain $\xi_j\sim N(0,\sigma_j)$ and constants $a>0$ and 
$C_0\le 2^7/(1-1/4)$ such that 
\begin{align}\label{lm-anti-bd} 
\P\Big\{a\le \max_{1\le j\le p} \xi_j \le a+\eps\Big\} 
\ge \frac{\eps}{\overline\sigma}\bigg(\frac{2 + \sqrt{2\log p}}{C_0}\bigg) 
\end{align}
for all $\eps$ satisfying $0\le (\eps/{\overline\sigma})\big(2 + \sqrt{2\log p} \big) \le 1/8$. 
Moreover, \eqref{lm-anti-bd} also holds for certain independent $\xi_j \sim N(\mu_j, \sigma_j)$ with possibly different nonzero $\mu_j$ and the same $\{a, C_0\}$. 
\end{theorem}}

Anti-concentration of the maxima of Gaussian vectors have been considered in the literature;
For example, \cite{nazarov2003maximal}, 
\cite{klivans2008learning} and \cite{chernozhukov2015comparison}. 
These results provides $C_0(2+\sqrt{2\log p})/\sigma_{(1)}$ as an upper bound for 
\eqref{lm-anti-2} or $C_0(2+\sqrt{2\log p})\eps/\sigma_{(1)}$ for \eqref{lm-anti-3}. 
A main advantage of Theorem \ref{anti-concentration} is the use of potentially much large 
${\overline\sigma}$ instead of $\sigma_{(1)}$. 
For example, when $1/\sigma_{(1)} \ge (1+\sqrt{2\log p})/\sigma_{(m)}$ for all $1\le m\le p$, 
we have ${\overline \sigma} = (2+ \sqrt{2\log p})(\sigma_{(1)}/2)$ and therefore 
the right-hand side of  \eqref{lm-anti-3} becomes $2 \epsilon/\sigma_{(1)}$. Moreover, 
Theorem \ref{anti-concentration} is sharp up to the constant factor $C_0$. 
The anti-concentration inequality for general $\xi_j\sim N(\mu_j,\sigma_j^2)$ is needed to study
the mixed wild bootstrap $T_n^{**}$ under the conditional probability $\P^{**}$,
in view of (\ref{conditional}).

\begin{theorem}\label{anti-mix}
Let $X_i =(X_{i,1},\ldots,X_{i,p})^T \in \R^p$ be independent centered random vectors with $p >1$ 
and $T_n^{**}$ the mixed wild bootstrap given by (\ref{multiplier-mix}) and (\ref{mixed-wild-1}). 
Let $\sigma_j^2 = \sum_{i=1}^n \E X_{i,j}^2/n$ 
and $\{\sigma_{(j)}, 1\le j\le p, {\overline\sigma}\}$ be as in \eqref{sigmas}. 
Suppose $\P\{\|\bX\|_{\max} \le a_n\}=1$ for certain constants $a_n$ satisfying 
\begin{align}\label{conc-anti-mix-0}
\max_{1\le j \le p} \frac{\log\big(j^2 \,{\overline \sigma}/(\eps \sqrt{\log p}) \big)}
{\sigma_{(j)}^2} \le p_0n/(8a_n^2), 
\end{align}
Then, with the $(a_0,b_0,p_0)$ in (\ref{multiplier-mix}) 
\begin{align}\label{conc-anti-mix}
\omega_n^{{(\P)}}\big(\eps;T_n^{**}\big)
= \sup_{t\in \R}\P\Big\{ t \le T_n^{**} \le t+\eps\Big\}
\le C_{a_0,b_0,p_0}\frac{\eps}{{\overline\sigma}}\sqrt{\log p}.
\end{align}
\end{theorem}

If we use the mixed wild bootstrap (\ref{mixed-wild-1}) to approximate the distribution of $T_n$,
Theorem~\ref{anti-mix} and the comparison theorems in Section 3 can be directly applied to establish
the consistency of the bootstrap via (\ref{eta_n-boot}).
However, for studying the consistency of bootstrap methods in general through (\ref{anti-conc-1}),
we desire an anti-concentration inequality for the original data.
This can be done by comparing the distributions of $T_n^{**}$ and $T_n$, resulting in the following theorem.

\begin{theorem}\label{anti-general}
Let $X_i \in \R^p$ be independent with $p>1$, $\E X_i=0$,
$M_m$ and ${\overline \sigma}$ be as in (\ref{M_m}) and \eqref{sigmas} respectively, $b_n>0$ and $\omega_n^{{(\P)}}(\eps;T_n)$ be as in (\ref{KS-bound}) with the $T_n$ in (\ref{T_n}).
Let $a_n = {c_1\sqrt{n}}/(b_n\,\log p)$ for some constant $c_1>0$.
Then, for a certain positive constant $C_{c_1}$,
\begin{align}\label{anti}
\omega_n^{{(\P)}}(1/b_n;T_n)
\le
\frac{C_0}{b_n{\overline \sigma}} \sqrt{\log p}
+ C_{c_1}\kappa_{n,4}
+ 2\,\P\bigg\{\max_{1\le j\le p}\bigg|\sum_{i=1}^n \frac{X_{i,j}I_{\{|X_{i,j}|> a_n 
\}}}{\sqrt{n}}\bigg|
>\frac{1}{8b_n}\bigg\}.
\end{align}
\end{theorem}

We have derived comparison theorems up to a general order $m^*\ge 3$ under the
moment matching condition (\ref{moment-match}). This includes $m^*>4$ for the Rademacher
wild bootstrap for symmetric $X_i$. However, as the Rademacher multiplier does not have a
Gaussian component, we settle for $m^*=4$ in the above theorem.
If the Gaussian wild bootstrap is used as a vehicle to prove Theorem~\ref{anti-general},
(\ref{moment-match}) holds only for $m^*=3$ and the term
$C_{c_1}\kappa_{n,4}=C_{c_1}b_n^4(\log p)^3n^{-1}M_4^4$ will have to be replaced by
$C_{c_1}\kappa_{n,3} = C_{c_1}b_n^3(\log p)^2n^{-1/2}M_3^3$,
leading to the condition $\log p \ll n^{1/7}$ for $b_n\gtrsim \sqrt{\log p}$
as in \cite{chernozhukov2015comparison}.

\medskip
\begin{figure}
  \includegraphics[width=\textwidth]{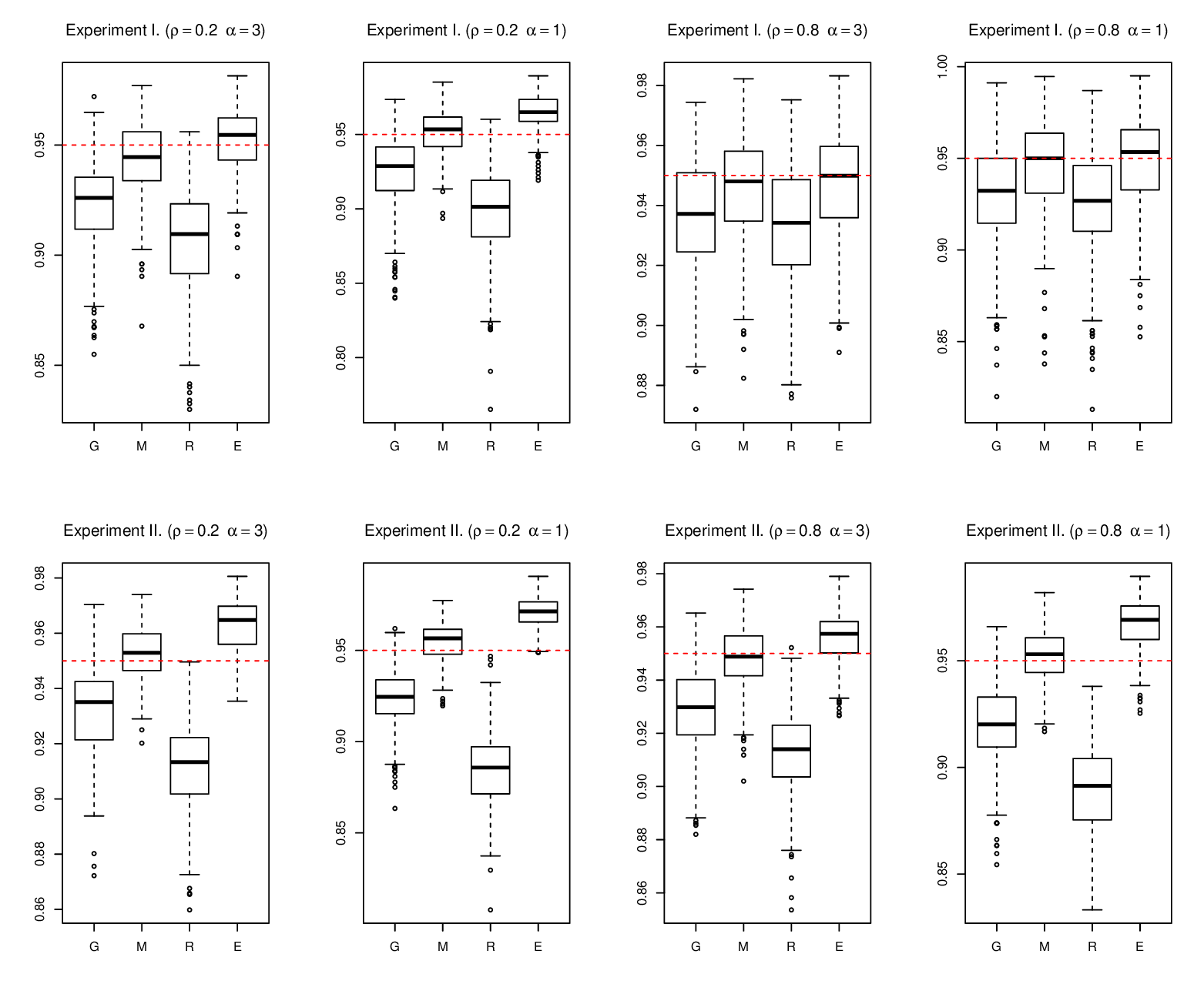}
  \centering
  \caption{Simulated relative frequency of the simultaneous coverage of 500
  95\% simultaneous confidence intervals for each bootstrap scheme:
  G, M and R respectively represent the Gaussian, Mammen and Rademacher wild bootstrap,
  while E represents Efron's empirical bootstrap.
 }

  \label{fig1}
\end{figure}

{\bf 5. Simulation results.}
We study the performance of different bootstrap procedures in two experiments.
In both experiments, we generate vectors $X_i =(X_{i,1},\ldots,X_{i,p})^T$ in a Gaussian copula model,
where $F(X_{i,j}) = \Phi(Y_{i,j})$ and $Y_i = (Y_{i,1},\ldots,Y_{i,p})^T$ are i.i.d.\,$N(0,\Sigma)$
with $N(0,1)$ marginal distributions, $n=200$, $p=400$, and
$F$ represents the gamma distribution with unit scale and shape parameter $ {\upalpha} = \E X_{i,j} \in \{1,3\}$.
We pick $\Sigma_{j,k} = \Cov(Y_{i,j},Y_{i,k}) = \rho + (1-\rho) I_{\{j=k\}}$ in Experiment I,
and $\Sigma_{j,k} = \rho^{|j-k|}$ in Experiment II, with $\rho \in \{0.2,0.8\}$.
Four bootstrap methods are considered: the Gaussian wild bootstrap with $W_i\sim N(0,1)$,
Mammen's wild bootstrap, the Rademacher wild bootstrap with $\P\{W_i = \pm 1\}=1/2$, and Efron's
empirical bootstrap. Note that the skewness for $X_{i,j}$ is ${2/\sqrt{{\upalpha}}}$, e.g. 2 for ${\upalpha}=1$ and $2/\sqrt{3}$ for ${\upalpha}=3$. Thus,
in this setting, the Gaussian multiplier and Rademacher wild bootstrap methods do not match the third moment of the original data. Our theorems in Section 2 therefore assert that Mammen's wild bootstrap and empirical bootstrap have better approximation properties.
This theoretical claim is supported by our simulation results.

Since $\E X_i$ is unknown, the wild bootstrap is defined as $X_i^* = W_i(X_i - \overline{X})$.
We compare the distribution of $T_n = \max_j \sum_{i=1}^n (X_{i,j}-\E X_{i,j})/\sqrt{n}$ against their
bootstrapped versions.
The true distribution of $T_n$ is evaluated based on 5000 simulations.
The results for the four bootstrap schemes are based on 500 copies of $\bX$, and 500 copies of $\bX^*$ for each observation of $\bX$.

Figures 1 plots the simulated relative frequency of the simultaneous coverage of 95\% bootstrap
simultaneous confidence intervals
for each bootstrap scheme in the four combinations of $(\rho,{\upalpha})$ in Experiments I and II.
This is closely related to the risk $\big|\P\{T_n > t_\alpha^*\} - \alpha\big|$.
The results for the Kolmogorov-Smirnov distance are shown in Figure~2 which contains
8 boxplots of the Kolmogorov-Smirnov distances between the true $T_n$ and bootstrapped $T_n^*$.

\begin{figure}
  \includegraphics[width=1\textwidth]{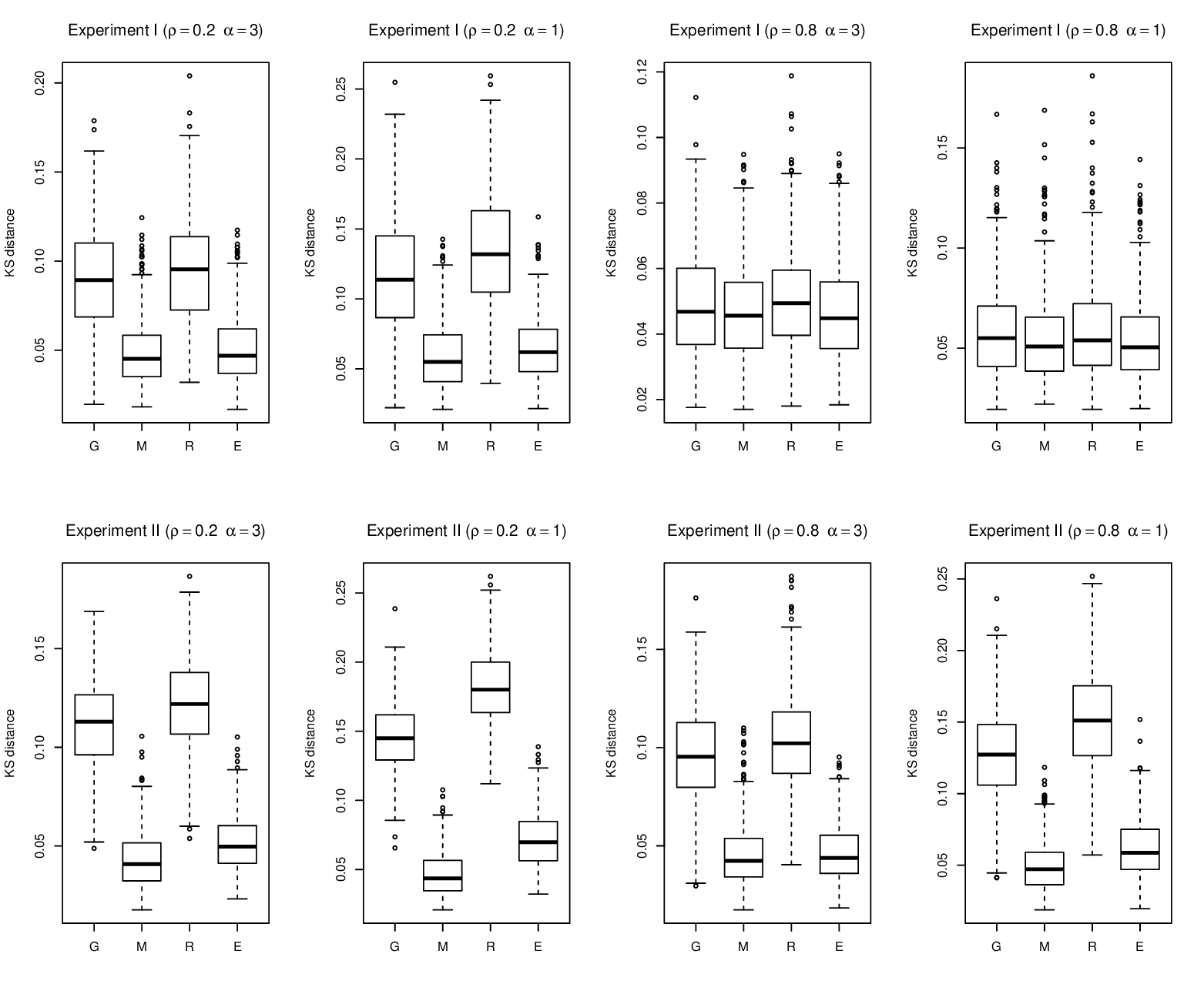}
  \centering
  \caption{The Kolmogorov-Smirnov distances of 500 runs for each bootstrap scheme:
  G, M, R and E respectively represent the Gaussian, Mammen, Rademacher and
  empirical bootstrap schemes.}
  \label{fig2}
\end{figure}

Corresponding to our theoretical results, this simulation study demonstrates that Mammen's wild bootstrap
is the best among all four schemes, empirical bootstrap is a close second,
while Gaussian and Rademacher wild bootstrap methods are clearly worse.
Because of the skewness of the Gamma distribution, an explanation of the
poor performance of the Gaussian and Rademacher wild bootstrap methods is the lack of the third moment
match as our theoretical results indicate.
We would like to mention that the difference among bootstrap procedures in two settings
(Experiment I, $\rho=0.8$, ${\upalpha}=3$ or $1$) are not as significant as the others,
possibly due to the smaller effective dimensionality caused by high correlation.
Nevertheless, Mammen's wild bootstrap and empirical bootstrap still perform slightly better.

In addition to the plots,
Table 1 provides the mean and standard deviation of the Kolmogorov-Smirnov distance
between the bootstrap estimates and the true cumulative distribution function of $T_n$,
and Table 2 provides the mean and standard deviation of the coverage probabilities of
95\% simultaneous confidence intervals with each bootstrap scheme.
These tables depicts the same picture as the plots.

\begin{table}[htb]
\begin{tabular}{c|c|c|c|c|c|c|c|c}
\hline
\multirow{2}{*}{Setting} & \multicolumn{2}{c|}{Gaussian} & \multicolumn{2}{c|}{Mammen} & \multicolumn{2}{c|}{Rademacher}& \multicolumn{2}{c}{Empirical}\\ \cline{2-9}
      & Mean & Std & Mean & Std& Mean & Std & Mean & Std \\ \hline
  I, $ \rho=0.2, {\upalpha}=3$ & 0.08996 &  0.02907 & 0.04893 & 0.01883 & 0.09484 & 0.02916 & 0.05088 & 0.01873\\
  I, $ \rho=0.2, {\upalpha}=1$ & 0.11660 & 0.03958 & 0.05964 & 0.02377 & 0.13428 & 0.04088 & 0.06457 & 0.02231 \\
  I, $\rho=0.8, {\upalpha}=3$ & 0.04910 & 0.01610 & 0.04699 & 0.01510 & 0.05091 & 0.01587 & 0.04690 & 0.01503 \\
  I, $ \rho=0.8, {\upalpha}=1$ & 0.05861 & 0.02364 & 0.05443 & 0.02198 & 0.05880 & 0.02432 & 0.05452 & 0.02107 \\ \hline
  II, $ \rho=0.2, {\upalpha}=3$ & 0.11106 & 0.02299 & 0.04324 & 0.01443 & 0.12176 & 0.02254 & 0.05105 & 0.01397 \\
  II, $ \rho=0.2, {\upalpha}=1$ & 0.14542 & 0.02451 & 0.04677 & 0.01622 & 0.18143 & 0.02654 & 0.07190 & 0.02053 \\
  II, $\rho=0.8, {\upalpha}=3$  & 0.09558 & 0.02485 & 0.04575 & 0.01629 & 0.10335 & 0.02493 & 0.04667 & 0.01488  \\
  II, $ \rho=0.8, {\upalpha}=1$ & 0.12780 & 0.03229 & 0.04998 & 0.01839 & 0.15043 & 0.03404 & 0.06249 & 0.02055  \\ \hline
\end{tabular}
\caption{The Kolmogorov-Smirnov distances between the
bootstrapped $T_n^*$ and true $T_n$}
\end{table}

\begin{table}[!h]
\begin{tabular}{ c|c|c|c|c|c|c|c|c }
\hline
\multirow{2}{*}{Setting} & \multicolumn{2}{c|}{Gaussian} & \multicolumn{2}{c|}{Mammen} & \multicolumn{2}{c|}{Rademacher}& \multicolumn{2}{c}{Empirical}\\ \cline{2-9}
      & Mean & Std & Mean & Std& Mean & Std & Mean & Std \\ \hline
  I, $ \rho=0.2, {\upalpha}=3$ & 0.9232 & 0.01938& 0.9446 & 0.01544 & 0.9072 & 0.2199 & 0.9527 & 0.01422\\
  I, $ \rho=0.2, {\upalpha}=1$ & 0.9251 & 0.02308 & 0.9517 & 0.01422 & 0.8975 & 0.02938 & 0.9646 & 0.01131  \\
  I, $\rho=0.8, {\upalpha}=3$ & 0.9364 & 0.01876 & 0.9457 & 0.01706 & 0.9331 & 0.01912 & 0.9471 & 0.01649 \\
  I, $ \rho=0.8, {\upalpha}=1$ & 0.9303 & 0.02671 & 0.9458 & 0.02447 & 0.9251 & 0.02785 & 0.9486 & 0.02357 \\ \hline
  II, $ \rho=0.2, {\upalpha}=3$ & 0.9323 & 0.01513 & 0.9527 & 0.00970 & 0.9124 & 0.01563 & 0.9628 & 0.00876 \\
  II, $ \rho=0.2, {\upalpha}=1$ & 0.9230 & 0.01613 & 0.9545 & 0.00955 & 0.8853 & 0.01890 & 0.9707 & 0.00721 \\
  II, $\rho=0.8, {\upalpha}=3$  & 0.9291 & 0.01456 & 0.9479 & 0.01061 & 0.9129 & 0.01540 & 0.9562 & 0.00872  \\
  II, $ \rho=0.8, {\upalpha}=1$ & 0.9196 & 0.01850& 0.9524 & 0.01172  & 0.8894 & 0.02079 & 0.9673 & 0.01116  \\ \hline
\end{tabular}
\caption{Relative frequency of bootstrap coverage of 95\% simultaneous confidence intervals}
\end{table}

It's worth mentioning that the empirical bootstrap does not always perform worse than Mammen's wild bootstrap
(Figure 1, Experiment I, $\rho=0.2, {\upalpha}=3$).
Recall that we discuss in Section 2 that the empirical bootstrap doesn't offer exact moments match,
and the fluctuation of the difference between true moments and empirically bootstrapped ones leads
to a slightly weaker consistency statement in Theorem \ref{cb-empi}.
However, the difference between the 4-th moments,
\bes
\mu^{{(4)}} - \nu^{(4)}
= \frac{1}{n}\sum_{i=1}^n \E X_i^{\otimes 4} - \frac{1}{n}\sum_{i=1}^n \E^* (X^*_i)^{\otimes 4},
\ees
for the empirical bootstrap can be much smaller than that for Mammen's.
This may provide an explanation of the performance of the Mammen and empirical
bootstraps in these two settings.

\begin{supplement}[id=supp]
 \stitle{Supplement to ``Beyond Gaussian Approximation: Bootstrap for Maxima of Sums of Independent Random Vectors"}
 \sdescription{This supplement contains proofs of all the theoretical results stated in the main body of the paper.}
\end{supplement}

\bibliographystyle{imsart-nameyear}
\bibliography{bootstrap-aos}

\begin{thebibliography}{45}

\bibitem[\protect\citeauthoryear{Belloni, Chernozhukov and
  Hansen}{2014}]{belloni2014inference}
\begin{barticle}[author]
\bauthor{\bsnm{Belloni},~\bfnm{Alexandre}\binits{A.}},
  \bauthor{\bsnm{Chernozhukov},~\bfnm{Victor}\binits{V.}} \AND
  \bauthor{\bsnm{Hansen},~\bfnm{Christian}\binits{C.}}
(\byear{2014}).
\btitle{Inference on treatment effects after selection among high-dimensional
  controls}.
\bjournal{The Review of Economic Studies}
\bvolume{81}
\bpages{608--650}.
\end{barticle}
\endbibitem

\bibitem[\protect\citeauthoryear{Belloni, Chernozhukov and
  Kato}{2015}]{bellonietal15}
\begin{barticle}[author]
\bauthor{\bsnm{Belloni},~\bfnm{A.}\binits{A.}},
  \bauthor{\bsnm{Chernozhukov},~\bfnm{V.}\binits{V.}} \AND
  \bauthor{\bsnm{Kato},~\bfnm{K.}\binits{K.}}
(\byear{2015}).
\btitle{Uniform post-selection inference for least absolute deviation
  regression and other {Z}-estimation problems}.
\bjournal{Biometrika}
\bvolume{102}
\bpages{77--94}.
\end{barticle}
\endbibitem

\bibitem[\protect\citeauthoryear{Bentkus}{1986}]{bentkus1986dependence}
\begin{barticle}[author]
\bauthor{\bsnm{Bentkus},~\bfnm{V}\binits{V.}}
(\byear{1986}).
\btitle{Dependence of the Berry-Esseen estimate on the dimension}.
\bjournal{Lithuanian Mathematical Journal}
\bvolume{26}
\bpages{110--114}.
\end{barticle}
\endbibitem

\bibitem[\protect\citeauthoryear{Bentkus}{2003}]{bentkus2003dependence}
\begin{barticle}[author]
\bauthor{\bsnm{Bentkus},~\bfnm{Vidmantas}\binits{V.}}
(\byear{2003}).
\btitle{On the dependence of the Berry--Esseen bound on dimension}.
\bjournal{Journal of Statistical Planning and Inference}
\bvolume{113}
\bpages{385--402}.
\end{barticle}
\endbibitem

\bibitem[\protect\citeauthoryear{Blanchet, Kang and
  Murthy}{2019}]{blanchet2019robust}
\begin{barticle}[author]
\bauthor{\bsnm{Blanchet},~\bfnm{Jose}\binits{J.}},
  \bauthor{\bsnm{Kang},~\bfnm{Yang}\binits{Y.}} \AND
  \bauthor{\bsnm{Murthy},~\bfnm{Karthyek}\binits{K.}}
(\byear{2019}).
\btitle{Robust wasserstein profile inference and applications to machine
  learning}.
\bjournal{Journal of Applied Probability}
\bvolume{56}
\bpages{830--857}.
\end{barticle}
\endbibitem

\bibitem[\protect\citeauthoryear{Cai, Liu and Xia}{2013}]{cai2013two}
\begin{barticle}[author]
\bauthor{\bsnm{Cai},~\bfnm{Tony}\binits{T.}},
  \bauthor{\bsnm{Liu},~\bfnm{Weidong}\binits{W.}} \AND
  \bauthor{\bsnm{Xia},~\bfnm{Yin}\binits{Y.}}
(\byear{2013}).
\btitle{Two-sample covariance matrix testing and support recovery in
  high-dimensional and sparse settings}.
\bjournal{Journal of the American Statistical Association}
\bvolume{108}
\bpages{265--277}.
\end{barticle}
\endbibitem

\bibitem[\protect\citeauthoryear{Chang et~al.}{2017}]{chang2016comparing}
\begin{barticle}[author]
\bauthor{\bsnm{Chang},~\bfnm{Jinyuan}\binits{J.}},
  \bauthor{\bsnm{Zhou},~\bfnm{Wen}\binits{W.}},
  \bauthor{\bsnm{Zhou},~\bfnm{Wen-Xin}\binits{W.-X.}} \AND
  \bauthor{\bsnm{Wang},~\bfnm{Lan}\binits{L.}}
(\byear{2017}).
\btitle{Comparing large covariance matrices under weak conditions on the
  dependence structure and its application to gene clustering}.
\bjournal{Biometrics}
\bvolume{73}
\bpages{31--41}.
\end{barticle}
\endbibitem

\bibitem[\protect\citeauthoryear{Chatterjee}{2006}]{chatterjee2006generalization}
\begin{barticle}[author]
\bauthor{\bsnm{Chatterjee},~\bfnm{Sourav}\binits{S.}}
(\byear{2006}).
\btitle{A generalization of the Lindeberg principle}.
\bjournal{The Annals of Probability}
\bvolume{34}
\bpages{2061--2076}.
\end{barticle}
\endbibitem

\bibitem[\protect\citeauthoryear{Chen et~al.}{2018}]{chen2018gaussian}
\begin{barticle}[author]
\bauthor{\bsnm{Chen},~\bfnm{Xiaohui}\binits{X.}} \betal{et~al.}
(\byear{2018}).
\btitle{Gaussian and bootstrap approximations for high-dimensional U-statistics
  and their applications}.
\bjournal{The Annals of Statistics}
\bvolume{46}
\bpages{642--678}.
\end{barticle}
\endbibitem

\bibitem[\protect\citeauthoryear{Chen, Genovese and
  Wasserman}{2015}]{chen2015asymptotic}
\begin{barticle}[author]
\bauthor{\bsnm{Chen},~\bfnm{Yen-Chi}\binits{Y.-C.}},
  \bauthor{\bsnm{Genovese},~\bfnm{Christopher~R}\binits{C.~R.}} \AND
  \bauthor{\bsnm{Wasserman},~\bfnm{Larry}\binits{L.}}
(\byear{2015}).
\btitle{Asymptotic theory for density ridges}.
\bjournal{The Annals of Statistics}
\bvolume{43}
\bpages{1896--1928}.
\end{barticle}
\endbibitem

\bibitem[\protect\citeauthoryear{Chen, Genovese and
  Wasserman}{2016}]{chen2016density}
\begin{barticle}[author]
\bauthor{\bsnm{Chen},~\bfnm{Yen-Chi}\binits{Y.-C.}},
  \bauthor{\bsnm{Genovese},~\bfnm{Christopher~R}\binits{C.~R.}} \AND
  \bauthor{\bsnm{Wasserman},~\bfnm{Larry}\binits{L.}}
(\byear{2016}).
\btitle{Density level sets: Asymptotics, inference, and visualization}.
\bjournal{Journal of the American Statistical Association}
\bvolume{just-accepted}.
\end{barticle}
\endbibitem

\bibitem[\protect\citeauthoryear{Chernozhukov, Chetverikov and
  Kato}{2013}]{chernozhukov2013}
\begin{barticle}[author]
\bauthor{\bsnm{Chernozhukov},~\bfnm{Victor}\binits{V.}},
  \bauthor{\bsnm{Chetverikov},~\bfnm{Denis}\binits{D.}} \AND
  \bauthor{\bsnm{Kato},~\bfnm{Kengo}\binits{K.}}
(\byear{2013}).
\btitle{Gaussian approximations and multiplier bootstrap for maxima of sums of
  high-dimensional random vectors}.
\bjournal{Annals of Statistics}
\bvolume{41}
\bpages{2786--2819}.
\bdoi{10.1214/13-AOS1161}
\end{barticle}
\endbibitem

\bibitem[\protect\citeauthoryear{Chernozhukov, Chetverikov and
  Kato}{2015}]{chernozhukov2015comparison}
\begin{barticle}[author]
\bauthor{\bsnm{Chernozhukov},~\bfnm{Victor}\binits{V.}},
  \bauthor{\bsnm{Chetverikov},~\bfnm{Denis}\binits{D.}} \AND
  \bauthor{\bsnm{Kato},~\bfnm{Kengo}\binits{K.}}
(\byear{2015}).
\btitle{Comparison and anti-concentration bounds for maxima of Gaussian random
  vectors}.
\bjournal{Probability Theory and Related Fields}
\bvolume{162}
\bpages{47--70}.
\end{barticle}
\endbibitem

\bibitem[\protect\citeauthoryear{Chernozhukov, Chetverikov and
  Kato}{2017}]{chernozhukov2017central}
\begin{barticle}[author]
\bauthor{\bsnm{Chernozhukov},~\bfnm{Victor}\binits{V.}},
  \bauthor{\bsnm{Chetverikov},~\bfnm{Denis}\binits{D.}} \AND
  \bauthor{\bsnm{Kato},~\bfnm{Kengo}\binits{K.}}
(\byear{2017}).
\btitle{Central limit theorems and bootstrap in high dimensions}.
\bjournal{The Annals of Probability}
\bvolume{45}
\bpages{2309--2352}.
\end{barticle}
\endbibitem

\bibitem[\protect\citeauthoryear{Dezeure, B{\"u}hlmann and
  Zhang}{2017}]{dezeure2016high}
\begin{barticle}[author]
\bauthor{\bsnm{Dezeure},~\bfnm{Ruben}\binits{R.}},
  \bauthor{\bsnm{B{\"u}hlmann},~\bfnm{Peter}\binits{P.}} \AND
  \bauthor{\bsnm{Zhang},~\bfnm{Cun-Hui}\binits{C.-H.}}
(\byear{2017}).
\btitle{High-dimensional simultaneous inference with the bootstrap}.
\bjournal{Test}
\bvolume{26}
\bpages{685--719}.
\end{barticle}
\endbibitem

\bibitem[\protect\citeauthoryear{Efron}{1979}]{efron1979}
\begin{barticle}[author]
\bauthor{\bsnm{Efron},~\bfnm{Bradley}\binits{B.}}
(\byear{1979}).
\btitle{Bootstrap methods: Another look at the jackknife.}
\bjournal{Annals of Statistics}
\bvolume{7}
\bpages{1--26}.
\end{barticle}
\endbibitem

\bibitem[\protect\citeauthoryear{Fan and Lv}{2008}]{fanlv07}
\begin{barticle}[author]
\bauthor{\bsnm{Fan},~\bfnm{J.}\binits{J.}} \AND
  \bauthor{\bsnm{Lv},~\bfnm{J.}\binits{J.}}
(\byear{2008}).
\btitle{Sure independence screening for ultra-high dimensional feature space
  (with discussion)}.
\bjournal{Journal of the Royal Statistical Society Series B}
\bvolume{70}
\bpages{849--911}.
\end{barticle}
\endbibitem

\bibitem[\protect\citeauthoryear{Fan and Zhou}{2016}]{fan2016guarding}
\begin{barticle}[author]
\bauthor{\bsnm{Fan},~\bfnm{Jianqing}\binits{J.}} \AND
  \bauthor{\bsnm{Zhou},~\bfnm{Wen-Xin}\binits{W.-X.}}
(\byear{2016}).
\btitle{Guarding against Spurious Discoveries in High Dimensions}.
\bjournal{Journal of Machine Learning Research}
\bvolume{17}
\bpages{1--34}.
\end{barticle}
\endbibitem

\bibitem[\protect\citeauthoryear{Freedman}{1975}]{freedman1975tail}
\begin{barticle}[author]
\bauthor{\bsnm{Freedman},~\bfnm{David~A}\binits{D.~A.}}
(\byear{1975}).
\btitle{On tail probabilities for martingales}.
\bjournal{the Annals of Probability}
\bpages{100--118}.
\end{barticle}
\endbibitem

\bibitem[\protect\citeauthoryear{Gin\'e and Zinn}{1990}]{gine1990}
\begin{barticle}[author]
\bauthor{\bsnm{Gin\'e},~\bfnm{Evarist}\binits{E.}} \AND
  \bauthor{\bsnm{Zinn},~\bfnm{Joel}\binits{J.}}
(\byear{1990}).
\btitle{Bootstrapping General Empirical Measures}.
\bjournal{Annals of Probability}
\bvolume{18}
\bpages{851--869}.
\bdoi{10.1214/aop/1176990862}
\end{barticle}
\endbibitem

\bibitem[\protect\citeauthoryear{Gotze}{1991}]{gotze1991rate}
\begin{barticle}[author]
\bauthor{\bsnm{Gotze},~\bfnm{F}\binits{F.}}
(\byear{1991}).
\btitle{On the rate of convergence in the multivariate CLT}.
\bjournal{The Annals of Probability}
\bpages{724--739}.
\end{barticle}
\endbibitem

\bibitem[\protect\citeauthoryear{Hall}{1988}]{hall1988theoretical}
\begin{barticle}[author]
\bauthor{\bsnm{Hall},~\bfnm{Peter}\binits{P.}}
(\byear{1988}).
\btitle{Theoretical comparison of bootstrap confidence intervals}.
\bjournal{The Annals of Statistics}
\bpages{927--953}.
\end{barticle}
\endbibitem

\bibitem[\protect\citeauthoryear{Hall and
  Presnell}{1999}]{hall1999intentionally}
\begin{barticle}[author]
\bauthor{\bsnm{Hall},~\bfnm{Peter}\binits{P.}} \AND
  \bauthor{\bsnm{Presnell},~\bfnm{Brett}\binits{B.}}
(\byear{1999}).
\btitle{Intentionally biased bootstrap methods}.
\bjournal{Journal of the Royal Statistical Society: Series B (Statistical
  Methodology)}
\bvolume{61}
\bpages{143--158}.
\end{barticle}
\endbibitem

\bibitem[\protect\citeauthoryear{Horowitz}{2019}]{horowitz2019bootstrap}
\begin{barticle}[author]
\bauthor{\bsnm{Horowitz},~\bfnm{Joel~L}\binits{J.~L.}}
(\byear{2019}).
\btitle{Bootstrap methods in econometrics}.
\bjournal{Annual Review of Economics}
\bvolume{11}.
\end{barticle}
\endbibitem

\bibitem[\protect\citeauthoryear{Klivans, O'Donnell and
  Servedio}{2008}]{klivans2008learning}
\begin{binproceedings}[author]
\bauthor{\bsnm{Klivans},~\bfnm{Adam~R}\binits{A.~R.}},
  \bauthor{\bsnm{O'Donnell},~\bfnm{Ryan}\binits{R.}} \AND
  \bauthor{\bsnm{Servedio},~\bfnm{Rocco~A}\binits{R.~A.}}
(\byear{2008}).
\btitle{Learning geometric concepts via Gaussian surface area}.
In \bbooktitle{Foundations of Computer Science, 2008. FOCS'08. IEEE 49th Annual
  IEEE Symposium on}
\bpages{541--550}.
\bpublisher{IEEE}.
\end{binproceedings}
\endbibitem

\bibitem[\protect\citeauthoryear{Lindeberg}{1922}]{lindeberg1922neue}
\begin{barticle}[author]
\bauthor{\bsnm{Lindeberg},~\bfnm{Jarl~Waldemar}\binits{J.~W.}}
(\byear{1922}).
\btitle{Eine neue Herleitung des Exponentialgesetzes in der
  Wahrscheinlichkeitsrechnung}.
\bjournal{Mathematische Zeitschrift}
\bvolume{15}
\bpages{211--225}.
\end{barticle}
\endbibitem

\bibitem[\protect\citeauthoryear{Liu}{1988}]{liu1988bootstrap}
\begin{barticle}[author]
\bauthor{\bsnm{Liu},~\bfnm{Regina~Y}\binits{R.~Y.}}
(\byear{1988}).
\btitle{Bootstrap procedures under some non-iid models}.
\bjournal{The Annals of Statistics}
\bvolume{16}
\bpages{1696--1708}.
\end{barticle}
\endbibitem

\bibitem[\protect\citeauthoryear{Mammen}{1993}]{mammenwild1993}
\begin{barticle}[author]
\bauthor{\bsnm{Mammen},~\bfnm{Enno}\binits{E.}}
(\byear{1993}).
\btitle{Bootstrap and Wild Bootstrap for High Dimensional Linear Models}.
\bjournal{Annals of Statistics}
\bvolume{21}
\bpages{255-285}.
\end{barticle}
\endbibitem

\bibitem[\protect\citeauthoryear{Nagaev}{1976}]{nagaev1976estimate}
\begin{binproceedings}[author]
\bauthor{\bsnm{Nagaev},~\bfnm{SV}\binits{S.}}
(\byear{1976}).
\btitle{An estimate of the remainder term in the multidimensional central limit
  theorem}.
In \bbooktitle{Proceedings of the Third Japan—USSR Symposium on Probability
  Theory}
\bpages{419--438}.
\bpublisher{Springer}.
\end{binproceedings}
\endbibitem

\bibitem[\protect\citeauthoryear{Nazarov}{2003}]{nazarov2003maximal}
\begin{barticle}[author]
\bauthor{\bsnm{Nazarov},~\bfnm{Fedor}\binits{F.}}
(\byear{2003}).
\btitle{On the Maximal Perimeter of a Convex Set in {$\mathbb{R}^n$} with
  Respect to a Gaussian Measure}.
\bjournal{Geometric Aspects of Functional Analysis}
\bpages{169--187}.
\end{barticle}
\endbibitem

\bibitem[\protect\citeauthoryear{Ning and Liu}{2017}]{ning2017general}
\begin{barticle}[author]
\bauthor{\bsnm{Ning},~\bfnm{Yang}\binits{Y.}} \AND
  \bauthor{\bsnm{Liu},~\bfnm{Han}\binits{H.}}
(\byear{2017}).
\btitle{A general theory of hypothesis tests and confidence regions for sparse
  high dimensional models}.
\bjournal{The Annals of Statistics}
\bvolume{45}
\bpages{158--195}.
\end{barticle}
\endbibitem

\bibitem[\protect\citeauthoryear{Pr{\ae}stgaard and
  Wellner}{1993}]{praestgaard1993exchangeably}
\begin{barticle}[author]
\bauthor{\bsnm{Pr{\ae}stgaard},~\bfnm{Jens}\binits{J.}} \AND
  \bauthor{\bsnm{Wellner},~\bfnm{Jon~A}\binits{J.~A.}}
(\byear{1993}).
\btitle{Exchangeably weighted bootstraps of the general empirical process}.
\bjournal{The Annals of Probability}
\bpages{2053--2086}.
\end{barticle}
\endbibitem

\bibitem[\protect\citeauthoryear{Sazonov}{1981}]{sazonov1981normal}
\begin{bbook}[author]
\bauthor{\bsnm{Sazonov},~\bfnm{V.~V.}\binits{V.~V.}}
(\byear{1981}).
\btitle{Normal Approximation: Some Recent Advances}.
\bseries{Lecture Notes in Computer Science}
\bvolume{no. 879}.
\bpublisher{Springer-Verlag}.
\end{bbook}
\endbibitem

\bibitem[\protect\citeauthoryear{Senatov}{1980}]{senatov1980several}
\begin{barticle}[author]
\bauthor{\bsnm{Senatov},~\bfnm{Vladimir~Vasil'evich}\binits{V.~V.}}
(\byear{1980}).
\btitle{Several uniform estimates of the rate of convergence in the
  multidimensional central limit theorem}.
\bjournal{Teoriya Veroyatnostei i ee Primeneniya}
\bvolume{25}
\bpages{757--770}.
\end{barticle}
\endbibitem

\bibitem[\protect\citeauthoryear{Shao and Tu}{2012}]{shao2012jackknife}
\begin{bbook}[author]
\bauthor{\bsnm{Shao},~\bfnm{Jun}\binits{J.}} \AND
  \bauthor{\bsnm{Tu},~\bfnm{Dongsheng}\binits{D.}}
(\byear{2012}).
\btitle{The jackknife and bootstrap}.
\bpublisher{Springer Science \& Business Media}.
\end{bbook}
\endbibitem

\bibitem[\protect\citeauthoryear{Singh}{1981}]{singh1981asymptotic}
\begin{barticle}[author]
\bauthor{\bsnm{Singh},~\bfnm{Kesar}\binits{K.}}
(\byear{1981}).
\btitle{On the asymptotic accuracy of Efron's bootstrap}.
\bjournal{The Annals of Statistics}
\bpages{1187--1195}.
\end{barticle}
\endbibitem

\bibitem[\protect\citeauthoryear{Slepian}{1962}]{slepian1962one}
\begin{barticle}[author]
\bauthor{\bsnm{Slepian},~\bfnm{David}\binits{D.}}
(\byear{1962}).
\btitle{The One-Sided Barrier Problem for Gaussian Noise}.
\bjournal{Bell System Technical Journal}
\bvolume{41}
\bpages{463--501}.
\end{barticle}
\endbibitem

\bibitem[\protect\citeauthoryear{Steiger}{1969}]{steiger1969best}
\begin{barticle}[author]
\bauthor{\bsnm{Steiger},~\bfnm{WL}\binits{W.}}
(\byear{1969}).
\btitle{A best possible Kolmogoroff-type inequality for martingales and a
  characteristic property}.
\bjournal{The Annals of Mathematical Statistics}
\bpages{764--769}.
\end{barticle}
\endbibitem

\bibitem[\protect\citeauthoryear{Stein}{1981}]{stein1981estimation}
\begin{barticle}[author]
\bauthor{\bsnm{Stein},~\bfnm{Charles~M}\binits{C.~M.}}
(\byear{1981}).
\btitle{Estimation of the mean of a multivariate normal distribution}.
\bjournal{The annals of Statistics}
\bpages{1135--1151}.
\end{barticle}
\endbibitem

\bibitem[\protect\citeauthoryear{Wu}{1986}]{wu1986jackknife}
\begin{barticle}[author]
\bauthor{\bsnm{Wu},~\bfnm{Chien-Fu~Jeff}\binits{C.-F.~J.}}
(\byear{1986}).
\btitle{Jackknife, bootstrap and other resampling methods in regression
  analysis}.
\bjournal{Annals of Statistics}
\bvolume{14}
\bpages{1261--1295}.
\end{barticle}
\endbibitem

\bibitem[\protect\citeauthoryear{Zhang and Cheng}{2017}]{ZhangCheng2016}
\begin{barticle}[author]
\bauthor{\bsnm{Zhang},~\bfnm{Xianyang}\binits{X.}} \AND
  \bauthor{\bsnm{Cheng},~\bfnm{Guang}\binits{G.}}
(\byear{2017}).
\btitle{Simultaneous inference for high-dimensional linear models}.
\bjournal{Journal of the American Statistical Association}
\bpages{1--12}.
\end{barticle}
\endbibitem

\bibitem[\protect\citeauthoryear{Zhang and Wu}{2017a}]{zhang2017gaussian}
\begin{barticle}[author]
\bauthor{\bsnm{Zhang},~\bfnm{Danna}\binits{D.}} \AND
  \bauthor{\bsnm{Wu},~\bfnm{Wei~Biao}\binits{W.~B.}}
(\byear{2017}a).
\btitle{Gaussian approximation for high dimensional time series}.
\bjournal{The Annals of Statistics}
\bvolume{45}
\bpages{1895--1919}.
\end{barticle}
\endbibitem

\bibitem[\protect\citeauthoryear{Zhang and Wu}{2017b}]{zhangd2017gaussian}
\begin{barticle}[author]
\bauthor{\bsnm{Zhang},~\bfnm{Danna}\binits{D.}} \AND
  \bauthor{\bsnm{Wu},~\bfnm{Wei~Biao}\binits{W.~B.}}
(\byear{2017}b).
\btitle{Gaussian approximation for high dimensional time series}.
\bjournal{The Annals of Statistics}
\bvolume{45}
\bpages{1895--1919}.
\end{barticle}
\endbibitem

\bibitem[\protect\citeauthoryear{Zhang and Zhang}{2014}]{zhangzhang11}
\begin{barticle}[author]
\bauthor{\bsnm{Zhang},~\bfnm{Cun-Hui}\binits{C.-H.}} \AND
  \bauthor{\bsnm{Zhang},~\bfnm{Stephanie~S.}\binits{S.~S.}}
(\byear{2014}).
\btitle{Confidence intervals for low dimensional parameters in high dimensional
  linear models}.
\bjournal{Journal of the Royal Statistical Society, Series B}
\bvolume{76}
\bpages{217--242}.
\end{barticle}
\endbibitem

\bibitem[\protect\citeauthoryear{Zhilova}{2016}]{zhilova2016non}
\begin{barticle}[author]
\bauthor{\bsnm{Zhilova},~\bfnm{Mayya}\binits{M.}}
(\byear{2016}).
\btitle{Non-classical Berry-Esseen inequality and accuracy of the weighted
  bootstrap}.
\bjournal{arXiv preprint arXiv:1611.02686}.
\end{barticle}
\endbibitem

\end{thebibliography}

\newpage

\centerline{\bf \large Supplement to ``Beyond Gaussian Approximation:}
\centerline{\bf \large Bootstrap for Maxima of Sums of Independent Random Vectors"}
\bigskip

\noindent
This supplement contains proofs of all the theoretical results stated in the main body of the paper.

\bigskip
{\bf A1. Proofs of the results in Section 2.}

\bigskip
{\bf A1.1.  Proof of Lemma \ref{lm:cp-decomp}. }
Let $t_1$ be the $1-\alpha - \eta$ quantile of $T_n$,
\bes
\P\Big\{T_n < t_1 \Big\} \le 1-\alpha - \eta \le \P\Big\{T_n\le t_1 \Big\}.
\ees
Let $\eta_n^*(\eps_n,t)=\eta_n^{{(\P^*)}}(\eps_n, t; T_n^0, T_n^*)$.
By the definition of $\omega_n(\eps_n;T_n)$,
\bes
\P\big\{ t_1-\eps_n < T_n \le t_1\}
= \lim_{\delta\to 0+} \P\big\{ t_1-\eps_n + \delta < T_n < t_1+\delta\} \le \omega_n(\eps_n;T_n),
\ees
so that $1-\alpha -\eta \le \P\big\{T_n\le t_1 -\eps_n \big\}+\omega_n(\eps_n;T_n)$.
It follows that
\bes
&& (1-\alpha) - \P\{T_n\le t^*_\alpha \}
\cr &\le& (1-\alpha) - \P\{T_n\le t_1-\eps_n,t^*_\alpha \ge t_1 -\eps_n \}
\cr &=& (1-\alpha) - \P\{T_n\le t_1 -\eps_n \} + \P\{T_n\le t_1-\eps_n,t^*_\alpha  < t_1-\eps_n\}
\cr &\le& \eta + \omega_n(\eps_n;T_n) +  \P\{t^*_\alpha  <  t_1-\eps_n \}.
\ees
It follows from the definition of $\eta$ in (\ref{eta_n-boot}) and Theorem \ref{th-comp-empi} that
\bes
&& \P\big\{t^*_\alpha + \eps_n < t_1\big\}
\cr &\le& \P\Big[ \P^*\big\{T_n^* < t_1 - \eps_n\big\} > 1 - \alpha \ge \P\{T_n < t_1\} + \eta\Big]
\cr &\le & \P\Big[ \eta_n^*(\eps_n,t_1) > \eta \Big].
\ees
Hence, $(1-\alpha) - \P\{T_n\le t^*_\alpha \} \le \eta  + \omega_n(\eps_n;T_n)
+ \P\big\{ \eta_n^*(\eps_n,t_1) > \eta \big\}$.

Let $t_2$ be the $1-\alpha+\eta$ quantile of $T_n$.
When $1-\alpha+\eta<\P\{T_n\le t_2\}$,
\bes
&&  \P\{T_n \le  t^*_\alpha \} -\omega_n(\eps_n;T_n)-(1-\alpha +\eta)
\cr &\le& \P\{ t^*_\alpha \ge t_2 + \eps_n \} + \P\{T_n < t_2 -\eps_n \} -\omega_n(\eps_n;T_n) - \P \{T_n < t_2 \}
\cr &\le&  \P\{ t^*_\alpha \ge t_2 + \eps_n \}
\cr &\le&  \P\Big[ \P^*\big\{T_n^* < t_2 + \eps_n\big\} \le 1 - \alpha < \P\{T_n \le t_2\} - \eta \Big]
\cr &\le & \P\Big[ \eta_n^*(\eps_n,t_2+\eps_n) > \eta \Big].
\ees
When $1-\alpha+\eta=\P\{T_n\le t_2\}$,
\bes
&&  \P\{T_n \le  t^*_\alpha \} -\omega_n(\eps_n;T_n)-(1-\alpha +\eta)
\cr &\le& \P\{ t^*_\alpha > t_2 + \eps_n \} + \P\{T_n \le t_2 -\eps_n \} -\omega_n(\eps_n;T_n) - \P \{T_n \le t_2 \}
\cr &\le&  \P\{ t^*_\alpha > t_2 + \eps_n \}
\cr &\le&  \P\Big[ \P^*\big\{T_n^* < t_2 + \eps_n\big\} < 1 - \alpha = \P\{T_n \le t_2\} - \eta \Big]
\cr &\le & \P\Big[ \eta_n^*(\eps_n,t_2+\eps_n) > \eta \Big].
\ees
Thus, the conclusion holds in all cases. $\hfill\square$

\medskip
{\bf A1.2.  Proof of Theorem \ref{cb-empi}.} 
Without loss of generality, we assume $\E X_i =0$ for all $i$.

We first prove \eqref{th-empi-new-2}. It follows from Lemma \ref{lm:cp-decomp} and the definition of $\eta_n^*(\eps)$ in (\ref{eta_n-boot}) that
\bel{pf-empi-0-new}
\sup_{t\in \R}\Big|\P\Big\{T_n < t\Big\} - \P^*\Big\{T_n^* < t \Big\} \Big| 
\le \eta + \omega_n^{(\P)}(1/b_n;T_n)
\eel
with probability at least $\P\{ \eta_n^*(1/b_n) \le \eta \}$. We prove \eqref{th-empi-new-2} in the following two steps.

\emph{Step 1.} Let $\eta$ be the right-hand side of (\ref{th-comp-empi-2}) in Theorem \ref{th-comp-empi}. By (\ref{th-comp-empi-2}) and Theorem \ref{anti-general},
\bel{pf-empi-1-new}
&& \sup_{t\in \R}\Big|\P\Big\{T_n < t\Big\} - \P^*\Big\{T_n^* < t \Big\} \Big| 
\\ \nonumber &\le& C_{c_1,c_2} b_n^2(\log(p/{{\eps_n}}))^{3/2}M_4^2/n^{1/2}
+ \P\Big\{\|\bX\|_{\max}>\atil_n\Big\} + 2 {{\eps_n}} 
\\ \nonumber  && \qquad \qquad \qquad \qquad + \,
\frac{C_0'}{b_n\linesigma}\sqrt{\log p}+ C_{c_1}\kappa_{n,4}+ 2\P\Big\{\|\bX\|_{\max}>\atil_n\Big\}
\\
\nonumber &\le& C_{c_1,c_2}' b_n^2(\log(p/{{\eps_n}}))^{3/2}M_4^2/n^{1/2}
+ \frac{C_0'}{b_n\linesigma}\sqrt{\log p} + 3\, \P\Big\{\|\bX\|_{\max}>\atil_n\Big\} + 2{{\eps_n}}
\eel
with probability at least $1 - \big( \P \{\|\bX\|_{\max} > \atil_n \} + 2{{\eps_n}}\big) $, provided the condition 
$\log(p/\eps_n)\le c_2 n$ of Theorem \ref{th-comp-empi} holds. 
Note the $a_n = c_1 \sqrt{n} /(b_n \log p)$ in Theorem \ref{anti-general} is greater than $\atil_n = c_1 \sqrt{n} /(b_n \log (p/\eps_n)$ here.

We balance the first two quantities on the right-hand side of \eqref{pf-empi-1-new} by letting
	\bes
	b_n M^{2/3}\linesigma^{1/3} = \Big(\frac{n^{1/2}\sqrt{\log p}}{(\log (p/\eps_n))^{3/2}}\Big)^{1/3} = \frac{n^{1/6}(\log p)^{1/6}}{(\log (p/\eps_n))^{1/2}}.
	\ees
	It follows from \eqref{pf-empi-1-new}, the $b_n M^{2/3}\linesigma^{1/3}$ above and $M \ge M_4$ that
	\bel{pf-empi-2-new}
	&& \sup_{t\in \R}\Big|\P\Big\{T_n < t\Big\} - \P^*\Big\{T_n^* < t \Big\} \Big|
	\cr
	&\le& C_{c_1,c_2}' b_n^2(\log(p/{{\eps_n}}))^{3/2}M_4^2/n^{1/2}
	+ \frac{C_0'}{b_n\linesigma}\sqrt{\log p} + 3\, \P\Big\{\|\bX\|_{\max}>\atil_n\Big\} + 2\eps_n
	\cr
	&\le& C' \bigg( \frac{(\log p)^2 (\log (p/\eps_n)^3)}{n} \frac{M^4}{\linesigma^4} \bigg)^{1/6}  + 3\, \P\Big\{\|\bX\|_{\max}>\atil_n\Big\} + 2{{\eps_n}}
	\eel
	with at least probability $1 - \big( 2{{\eps_n}} + \P\{\|\bX\|_{\max} > \atil_n\} \big) $.
	Here we set $c_1=c_2=1$. The condition $\log(p/\eps_n)\le c_2n = n$ is then satisfied since the above bound is trivial otherwise. The constant $C_{c_1, c_2}'$ can then be replaced with a constant that doesn't depend on $c_1$ or $c_2$. 

	When ${{\delta}} \le \gamma^*_{{{\delta}}, M}$, we let $\eps_n = {{\delta}} /4$ so that $\log(p/\eps_n) \le \log(4p/{{\delta}})$. Otherwise, we let $\eps_n = 1/(4n)$, which is smaller then $\gamma^*_{{{\delta}}, M}/4$ and therefore ${{\delta}}/4$, so that $\log(p/\eps_n) \le \log(4np)$. In either case, we have $\log(p/\eps_n) \le \log(4np/{{\delta}})$. By \eqref{th-empi-new-1},
	\bes
	\P\{\|\bX\|_{\max} > \atil_n\} &=& \P\Big\{\|\bX\|_{\max} > \frac{n^{1/3}M^{2/3}\linesigma^{1/3}}{(\log p)^{1/6}(\log (p/{{\eps_n}}))^{1/2}}  \Big\}
	\cr
	&\le&  \P\Big\{\|\bX\|_{\max} > \frac{n^{1/3}M^{2/3}\linesigma^{1/3}}{(\log p)^{1/6}(\log (4np/{{\delta}}))^{1/2}}  \Big\}
	\cr
	&\le& \min \bigg\{ \frac{{{\delta}}}{2}, \frac{1}{2}\bigg(\frac{(\log p)^2 (\log(np/{{\delta}}))^3}{n} \frac{M^4}{\linesigma^4} \bigg)^{1/6} \bigg\},
	\ees
	so that with at least probability $1 - {{\delta}}$,
	\bes
	\sup_{t\in \R}\Big|\P\Big\{T_n < t\Big\} - \P^*\Big\{T_n^* < t \Big\} \Big| \le C_0 \gamma^*_{{{\delta}}, M}.
	\ees 

	\emph{Step 2.} Let $\eta$ be the lower bound of $\eta_n^*(1/b_n)$ in (\ref{th-comp-empi-3}) of Theorem \ref{th-comp-empi}. By Theorem \ref{th-comp-empi}, Theorem \ref{anti-general} and the second inequality of \eqref{lm-truncate-new-1} in Lemma \ref{lm-truncate-new}, it also holds with at least probability $1-{{\delta}}$ that
	\bes
	&& \sup_{t\in \R}\Big|\P\Big\{T_n < t\Big\} - \P^*\Big\{T_n^* < t \Big\} \Big|
	\cr
	&\le& C_{c_1=1, c_2=1} \bigg( {{\delta}} + \frac{b_n^4(\log(p/{{\delta}}))^{3} \M_4^4}{ {{\delta}} n} \bigg)
	+ \frac{C_0'}{b_n\linesigma}\sqrt{\log p} + C_{c_1=1} \kappa_{n,4} 
	\cr
	&& \qquad \qquad \qquad +  2\,\P\bigg\{\max_{1\le j\le p}\bigg|\sum_{i=1}^n \frac{X_{i,j}I_{\{|X_{i,j}|> \sqrt{n}/(b_n \log p) \}}}{\sqrt{n}}\bigg| >\frac{1}{8b_n}\bigg\}.
	\cr
	&\le& C' \bigg( {{\delta}} + \frac{b_n^4(\log(p/{{\delta}}))^{3}}{ {{\delta}} n}  \M_4^4
	+ \frac{\sqrt{\log p}}{b_n\linesigma} \bigg) +  2 \frac{b_n^4(\log p)^{3} }{n} \M_{4,2}^4
	\cr
	&\le& C_0 \bigg[ {{\delta}} + \bigg( \frac{(\log p)^2 (\log (p/{{\delta}}))^3}{{{\delta}} n} \frac{\M_4^4}{\linesigma^4} \bigg)^{1/5} \bigg]
	\cr
	&\le& C_0 \bigg[ {{\delta}} + \big( \gamma^*_{{{\delta}}, \M_4} \big)^{6/5} \big/{{\delta}}^{1/5} \bigg].
	\ees
	This bound is effective when ${{\delta}} \le \gamma^*_{{{\delta}}, \M_4}$, making it dominated by the second term, $( \gamma^*_{{{\delta}}, \M_4} )^{6/5}/{{\delta}}^{1/5}$. When ${{\delta}}$ is greater, we simply take the bound when ${{\delta}} = \gamma^*_{{{\delta}}, \M_4}$, yielding $\gamma^*_{{{\delta}}, \M_4}$.
	The proof of \eqref{th-empi-new-2} is then complete.

\medskip
Finally, we prove \eqref{th-empi-new-3}. It follows from Lemma \ref{lm:cp-decomp} and the definition of $\eta_n^*(\eps)$ in (\ref{eta_n-boot}) that
\bel{pf-empi-0}
\Big| \P\{T_n\le t^*_\alpha \} - (1-\alpha) \Big|
\le \P\Big\{ \eta_n^*(1/b_n) > \eta \Big\} +\eta  + \omega_n(1/b_n;T_n).
\eel
By the $\eta$ in the Step 1 above and $\eps_n = 1/n$ , \eqref{th-empi-new-3} holds with $\gamma_n^*$ being the first component. The second component follows from the $\eta$ as in the Step 2 above and $\eps_n = {{\delta}} = \gamma^*_{{{\delta}}, \M_4}$. We omit the details. 
$\hfill\square$

\medskip
{\bf A1.3. Proof of Theorem \ref{cb-wild}.} 	
The proof is similar to that of Theorem \ref{cb-empi} as the major difference is just the replacement of the stronger maximum fourth moment condition $\M_4$ in 
Theorem~\ref{th-comp-empi} by the weaker $\M_{4,2}$ in Theorem \ref{th-comp-wild-a} when $p$ is large.
In this application of Theorem \ref{th-comp-wild-a}, we pick $\eps_n = \min\{{{\delta}}/4, \overline{\eps}_n \} \le \overline{\eps}_n = 1/n$ for the first component of $\gamma^*_{{{\delta}}, M}$ and 
$\eps_n = \min\{ {{\delta}}$, $\gamma^*_{{{\delta}}, \M_{4,2}} \}  \le \overline{\eps}_n 
= \gamma^*_{{{\delta}}, \M_{4,2}}$ for the second component. We omit further details. $\hfill\square$

\medskip
{\bf A1.4.  Proof of Theorem \ref{cb-rade}.}
For the Rademacher multiplier, (\ref{subgaussian}) holds with $\tau_0=1$.
Let $m^*=6$ and $\eta = \big\{\eta_n^{{(2)}}(1/b_n)\big\}^{2/3}$. It follows from
Lemma \ref{lm:cp-decomp}, (\ref{th-comp-wild-2}) of Theorem \ref{th-comp-wild}
and the definitions of  $\omega_n(\eps,T_n)$ and $\kappa_{n,6}$ that
\bes
&& \Big| \P\{T_n\le t^*_\alpha \} - (1-\alpha) \Big|
\cr &\le& \sup_{t \in \R} \P\Big\{ \eta_n^{{(\P^*)}}(1/b_n, t ; T_n^0, T_n^*) > \eta \Big\} +\eta  + \omega_n(1/b_n;T_n)
\cr &\le& 2\big\{\eta_n^{{(2)}}(1/b_n)\big\}^{2/3}
+ \sup_{t \in \R} \P\big\{ t - 1/b_n \le T_n \le t\big\}
\cr &\le& C_{{c_1}}\bigg\{\bigg(
\frac{b_n^2\log p}{n^{1/2}}M^2\bigg)^{2/3}+\kappa_{n,6}^{1/3}\bigg\}
+ \sup_{t \in \R} \P\bigg\{ t - \frac{\sqrt{\log p}}{b_nM} \le \sqrt{\log p}\,\frac{T_n}{M} \le t\bigg\}
\cr && +2\bigg[\E \min\bigg\{2,C_{c_1}\frac{b_n^6(\log p)^{6-1}}{n^3}
\max_{1\le j\le p}\sum_{i=1}^n |X_{i,j}|^6I_{\{|X_{i,j}|>a_n\}}\bigg\}\bigg]^{1/3},
\cr &\le& C_{{c_1}}\bigg[
\bigg\{\bigg(\frac{b_nM}{\sqrt{\log p}}\bigg)^2\bigg(\frac{\log p}{n^{1/4}}\bigg)^2\bigg\}^{2/3}
+\bigg\{\bigg(\frac{b_n M}{\sqrt{\log p}}\bigg)^6\bigg(\frac{\log p}{n^{1/4}}\bigg)^{8}\bigg\}^{1/3}\bigg]
\cr && + \sup_{t \in \R} \P\bigg\{ t - \frac{\sqrt{\log p}}{b_nM} \le \sqrt{\log p}\,\frac{T_n}{M} \le t\bigg\}
\cr && +\bigg[\E \min\bigg\{4,C_{c_1}
\bigg(\frac{b_n M}{\sqrt{\log p}}\bigg)^6\bigg(\frac{\log p}{n^{1/4}}\bigg)^{8}
\max_{1\le j\le p}\sum_{i=1}^n \frac{X_{i,j}^6}{M^6n}I_{\{|X_{i,j}|>a_n\}}\bigg\}\bigg]^{1/3},
\ees
with $a_n = c_1\sqrt{n}/(b_n\log p) = c_1M\sqrt{\log p}
\big(n^{1/4}/\log p\big)^2\big(\sqrt{\log p}/(b_n M)\big)$.
Let $b_n$ be the real number satisfying $\sqrt{\log p}/(b_n M) = c_0(\log p/n^{1/4})^{4/7}$, we have
\bes
&& \Big| \P\{T_n\le t^*_\alpha \} - (1-\alpha) \Big|
\cr &\le& C_{{c_1}}\bigg[
\bigg\{\frac{1}{c_0^2}\bigg(\frac{\log p}{n^{1/4}}\bigg)^{2-8/7}\bigg\}^{2/3}
+\bigg\{\frac{1}{c_0^6}\bigg(\frac{\log p}{n^{1/4}}\bigg)^{8-24/7}\bigg\}^{1/3}\bigg]
\cr && + \sup_{t \in \R} \P\bigg\{ t - c_0\bigg(\frac{\log p}{n^{1/4}}\bigg)^{4/7} \le \sqrt{\log p}\,\frac{T_n}{M} \le t\bigg\}
\cr && +\bigg[\E \min\bigg\{4,
\frac{C_{c_1}}{c_0^6}\bigg(\frac{\log p}{n^{1/4}}\bigg)^{8-24/7}
\max_{1\le j\le p}\sum_{i=1}^n \frac{X_{i,j}^6}{M^6n}I_{\{|X_{i,j}|>a_n\}}\bigg\}\bigg]^{1/3},
\cr &\le& C_{{c_1}}\bigg\{
\frac{1}{c_0^{4/3}}\bigg(\frac{\log p}{n^{1/4}}\bigg)^{4/7}
+\frac{1}{c_0^2}\bigg(\frac{\log p}{n^{1/4}}\bigg)^{32/21}\bigg\}
\cr && + \sup_{t \in \R} \P\bigg\{ t - c_0\bigg(\frac{\log p}{n^{1/4}}\bigg)^{4/7} \le \sqrt{\log p}\,\frac{T_n}{M} \le t\bigg\}
\cr && +\bigg[\E \min\bigg\{4,
\frac{C_{c_1}}{c_0^6}\bigg(\frac{\log p}{n^{1/4}}\bigg)^{32/7}
\max_{1\le j\le p}\sum_{i=1}^n \frac{X_{i,j}^6}{M^6n}I_{\{|X_{i,j}|>a_n\}}\bigg\}\bigg]^{1/3},
\ees
with $a_n = c_1c_0M\sqrt{\log p}\big(n^{1/4}/\log p\big)^{10/7}$.

Similarly by (\ref{KS-bound}) and (\ref{th-comp-wild-3}) of Theorem \ref{th-comp-wild}
\bes
\bigg(\E \sup_{t\in \R}\Big|\P\Big\{T_n < t\Big\} - \P^*\Big\{T_n^* < t \Big\} \Big|^2\bigg)^{1/2}
 \le (3/2^{1/3})\big\{\eta_n^{{(2)}}(1/b_n)\big\}^{2/3}+\omega_n(1/b_n;T_n).
\ees
This completes the proof.
$\hfill\square$

\medskip
{\bf A1.5.  Proof of Corollary \ref{eg-1}.} By (E.1),
\bes
\P\Big\{\|\bX  - \E \bX\|_{\max} > \frac{n^{1/3}M^{2/3}\linesigma^{1/3}}{(\log p)^{1/6}(\log(4np/{{\delta}}))^{1/2}} \Big\} &\le& 2(np) \exp \Big(-\frac{n^{1/3}M^{2/3}\linesigma^{1/3}}{B_n(\log p)^{1/6}(\log(4np/{{\delta}}))^{1/2}} \Big)
\cr
&\le& \min\{ {{\delta}}/2, 1/(2n) \}  \le {{\delta}}/(2n)
\ees
holds with 
\bes
\Big(\frac{M}{\linesigma}\Big)^{2/3} = \max \bigg\{ \Big( \frac{M_4}{\linesigma}\Big)^{2/3}, \frac{(\log p)^{1/6} (\log (4np/{{\delta}}))^{1/2} (\log (4n^2p/{{\delta}}))}{n^{1/3}} \frac{B_n}{\linesigma} \bigg\}.
\ees
The proof of (i) is complete by inserting the above quantity to $\gamma^*_{{{\delta}}, M}$.

Similarly, we let 
\bes
\Big(\frac{M}{\linesigma}\Big)^{2/3} = \max \bigg\{ \Big( \frac{M_4}{\linesigma}\Big)^{2/3}, \frac{(\log p)^{1/6}(\log (np))^{1/3}  (\log (4np/{{\delta}}))^{1/6} (\log (4n^2p/{{\delta}}))}{n^{1/3}} \frac{B_n}{\linesigma} \bigg\}
\ees
in (ii). $\hfill\square$

\medskip
{\bf A1.6} {\bf Proof of Corollary \ref{eg-2}.}
Observe that the third term on the right-hand side of (\ref{cb-rade-1}) 
is bounded by $\big[4\P\{\|\bX  - \E \bX\|_{\max} > a_n\}\big]^{1/3}$.
Under the sub-Gaussian tail probability condition in (E.2),
\bes
\P{\{\|\bX  - \E \bX\|_{\max} >a_n\}}  
\le 2(np) \exp(-a_n^2/B_n^2)
\le 2/(np)
\ees
when $a_n = c_1M\sqrt{\log p}\big(n^{1/4}/\log p\big)^{10/7} \ge B_n\sqrt{2\log(np)}$. Thus, we are allowed to take
\bes
M = M_6\vee\Big[B_n\sqrt{2\log(np)}\big/\big\{c_1\sqrt{\log p}\big(n^{1/4}/\log p\big)^{10/7}\big\}\Big].
\ees
On the other hand, by the anti-concentration inequality in Theorem \ref{anti-concentration} of Section 4,
\bes
\sup_{ t \in \R} \P\bigg\{ t - c_0\bigg(\frac{\log p}{n^{1/4}}\bigg)^{4/7}
\le \sqrt{\log p}\,\frac{T_n}{M} \le t\bigg| {\cal A} \bigg\}
\le C_{c_0}\bigg(\frac{\log p}{n^{1/4}}\bigg)^{4/7}\frac{M}{\linesigma}.
\ees
The conclusion follows as the above term dominates other terms in the error bound
and $c_0$ and $c_1$ can be treated as numerical constants. $\hfill\square$

\medskip
{\bf A1.7} {\bf Proof of Corollary \ref{eg-3}.}
By (E.3),
    \bes
    && \P\Big\{\|\bX  - \E \bX\|_{\max} > \frac{n^{1/3}M^{2/3}\linesigma^{1/3}}{(\log p)^{1/6}(\log(4np/{{\delta}}))^{1/2}} \Big\} 
    \cr
    &\le& \sum_{i=1}^n \P\bigg\{\|X_i  - \E X_i\|_{\infty}> \frac{n^{1/3}M^{2/3}\linesigma^{1/3}}{(\log p)^{1/6}(\log(4np/{{\delta}}))^{1/2}} \bigg\}
    \cr
    &\le& n \bigg(\frac{B_n(\log p)^{1/6}(\log(4np/{{\delta}}))^{1/2}}{n^{1/3}M^{2/3}\linesigma^{1/3}} \bigg)^q. 
    \ees
    Let $M$ be the smallest positive number satisfying both 
    $M \ge M_4$ and
    \bes
    n \bigg(\frac{B_n(\log p)^{1/6}(\log(4np/{{\delta}}))^{1/2}}{n^{1/3}M^{2/3}\linesigma^{1/3}} \bigg)^q
    \le \min\bigg\{ \frac{{{\delta}}}{2},  \frac{1}{2}\bigg(\frac{(\log p)^2 (\log(np/{{\delta}}))^3}{n} \frac{M^4}{\linesigma^4} \bigg)^{1/6} \bigg\},
    \ees
    that is,
    \bes
    \Big(\frac{M}{\linesigma} \Big)^{2/3} &=& \max \bigg\{ \Big(\frac{M_4}{\linesigma} \Big)^{2/3}, 2^{1/q} \frac{ (\log p)^{1/6}(\log (4np/{{\delta}}))^{1/2}}{{{\delta}}^{1/q}n^{1/3-1/q}} \frac{B_n}{\linesigma}, 
    \cr
    && \qquad \qquad 2^{1/q}\Big( \frac{(\log p)^{q-2}(\log (4np/{{\delta}}))^{3q}}{n^{2q-7}( \log (np/{{\delta}}))^3}\Big)^{\frac{1}{6(q+1)}}\Big( \frac{B_n}{\linesigma} \Big)^{q/(q+1)} \bigg\}.
    \ees
    It follows that
    \bes
    \gamma^*_{{{\delta}}, M} &\lesssim& \max \bigg\{\Big(\frac{(\log p)^2(\log(np/{{\delta}}))^3}{n} \frac{M_4^4}{\linesigma^4} \Big)^{1/6}, \frac{(\log p)^{1/2}(\log (np/{{\delta}}))}{{{\delta}}^{1/q}n^{1/2-1/q}}\frac{B_n}{\linesigma}, 
    \cr
    && \qquad \qquad \qquad \Big(\frac{(\log p)^{1/2}(\log (np/{{\delta}}))}{n^{1/2-1/q}} \frac{B_n}{\linesigma}\Big)^{\frac{q}{(q+1)}}\bigg\}.
    \ees

    We omit the proof of (ii) as it's similar. The proof is complete. $\hfill\square$

\bigskip
{\bf A2. Proofs  of the results in Section 3}

\bigskip
{\bf A2.1. Proof of Lemma \ref{k-nonspecific}.}
Let $\scrA_{k,i} =\{(A,B): A\cup B = (1:(i-1))\cup((i+1):n), |A|=k-1, |B| = n-k\}$ and
$\scrA_k =\{(A,B): A\cup B = 1:n, |A|=k, |B| = n-k\}$.
Let $X_A=\{X_i, i\in A\}$ and $X^*_B=\{X^*_i, i\in B\}$. We have
\bes
\sum_{k=1}^n \sum_{\sigma}I_{\{\sigma_k=i\}}f(\bU_{\sigma,k},\zeta_{k,i})
= \sum_{k=1}^n \sum_{(A,B)\in \scrA_{k,i}} c_{n,k} f(X_A,X^*_B,\zeta_{k,i})
\ees
where $c_{n,k} ={\#} \big\{\sigma: \sigma_\ell\in A\ \ \forall\ \ell<k,\sigma_k=i\big\} = (k-1)!(n-k)!$.
We observe that
\bes
&& \sum_{(A,B)\in \scrA_k}f(X_A,X^*_B)
\cr &=& \sum_{(A,B)\in \scrA_k,i\in A}  f(X_A,X^*_B)
+ \sum_{(A,B)\in \scrA_k,i\in B} f(X_A,X^*_B)I_{\{k<n\}}
\cr &=& \sum_{(A,B)\in \scrA_{k,i}}  f(X_A,X^*_B,X_i)
+ \sum_{(A,B)\in \scrA_{k+1,i}} f(X_A,X^*_B,X^*_i)I_{\{k<n\}}.
\ees
Let $c_{n,k}=0$ for $k<1$ or $k>n$. As {$c_{n+1,k+1}=kc_{n,k}$ for $1\le k\le n$
and $c_{n+1,k+1} = (n-k)c_{n,k+1}$ for $0\le k \le n-1$,}
\bes
&& \sum_{k=1}^n \frac{c_{n+1,k+1}}{n+1}\sum_{(A,B)\in \scrA_k}f(X_A,X^*_B)
+ \frac{c_{n+1,1}}{n+1}f(X^*_1,\ldots,X_n^*)
\cr &=& \sum_{k=1}^n \frac{k c_{n,k}}{n+1}\sum_{(A,B)\in \scrA_{k,i}}  f(X_A,X^*_B,X_i)
+ \sum_{k=0}^{n-1}\frac{(n-k)c_{n,k+1}}{n+1}\sum_{(A,B)\in \scrA_{k+1,i}} f(X_A,X^*_B,X^*_i)
\cr &=& \sum_{k=1}^n c_{n,k}\sum_{(A,B)\in \scrA_{k,i}}\Big\{\frac{k }{n+1} f(X_A,X^*_B,X_i)
+ \frac{(n+1-k)}{n+1}f(X_A,X^*_B,X^*_i)\Big\}
\cr &=& {\sum_{k=1}^n c_{n,k}\sum_{(A,B)\in \scrA_{k,i}}\E\Big[f(X_A,X^*_B,\zeta_{k,i})\Big|X_i,X_i^*, i\le n\Big]}
\cr &=& {\E\bigg[\sum_{k=1}^n \sum_{\sigma,\sigma_k=i} f(\bU_{\sigma,k},\zeta_{k,i})\bigg|X_i,X_i^*, i\le n\bigg].}
\cr &=& {n!\A_{\sigma,k}\Big(I_{\{\sigma_k=i\}}f(\bU_{\sigma,{k}},\zeta_{{k},{i}})\Big)}
\ees
{The proof is complete as the left-hand side above does not depend on $i$.}
$\hfill\square$

\medskip
{\bf A2.2. Properties of the smooth max function}
We study here
\bes
F_\beta(z) = \beta^{-1} \log\Big( e^{\beta z_1}+\cdots+e^{\beta z_p}\Big),
\quad z=(z_1,...,z_p)^T.
\ees
The lemmas below are straightforward extensions of similar calculations
in \cite{chernozhukov2013} to higher order derivatives. For $z=(z_1,...,z_p)^T$ let
\bes
\pi_j(z) = \frac{e^{\beta z_j}}{\sum_{k=1}^p e^{\beta z_k}},
\ees
and for positive integers $m$ define
\bes
\pi^{(m)} = \frac{(\pa/\pa z)^{\otimes m}\sum_{k=1}^p e^{\beta z_k}}
{\beta^m \sum_{k=1}^p e^{\beta z_k}}
= \diag\Big(\pi_1(z),\ldots,\pi_p(z)\Big)
\ees
as $m$-dimensional diagonal tensors in $\R^{p\times \cdots\times p}$, and
\bes
K_m = \Big\{(k_1,\ldots,k_m): k_j\ge 0, k_1+2k_2+\cdots+ mk_m = m\Big\},
\ees
where $k_j$ are integers. For $m$-dimensional tensors $B$ in $\R^{p\times \cdots\times p}$, define
\bes
\hbox{\rm Sym}(B)
= \bigg(\frac{1}{m!}\sum_\sigma B_{j_{\sigma_1},\ldots,j_{\sigma_m}}\bigg)_{p\times \cdots\times p}
\ees
where the summation is taken over all permutations of $\{1,\ldots,m\}$.
The following lemma gives the derivatives of $F_\beta$ and relates $F_\beta(z)$ to $\|z\|_\infty$.

\begin{lemma}\label{Fbeta}
Let $F_\beta^{(m)}(z) =(\pa/\pa z)^{\otimes m}F_\beta(z)$.
For all $z=(z_1,...,z_p)^T$ integers $m\ge 1$,
\bes
\beta^{1-m} F_\beta^{(m)}(z)
= \sum_{(k_1,\ldots,k_m)\in K_m}\frac{m!(k-1)!(-1)^{k-1}}{k_1!\cdots k_m!}
\hbox{\rm Sym}\bigg(\otimes_{1\le j\le m, k_j > 0} \bigg(\frac{\pi^{(j)}(z)}{j!}\bigg)^{\otimes k_j} \bigg)
\ees
where $k=k_1+\cdots+k_m$.
Consequently, for $C_m = \sum_{(k_1,\ldots,k_m)\in K_m}m!(k-1)!/
\otimes_{j=1}^m \{k_j!(j!)^{k_j}\}$,
\bes
\big\|\beta^{1-m}F_\beta^{(m)}(z)\big\|_1 \le C_m.
\ees
In particular, with $F_\beta^{(1)} = F_\beta^{(1)}(z)$ and $\pi^{(m)} = \pi^{(m)}(z)$
\bes
F_\beta^{(1)} &=& \pi^{(1)}
\cr \beta^{-1} F_\beta^{(2)} &=& \pi^{(2)} - \pi^{(1,1)}
\cr \beta^{-2} F_\beta^{(3)} &=& \pi^{(3)} - 3\pi^{(2,1)} + 2\pi^{(1,1,1)}
\cr \beta^{-3} F_\beta^{(4)} &=& \pi^{(4)} - 4\pi^{(3,1)} - 3\pi^{(2,2)}
+ 12\pi^{(2,1,1)} - 6\pi^{(1,1,1,1)}
\ees
with $\pi^{(k_1,\ldots,k_m)} = ${\rm Sym}$\big(\otimes_{1\le j\le m, k_j > 0}\pi^{(k_j)}\big)$, and
$C_1=1$, $C_2=2$, $C_3=6$ and $C_4=26$.
\end{lemma}

We omit the proof of Lemma \ref{Fbeta} as it is an immediately consequence of
the Faa di Bruno formula. The Faa di Bruno formula also yields the following lemma. Let
\bes
G_\beta^{(m)}(z)
= \sum_{(k_1,\ldots,k_m)\in K_m}\frac{m!(k-1)!}{k_1!\cdots k_m!}
\hbox{\rm Sym}\bigg(\otimes_{1\le j\le m, k_j > 0} \bigg(\frac{\pi^{(j)}(z)}{j!}\bigg)^{\otimes k_j} \bigg)
\ees
and for positive constants $b$ and $\beta$ define
\bes
H^{(m)}_{b,\beta}(z)
= \sum_{(k_1,\ldots,k_m)\in K_m}\frac{m!\|h^{(k)}\|_\infty {b}^k\beta^{m-k}}{k_1!\cdots k_m!}
\hbox{\rm Sym}\bigg(\otimes_{1\le j\le m, k_j > 0}
\bigg(\frac{G_{\beta}^{(j)}(z)}{j!}\bigg)^{\otimes k_j} \bigg)
\ees

\medskip
\begin{lemma}\label{f-derivatives}
Let $h(\cdot)$ be a smooth function and $z = \sum_{i=1}^n x_i/\sqrt{n}$. Then,
\bes
n^{m/2}\bigg(\frac{\pa}{\pa x_n}\bigg)^{\otimes m} h\big({b} F_{\beta}(z)\big)
= \sum_{(k_1,\ldots,k_m)\in K_m}\frac{m!{b}^kh^{(k)}\big({b} F_{\beta}(z)\big)}{k_1!\cdots k_m!}
\hbox{\rm Sym}\bigg(\otimes_{1\le j\le m, k_j > 0}
\bigg(\frac{F_{\beta}^{(j)}(z)}{j!}\bigg)^{\otimes k_j} \bigg)
\ees
where $h^{(k)}(t)=(d/dt)^kh(t)$ and $k = k_1+\cdots+k_m$.
Consequently,
\bes
\big|n^{m/2}(\pa/\pa x_n)^{\otimes m} h({b} F_{\beta}(z))\big| \le
H^{(m)}_{b,\beta}(z)
\ees
and with $C_{h,m} = \sum_{(k_1,\ldots,k_m)\in K_m} m!\|h^{(k)}\|_\infty
\prod_{j=1}^m (C_j/j!)^{k_j}/k_j!$
\bes
\|H^{(m)}_{b,\beta}(z)\|_1 \le \sum_{(k_1,\ldots,k_m)\in K_m}
\frac{m!\|h^{(k)}\|_\infty {b}^k\beta^{m-k}}{k_1!\cdots k_m!}
\prod_{j=1}^m \bigg(\frac{C_j}{j!}\bigg)^{k_j} \le C_{h,m}\max\big({b}^m,{b}\beta^{m-1}\big).
\ees
\end{lemma}

\medskip
\begin{lemma}\label{H-stability}
Let $H^{(m)}_{b,\beta}$ be as in Lemma \ref{f-derivatives}. We have
\bes
e^{-2m\|t\|_{\infty}\beta}H^{(m)}_{b,\beta}(z+t) \le H^{(m)}_{b,\beta}(z)
\le e^{2m\|t\|_{\infty}\beta}H^{(m)}_{b,\beta}(z+t).
\ees
\end{lemma}
{\sc Proof.} We have
\bes
\pi_j(z+t) = \frac{e^{(z_j+t_j)\beta}}{\sum_{s=1}^p e^{(z_s+t_s)\beta}} \le \frac{e^{z_j\beta}}{\sum_{s=1}^p e^{z_s\beta+ (t_s-t_j)\beta}} \le \frac{e^{z_j\beta}}{\sum_{s=1}^p e^{z_s\beta}} \ {e^{\max_{s} \{(t_j-t_s)\beta\}}} \le e^{2 \|t\|_{\infty}\beta }\pi_j(z)
\ees
and similarly $\pi_j(z) \le e^{2\|t\|_{\infty}\beta} \pi_j(z+t) $. As each element of $H^{(m)}_{b,\beta}$
is a positive weighted sum of products of no more than $m$ such $\pi_{j}(z)$, the claim follows.
$\hfill\square$

\medskip
{\bf A2.3. Proof of Theorem \ref{comparison}.}
It follows from (\ref{Lindeberg-2}) that
\bes
\Delta_n(f) = \E\Big\{f(X_1,\ldots,X_n) - f(X_1^*,\ldots,X_n^*) \Big\}
= \sum_{m=2}^{m^*-1}\A_\sigma\big(\Delta_{n,m,\sigma}\big) + \A_\sigma\big(\Rem_\sigma\big)
\ees
with $\A_\sigma\big(\Rem_\sigma\big)  = \uppercase\expandafter{\romannumeral 1\relax}
+\uppercase\expandafter{\romannumeral 2\relax}$,
where
\bes
 \uppercase\expandafter{\romannumeral 1\relax} &=&
 n\, \A_{\sigma,k}\bigg(
\E \int_0^1 \left\langle {f}^{(m^*)}(\bU_{\sigma,{k}},tX_{\sigma_k}),
\frac{(1-t)^{(m^*-1)}}{(m^*-1)!}X_{\sigma_k}^{\otimes m^*}\right\rangle dt\bigg) \\
\uppercase\expandafter{\romannumeral 2\relax} &=& -  n\, \A_{\sigma,k}\bigg(
\E \int_0^1 \left\langle {f}^{(m^*)}(\bU_{\sigma,{k}},tX_{\sigma_k}^*),
\frac{(1-t)^{(m^*-1)}}{(m^*-1)!}(X_{\sigma_k}^*)^{\otimes m^*}\right\rangle dt\bigg).
\ees

Let $\zeta_{{k},{i}}$ be as in Lemma \ref{k-nonspecific}.
By (\ref{approx-Delta_{n,m}}) and the definition of $F^{{(m)}}$,
the leading term can be written as
\bel{decomp3}
\sum_{m=1}^{m^*-1}\A_\sigma\big(\Delta_{n,m,\sigma}\big)
&=&  \sum_{m=2}^{m^*-1}\frac{n}{m!} \, \A_{\sigma,k}
\left\langle \E {f}^{(m,0)}(\bU_{\sigma,{k}},\zeta_{{k},{\sigma_k}}),
\mu^{{(m)}}-\nu^{{(m)}}\right\rangle
+ \uppercase\expandafter{\romannumeral 3\relax}
\cr &=& \sum_{m=2}^{m^*-1} \left\langle F^{{(m)}}, \mu^{{(m)}}-\nu^{{(m)}}\right\rangle + \uppercase\expandafter{\romannumeral 3\relax}
\eel
where
\bes
\uppercase\expandafter{\romannumeral 3\relax}
= - \sum_{m=2}^{m^*-1}\frac{n}{m!} \, \A_{\sigma,k}
\left\langle {f}^{(m,0)}(\bU_{\sigma,{k}},{\zeta_{{k},{\sigma_k}}}) - \E {f}^{(m,0)}(\bU_{\sigma,{k}},0),
\E X_{\sigma_k}^{\otimes m} - \E (X_{\sigma_k}^* )^{\otimes m}\right\rangle
\ees
Hence, we could re-write $\Delta_n(f)$ as
\bes
\Delta_n(f) = \sum_{m=2}^{m^*-1} \left\langle F^{{(m)}}, \mu^{{(m)}}-\nu^{{(m)}}\right\rangle + \Rem
\ees
with the remainder term $\Rem = \uppercase\expandafter{\romannumeral 1\relax}+\uppercase\expandafter{\romannumeral 2\relax}+\uppercase\expandafter{\romannumeral 3\relax}$.

\medskip
It remains to bound all three terms of the above $\Rem$.
By (\ref{cond-1a}), (\ref{cond-1b}), Lemma \ref{k-nonspecific} and
the independence of $\bU_{\sigma,k}$ and $X_{\sigma_k}$ for each $\sigma$,
\bes
|\uppercase\expandafter{\romannumeral 1\relax}|
&\le& n\, \A_{\sigma,k}\bigg(
\E \int_0^1 \left\langle \Big|{f}^{(m^*)}(\bU_{\sigma,{k}},tX_{\sigma_k})\Big|,
\frac{(1-t)^{(m^*-1)}}{(m^*-1)!}\big|X_{\sigma_k}\big|^{\otimes m^*}\right\rangle dt\bigg)
\cr &\le&n\,   \A_{\sigma,k}\bigg(
\int_0^1 \left\langle \E \fbar^{{(m^*)}}(\bU_{\sigma,{k}},0),
\frac{(1-t)^{(m^*-1)}}{(m^*-1)!}\E |X_{\sigma_k}|^{\otimes m^*}g\big(\|X_{\sigma_k}\|/u_n\big)\right\rangle dt \bigg)
\cr &=& n\,  \A_{\sigma,k}
\left\langle \frac{\E\big\{\fbar^{{(m^*)}}(\bU_{\sigma,{k}},0)/g\big(\|\zeta_{{k},{\sigma_k}}\|/u_n\big)\big\}}
{\E \big\{1/g\big(\|\zeta_{{k},{\sigma_k}}\|/u_n\big)\big\}},
\frac{1}{m^*!}\E |X_{\sigma_k}|^{\otimes m^*}g\big(\|X_{\sigma_k}\|/u_n\big)\right\rangle
\cr &\le& n\,  \A_{\sigma,k}
\left\langle \E\big\{{f}^{(m^*)}_{\max}(\bU_{\sigma,{k}},\zeta_{{k},{\sigma_k}})\big\},
\frac{\E |X_{\sigma_k}|^{\otimes m^*}g\big(\|X_{\sigma_k}\|/u_n\big)}{m^*! G_{\sigma_k}}\right\rangle
\cr &=& \left\langle n\,  \A_{\sigma,k}\Big(
\E {f}^{(m^*)}_{\max}(\bU_{\sigma,{k}},\zeta_{{k},{\sigma_k}})\Big),
\sum_{{i}=1}^n \frac{\E |X_{i}|^{\otimes m^*}g\big(\|X_{i}\|/u_n\big)}{nG_{i}}\right\rangle
\cr &\le & \left\langle F_{\max}^{{(m^*)}},\mu_{\max}^{{(m^*)}}\right\rangle.
\ees
The second and third inequalities above follows from (\ref{cond-1a}) and (\ref{cond-1b}) respectively,
the second equality follows from Lemma \ref{k-nonspecific}, and the last inequality follows from
the H\"{o}lder inequality. Similarly,
\bes
|\uppercase\expandafter{\romannumeral 2\relax}| \le \left\langle F_{\max}^{{(m^*)}},\mu_{\max}^{{(m^*)}}\right\rangle.
\ees

For the third term of $\Rem$, we note that ${\zeta_{{k},{i}}}$ is also centered, so that by Taylor's expansion
\bes
&& |\uppercase\expandafter{\romannumeral 3 \relax}|
\cr &=& \Bigg|\sum_{m=2}^{m^*-1}\frac{n}{m!} \A_{\sigma,k}
\left\langle \E {f}^{(m,0)}(\bU_{\sigma,{k}},{\zeta_{{k},{i}}}) - \E {f}^{(m,0)}(\bU_{\sigma,{k}},0),
\E X_{i}^{\otimes m} - \E (X_{i}^* )^{\otimes m}\right\rangle\Bigg|
\cr &=& \bigg|\sum_{m=2}^{m^*-1}\frac{n}{m!} \A_{\sigma,k}
\E \int_0^1  \bigg\langle {f}^{(m,m^*-m)}(\bU_{\sigma,{k}}, t {\zeta_{{k},{i}}}),
\cr &&\qquad\qquad
\frac{(1-t)^{m^*-1-m}}{(m^*-1-m)!}{{\zeta_{{k},{i}}}}^{\otimes(m^*-m)}\otimes
\Big(\E X_{i}^{\otimes m} - \E(X_{i}^*)^{\otimes m}\Big)\bigg\rangle dt \bigg|
\cr &\le& \sum_{m=2}^{m^*-1}{m^*\choose m}\frac{n}{m^*!}
\A_{\sigma,k}\left\langle {\E \bar{f}^{(m^*)}(\bU_{\sigma,{k}}, 0)},
\E \big|\zeta_{{k},{i}}^{\otimes(m^*-m)}g\big(\|\zeta_{{k},{i}}\|/u_n\big)\big|\otimes
\big|\E X_{i}^{\otimes m} - \E(X_{i}^*)^{\otimes m}\big|\right\rangle
\cr &=& \sum_{m=2}^{m^*-1} {m^*\choose m}\frac{n}{m^*!}\A_{\sigma,k}
\left\langle \E\, \frac{\bar{f}^{(m^*)}(\bU_{\sigma,{k}}, 0)}{g(\|\zeta_{{k},{i}}\|/u_n)},
\frac{\E \big|\zeta_{{k},{i}}^{\otimes(m^*-m)}g\big(\|\zeta_{{k},{i}}\|/u_n\big)\big|\otimes
\big|\E X_{i}^{\otimes m} - \E(X_{i}^*)^{\otimes m}\big|}{\E\{ 1\big/g(\|\zeta_{{k},{i}}\|/u_n)\}}
\right\rangle
\cr &\le& \sum_{m=2}^{m^*-1} {m^*\choose m}\frac{n}{m^*!}\A_{\sigma,k}
\Bigg\langle \E\,f_{\max}^{{(m^*)}}(\bU_{\sigma,{k}},\zeta_{{k},{i}}),
\frac{\E \big|X_{i}^{\otimes(m^*-m)}g\big(\|X_{i}\|/u_n\big)\big|\otimes
\big|\E X_{i}^{\otimes m} - \E(X_{i}^*)^{\otimes m}\big|}{G_{i}}
\Bigg\rangle
\cr && + \sum_{m=2}^{m^*-1}{m^*\choose m}\frac{n}{m^*!}
\A_{\sigma,k}\Bigg\langle \E\,f_{\max}^{{(m^*)}}(\bU_{\sigma,{k}},\zeta_{{k},{i}}),
\cr &&\qquad\qquad
\frac{\E \big|(X_{i}^*)^{\otimes(m^*-m)}g\big(\|X_{i}^*\|/u_n\big)\big|\otimes
\big|\E X_{i}^{\otimes m} - \E(X_{i}^*)^{\otimes m}\big|}{G_{i}}
\Bigg\rangle
\cr & = & \sum_{m=2}^{m^*-1} {m^*\choose m}
\bigg\langle F_{\max}^{{(m^*)}},\sum_{{i}=1}^n
\frac{\E \big|X_{i}^{\otimes(m^*-m)}g\big(\|X_{i}\|/u_n\big)\big|\otimes
\big|\E X_{i}^{\otimes m} - \E(X_{i}^*)^{\otimes m}\big|}{nG_{i}}
\bigg\rangle
\cr && + \sum_{m=2}^{m^*-1} {m^*\choose m}
\bigg\langle F_{\max}^{{(m^*)}},\sum_{{i}=1}^n
\frac{\E \big|(X_{i}^*)^{\otimes(m^*-m)}g\big(\|X_{i}^*\|/u_n\big)\big|\otimes
\big|\E X_{i}^{\otimes m} - \E(X_{i}^*)^{\otimes m}\big|}{nG_{i}}
\bigg\rangle
\cr & \le & \sum_{m=2}^{m^*-1}{m^*\choose m}\left\langle F_{\max}^{{(m^*)}},4\mu_{\max}^{{(m^*)}}\right\rangle.
\ees
Again, the second and third inequalities above follows from (\ref{cond-1a}) and (\ref{cond-1b}) respectively,
the last equality follows from Lemma \ref{k-nonspecific}, and the last inequality follows from
the H\"{o}lder inequality.
The conclusion follows.

\medskip
{\bf A2.4. Proof of Theorem \ref{th-comp-gene}.}
Let $h_0$ be a smooth decreasing function taking value 1 in $(-\infty,-1]$ and
0 in $[0,\infty)$. Let $h_t(\cdot) = h_0(\cdot - t)$ be the location shift of $h_0$ and
$\beta_n = 2b_n\log p$. We have $h_{2b_nt_0}(2b_nt) = 1$ for $t\le t_0-1/(2b_n)$ and
$h_{2b_nt_0}(2b_nt)=0$ for $t > t_0$.
Let $z_{\max} = \max(z_1,\ldots,z_p)$ and $F_\beta$ be as in (\ref{smooth-max}).
Because $z_{\max} \le F_\beta(z) \le z_{\max} + \beta_n^{-1}\log p  = z_{\max} + 1/(2b_n)$,
\bes
I\{z_{\max} \le t_0-1/b_n\}\le I\{F_{\beta_n}(z) \le t_0-1/(2b_n)\}
\le h\big(2b_nF_{\beta_n}(z)\big) \le h\big(2b_n z_{\max}\big) \le I\{z_{\max} < t_0\}
\ees
with $h(t) =h_{2b_nt_0}(t)$.
Thus, by the definition of $\eta_n(\cdot)$ in (\ref{Prokhorov})
%
\bel{pf-th-comp-gene-1}
\eta_n(1/b_n) &=& \sup_{t \in \R} \max\Big[\P\Big\{T_n^*\le t - 1/b_n\Big\} - \P\Big\{T_n < t \Big\},
\P\Big\{T_n\le t - 1/b_n\Big\} - \P\Big\{T_n^* < t \Big\},0\Big]
\cr &\le& \sup_{t \in \R }\Big\{\Big|\E\Big(f(X_1^*,\ldots,X_n^*)-f(X_1^0,\ldots,X_n^0)\Big)\Big|:
f = h_t\circ (2b_nF_{\beta_n})\Big\}.
\eel
By the definition of $F^{{(m)}}$ and $F_{\max}^{{(4)}}$ and Lemmas \ref{f-derivatives} and \ref{H-stability},
$n^{m/2-1}\|F^{{(m)}}\|_1/(b_n\beta_n^{m-1})$ and \\$n^{m/2-1}\|F^{{(m)}}_{\max}\|_1/(b_n\beta_n^{m-1})$
are all bounded by constants depending on $m$ only,
so that (\ref{th-comp-gene-1}) follows directly from an application of
Theorem~\ref{comparison} to the right-hand side of (\ref{pf-th-comp-gene-1}).

For (\ref{th-comp-gene-2}) we apply Theorem \ref{comparison} to
$\Ttil_n = \max_j \sum_{i=1}^n \Xtil_{i,j}/\sqrt{n}$ and
$\Ttil_n^* = \max_j \sum_{i=1}^n \Xtil_{i,j}^*/\sqrt{n}$.
It follows from (\ref{triangle}) and the definition of $\Omega_0$ and $\Omega_0^*$ in (\ref{Omega_0}) that
\bel{pf-th4-2}
&& \eta_n^{{(\P)}}\big(1/b_n;T_n,T_n^*\big)
\cr &\le& \eta_n^{{(\P)}}\big((1/(4b_n))-;T_n,\Ttil_n\big) + \eta_n^{{(\P)}}\big(1/(2b_n);\Ttil_n,\Ttil_n^*\big)
+ \eta_n^{{(\P)}}\big((1/(4b_n))-;\Ttil_n^*,T_n^*\big)
\cr &\le& \P\big\{\Omega_0\big\} + \eta_n^{{(\P)}}\big(1/(2b_n);\Ttil_n,\Ttil_n^*\big)
+ \P\big\{\Omega_0^*\big\}
\eel
Thus, (\ref{th-comp-gene-2}) follows from (\ref{th-comp-gene-1}) due to the boundedness
of $\|\tbX\|_{\max}/u_n$ and $\|\tbX^*\|_{\max}/u_n$. $\hfill\square$

\medskip
{\bf A2.5. Proof of Lemma \ref{lm-truncate-new}.} 
We write $X_{i,j}-\Xtil_{i,j} = X_{i,j,1} + X_{i,j,2}$ with
\bes
\quad X_{i,j,1} = X_{i,j}I_{\{|X_{i,j}|> \atil_n\}}-\E X_{i,j}I_{\{|X_{i,j}|> \atil_n\}},\quad
X_{i,j,2} = X_{i,j}I_{\{a_n<|X_{i,j}|\le \atil_n\}}
- \E X_{i,j}I_{\{a_n<|X_{i,j}|\le \atil_n\}}.
\ees
The second inequality in (\ref{lm-truncate-new-1}) follows from
\bel{pf-lm-truncate-1}
\P\big\{\Omegatil_0\big\}
&\le & \E \max_{1\le j\le p}\frac{8b_n}{n^{1/2}}\sum_{i=1}^n|X_{i,j}|\,I_{\{|X_{i,j}|> \atil_n\}}
\cr &\le& C_0 b_nn^{1/2}\atil_n^{1-m}\E\bigg(\max_{1\le j\le p}\sum_{i=1}^n \frac{|X_{i,j}|^m}{n}\bigg)
\cr &\le& C_{m,c_1} b_n^m(\log (p/{{\eps_n}}))^{m-1}\M_{m,2}^m n^{1-m/2}.
\eel

By the definition of $\Omega_0$ and $\Omegatil_0$,
\bel{pf-lm-truncate-2}
&& \max_{1\le j\le p}\bigg(\bigg|\frac{16b_n}{n^{1/2}}\sum_{i=1}^n \E X_{i,j}I_{\{|X_{i,j}| > \atil_n\}}\bigg|\wedge
\bigg|\frac{16b_n}{n^{1/2}}\sum_{i=1}^n X_{i,j,2}\bigg|\bigg) \ge 1
\ \hbox{ in }\ \Omega_0\setminus \Omegatil_0.
\eel
Because $a_n\ge M_m(n/\log(p/{{\eps_n}}))^{1/m}$,
the variance of $n^{-1/2}\sum_{i=1}^n X_{i,j,2}$ is bounded by
\bes
v_{n,j} = \E \sum_{i=1}^n \frac{X_{i,j}^2}{n}I_{\{a_n<|X_{i,j}|\le \atil_n\}}
\le M_m^m/a_n^{m-2} \le M_m^2\big[\{\log(p/{{\eps_n}})\}/n\big]^{1-2/m}.
\ees
Thus, by the Bennett inequality,
\bes
\P\bigg\{\max_{1\le j\le p}\bigg|\frac{16b_n}{n^{1/2}}\sum_{i=1}^n X_{i,j,2}\bigg|^2 \ge 1\bigg\}
\le 2p\,I_{\{a_n<\atil_n\}}\exp\Bigg[ - \frac{n^{1/2}\rho(u)/u}{(16b_n)(2\atil_n)}\Bigg],
\ees
where $u = \{(2\atil_n)n^{1/2}/(16b_n)\}/\{nM_m^2\big[\{\log(p/{{\eps_n}})\}/n\big]^{1-2/m}\}$
and $\rho(u)=(1+u)\log(1+u)-u$. Because 
\bes
\frac{1}{u^{m/2}}
&=& \bigg[\frac{n^{1/2}M_m^2\big[\{\log(p/{{\eps_n}})\}/n\big]^{1-2/m}}{(2\atil_n)/(16b_n)}\bigg]^{m/2}
\cr
&=& (8/c_1)^{m/2} b_n^m (\log (p/\eps_n))^{m-1}M_m^m/n^{m/2-1}
\cr 
&\le& (8/c_1)^{m/2}/C_{m,c_1}, 
\ees
$u$ is large for large $C_{m,c_1}$. It follows that for sufficiently large $C_{m,c_1}$ 
\bes
\frac{n^{1/2}\rho(u)/u}{(16b_n)(2\atil_n)}
= \frac{\log(p/{{\eps_n}})}{32c_1} \rho(u)/u \ge \log(2p/\eps_{n}).
\ees
Consequently,
\bel{pf-lm-truncate-3}
\P\bigg\{\max_{1\le j\le p}\bigg|\frac{16b_n}{n^{1/2}}\sum_{i=1}^n X_{i,j,2}\bigg|^2 \ge 1\bigg\}
\le \eps_{n}.
\eel
Moreover,
\bel{pf-lm-truncate-4}
\max_{1\le j\le p}\bigg|\frac{16b_n}{n^{1/2}}\sum_{i=1}^n \E X_{i,j}I_{\{|X_{i,j}| > \atil_n\}}\bigg|
\le \frac{16b_nn^{1/2}M_{m}^{m}}{(\atil_n)^{m-1}}
= \frac{16b_n^m {(\log (p/\eps_n))}^{m-1}M_{m}^{m}}{c_1^{m-1}n^{m/2-1}}
\le \frac{16
{/C_{m,c_1}}}{c_1^{m-1}} < 1
\eel
for a sufficiently large $C_{m,c_1}$.
Thus, (\ref{lm-truncate-new-1}) follows from (\ref{pf-lm-truncate-2}), (\ref{pf-lm-truncate-3})
and (\ref{pf-lm-truncate-1}).

\medskip
{\bf A2.4. Proof of Theorem \ref{th-comp-empi}}

The following lemma is needed in the proof of Theorem \ref{th-comp-empi}.

\begin{lemma}\label{lm-truncate-empi}
Let $M_4$ be as in (\ref{M_m}) 
and $M_4\{n/\log(p/\eps_n)\}^{1/4} \le a_n \le\atil_n = c_1n^{1/2}/\{b_n\log(p/\eps_n ) $ with $c_1>0$ .
Let $\tbX$ be as in (\ref{Xtil-a_n}),
$X_i^*$ and $\Xtil_i^*$ be the empirical bootstrap of $X_i$ and $\Xtil_i$ respectively,
and $\Omega_0^*$ as in (\ref{Omega_0}).
Suppose $b_n^4 (\log(p/\eps_n))^3 M_4^4/n \le 1/C_{c_1}$ for a sufficiently large $C_{c_1}$. Then,
\bel{lm-truncate-empi-1}
\P\big\{\P^*\big\{\Omega_0^*\big\} > \eps_n\big\} \le \eps_{n} +\P\big\{\|\bX\|_{\max} \ge \atil_n \}.
\eel
Moreover, with $\M_4$ as in \eqref{M_m},
\bel{lm-truncate-empi-2}
\P\bigg\{\P^*\big\{\Omega_0^*\big\} > \eps_n/2 + 
C_{c_1} \frac{b_n^4 (\log(p/{{\eps_n}}))^3 \M_4^4}{\eps_n \cdot n} \bigg\}
\le 2\eps_n
\eel
\end{lemma}

{\sc Proof of Lemma \ref{lm-truncate-empi}.}
We write $X_{i,j}-\Xtil_{i,j} = X_{i,j,1} + X_{i,j,2}$ with
\bes
\quad X_{i,j,1} = X_{i,j}I_{\{|X_{i,j}|> \atil_n\}}-\E X_{i,j}I_{\{|X_{i,j}|> \atil_n\}},\quad
X_{i,j,2} = X_{i,j}I_{\{a_n<|X_{i,j}|\le \atil_n\}}
- \E X_{i,j}I_{\{a_n<|X_{i,j}|\le \atil_n\}}.
\ees
Because $(X_i^*-\Xtil^*_i)$ are uniformly sampled
from $(X_i-\Xtil_i) - \sum_{i'=1}^n (X_{i'}-\Xtil_{i'})/n$, we have
$X_{i,j}^*-\Xtil^*_{i,j} = \Xtil_{i,j,1}^* + \Xtil_{i,j,2}^*$, where
$\Xtil_{i,j,k}^*$ are sampled from $X_{i,j,k} - \sum_{i'=1}^n X_{i',j,k}/n$. 
We assume $a_n < \atil_n$ as otherwise $X_{i,j,2}=0$ and the proof would be simpler.

Under $\P^*$, $\Xtil_{i,j,2}^*$ are i.i.d.\,variables with zero mean and
\bes
\E^* \big(\Xtil_{i,j,2}^*\big)^2 \le v^*_{n,j}
= n^{-1}\sum_{i=1}^n\Big(X_{i,j}I_{\{a_n<|X_{i,j}|\le \atil_n\}}- \E X_{i,j}I_{\{a_n<|X_{i,j}|\le \atil_n\}}\Big)^2.
\ees
Let $c_0$ be a small positive number, and $C_{c_1}$ is sufficiently large to satisfy $1/C_{c_1}^{1/2} \le c_0$. 
By the definition of $a_n$ and the condition $b_n^4 (\log(p/\eps_n))^3 M_4^4/n \le 1/C_{c_1}$,
\bel{pf-lm-truncate-empi-1}
\E v^*_{n,j} 
\le \frac{M_4^4}{a_n^2} \le \frac{\{\log(p/{{\eps_n}})\}^{1/2}M_4^2}{n^{1/2}}
\le \frac{1/C_{c_1}^{1/2}}{b_n^2\log (p/\eps_n)}
\le \frac{c_0}{b_n^2\log(p/{{\eps_n}})}.
\eel
By the Bennett inequality,
\bel{pf-lm-truncate-empi-2}
\P\bigg\{ v^*_{n,j} > \E v^*_{n,j} + \frac{c_0}{b_n^2\log(p/{{\eps_n}})} \bigg\}
\le \exp\Big( - \frac{(nc_0)\rho(u)/u}{b_n^2\log(p/{{\eps_n}})(2\atil_n)^2}\Big)I_{\{a_n \le \atil_n\}}
\eel
with $u = \{c_0/(b_n^2\log(p/{{\eps_n}}))\}(2\atil_n)^2/(2M_4)^4$ and $ \rho(u) = (1+u)\log(1+u)-u $.

As $\atil_n = c_1n^{1/2}/\{b_n(\log(p/{{\eps_n}}))\} $,
\bes
\frac{1}{u} = \frac{b_n^2\log(p/{{\eps_n}})(2M_4)^4}{c_0(2\atil_n)^2}
= \frac{4 b_n^4 (\log (p/{{\eps_n}}))^3 M_4^4}{c_0c_1^2 n} \le \frac{4}{C_{c_1}c_0c_1^2}
\ees
Thus, 
for sufficiently large $C_{c_1}$, 
$u$ is large and
\bes
\frac{nc_0\rho(u)/u}{b_n^2\log(p/{{\eps_n}})(2\atil_n)^2}
= \frac{c_0\log(p/{{\eps_n}})\rho(u)/u}{4c_1^2} \ge \log(p/{{\eps_n}}).
\ees
Consequently, by (\ref{pf-lm-truncate-empi-1}) and (\ref{pf-lm-truncate-empi-2}),
\bel{pf-lm-truncate-empi-3}
\P\bigg\{ \max_{1\le j\le p} v^*_{n,j} > \frac{2c_0}{b_n^2\log(p/{{\eps_n}})} \bigg\}
\le {{\eps_n}}.
\eel

Consider $\max_{1\le j\le p} v^*_{n,j} \le 2c_0/\{b_n^2\log(p/{{\eps_n}})\}$.
By the Bennett inequality,
\bes
\P^*\bigg\{\max_{1\le j\le p}\bigg|\sum_{i=1}^n \frac{\Xtil^*_{i,j,2}}{n^{1/2}}\bigg|\ge \frac{1}{8b_n} \bigg\}
\le 2p \exp\bigg[ - \frac{n^{1/2}\rho(u)/u}{(8b_n)(2\atil_n)}\bigg]
= 2p \exp\bigg[ - \frac{\log(p/{{\eps_n}})}{16c_1}\frac{\rho(u)}{u}\bigg],
\ees
with $u = \{(2\atil_n)n^{1/2}/(8b_n)\}/[n 2c_0/\{b_n^2\log(p/{{\eps_n}})\}] = c_1/(8c_0)$.
Thus, when $c_1$ is given and $c_0$ is sufficiently small, $u$ is large and
\bel{pf-lm-truncate-empi-4}
\max_{1\le j\le p} v^*_{n,j} \le \frac{2c_0}{b_n^2\log(p/{{\eps_n}})}
& \Rightarrow &
\P^*\bigg\{\max_{1\le j\le p}
\bigg|\sum_{i=1}^n \frac{\Xtil^*_{i,j,2}}{n^{1/2}}\bigg|\ge \frac{1}{8b_n} \bigg\} \le \frac{{{\eps_n}}}{2}
\cr & \Rightarrow &
\P^*\big\{\Omega_0^*\big\}
\le \frac{{{\eps_n}}}{2} + \P^*\bigg\{\max_{1\le j\le p}
\bigg|\sum_{i=1}^n \frac{\Xtil^*_{i,j,1}}{n^{1/2}}\bigg|\ge \frac{1}{8b_n} \bigg\}.
\eel

When $\|\bX\|_{\max}\le \atil_n$, $\Xtil^*_{i,j,1}$ are sampled from $ - \mu_{i,j}$ where
\bes
\mu_{i,j} = \E X_{i,j}I_{\{|X_{i,j}| > \atil_n\}} - \frac{1}{n}\sum_{k=1}^n\E X_{i,k}I_{\{|X_{i,k}| > \atil_n\}}.
\ees
As $\max_{i,j}|\mu_{i,j}|\le 2\atil_n$ and
$\sum_{i=1}^n\mu_{i,j}^2/n\le M_4^4/\atil_n^2 \le c_0/\{b_n^2\log(p/{{\eps_n}})\}$ by (\ref{pf-lm-truncate-empi-1}),
\bes
\P^*\bigg\{\max_{1\le j\le p}
\bigg|\sum_{i=1}^n \frac{\Xtil^*_{i,j,1}}{n^{1/2}}\bigg|\ge \frac{1}{8b_n} \bigg\}
\le \frac{{{\eps_n}}}{2}I_{\{\|\bX\|_{\max}\le \atil_n\}}
\ees
by a simpler version of the proof of (\ref{pf-lm-truncate-empi-4}).
This and (\ref{pf-lm-truncate-empi-3}) and (\ref{pf-lm-truncate-empi-4}) yield (\ref{lm-truncate-empi-1}).

Finally, because $\max_{1\le j\le n}|X^*_{i,j,1}|$ is uniformly sampled from
$\max_{1\le j\le n}|X_{i,j,1} - \sum_{i'=1}^n X_{i',j,1}/n|$,
\bes
\E \max_{1\le j\le p}
\bigg|\frac{8b_n}{n^{1/2}}\sum_{i=1}^n \Xtil^*_{i,j,1}\bigg|
\le C_0\, \E \frac{b_n}{n^{1/2}}\sum_{i=1}^n\max_{1\le j\le p}|X_{i,j}|I_{\{|X_{i,j}|>\atil_n\}}
\le C_0\, \frac{b_n n^{1/2}}{\atil_n^3} \M_4^ 4
\ees
with $\M_4 = \big\{\E \sum_{i=1}^n\max_{1\le j\le p}X_{i,j}^4/n\big\}^{1/4}$.
Thus, as $b_n n^{1/2}/\atil_n^3 = b_n^4\{\log(p/{{\eps_n}})\}^3/(c_1^3n)$, 
\bes
\P^*\bigg\{\max_{1\le j\le p}
\bigg|\sum_{i=1}^n \frac{\Xtil^*_{i,j,1}}{n^{1/2}}\bigg|\ge \frac{1}{8b_n} \bigg\}
\le  C_{c_1} \frac{b_n^4 (\log(p/{{\eps_n}}))^3 \M_4^4}{\eps_n \cdot n}
\ees
with at least probablity $1 - \eps_n$.
This and (\ref{pf-lm-truncate-empi-4}) give (\ref{lm-truncate-empi-2}).
$\hfill\square$

\medskip
{\sc Proof of Theorem \ref{th-comp-empi}.}
Let $a_n=M_4(n/\log(p/{{\eps_n}}))^{1/4}$. 
We assume without loss of generality that
$b_n^4(\log(p/\eps_n))^3M_4^4/n \le 1/C_{c_1}$ for a sufficiently large $C_{c_1}$ as the conclusion is trivial otherwise. Note that it implies $a_n \le \atil_n$ and Lemma \ref{lm-truncate-empi} is applicable. 

We apply (\ref{th-comp-gene-2}) with the $\tbX$ in (\ref{Xtil-a_n})
and its empirical bootstrap $\tbX^*$.
The main task is to find an upper bound for
\bes
\sum_{m=2}^4 n^{1-m/2}b_n^m(\log p)^{m-1}\Big\|\nutil^{{(m)}} - \mutil^{{(m)}}\Big\|_{\max}.
\ees
We have $\mutil^{{(m)}} =\frac{1}{n}\sum_{i=1}^n \E\Xtil_i^{\otimes m}$ and $\mutil^{{(1)}}=0$.
As $\nutil^{{(m)}}$ is the average moment tensor for $\tbX^*$ under $\P^*$,
\bel{pf-th-comp-empi-1}
\nutil^{{(m)}} - \mutil^{{(m)}} = \frac{1}{n}\sum_{i=1}^n \Big(\Xtil_i^{\otimes m} - \mutil^{{(m)}}\Big)
+ \sum_{k=1}^m {m\choose k}\hbox{Sym}
\bigg(\Big(- {\overline \Xtil} \Big)^{\otimes k}\sum_{i=1}^n \frac{\Xtil_i^{\otimes (m-k)}}{n} \bigg).
\eel
Because $\|\Xtil_i\|_\infty\le 2a_n$ and
$n^{-1}\sum_{i=1}^n \E\Xtil_{i,j}^{2m} \le {(2a_n)^{(2m-4)_+}(2M_4)^{4\wedge(2m)}}$,
it follows from Boole's and Bennett's inequalities that
\bel{pf-th-comp-empi-2}
&& \sum_{m=1}^4
\P\bigg\{\bigg\|\frac{1}{n}\sum_{i=1}^n \Xtil_i^{\otimes m} - \mutil^{{(m)}}\bigg\|_{\max} > (2a_n)^mB_{n,m} \bigg\}
\cr &\le& \sum_{m=1}^4 \frac{2p^m}{m!} \exp \bigg(-\frac{n{(2a_n)^{(2m-4)_+}(2M_4)^{4\wedge(2m)}}}{(2a_n)^{2m}}
\rho\bigg(\frac{(2a_n)^{2m} B_{n,m}}{{(2a_n)^{(2m-4)_+}(2M_4)^{4\wedge(2m)}}}\bigg)\bigg)
\cr &=& \sum_{m=1}^4\frac{2p^m}{m!}
\exp \Big(- n (M_4/a_n)^{4\wedge(2m)}\rho\big((a_n/M_4)^{4\wedge(2m)} B_{n,m}\big)\Big)
\eel
where $\rho(u)=(1+u)\log(1+u) -u$.
We note that $\rho(u) \approx u\log u$ if $u$ is large and $\rho(u) \approx u^2/2$ if $u$ is small.
As we want to bound the left-hand side above by $\kappa_{n,4}$, we pick $B_{n,m}$ to satisfy
\bel{pf-th-comp-empi-3}
n (M_4/a_n)^{4\wedge(2m)}\rho\big((a_n/M_4)^{4\wedge(2m)} B_{n,m}\big)
= \log\big(4p^m/{{\eps_n}}\big).
\eel
As $a_n = M_4(n/\log(p/{{\eps_n}}))^{1/4}$ and $\log(p/{{\eps_n}})/n \le c_2$, this implies
\bes
\rho\big((a_n/M_4)^{4\wedge(2m)} B_{n,m}\big)
\asymp \big\{\log(p/{{\eps_n}})/n\big\}^{1-1\wedge(m/2)} \le c_2^{1-1\wedge(m/2)}.
\ees
or equivalently
\bel{pf-th-comp-empi-4}
B_{n,m} \asymp \begin{cases} (M_4/a_n)^3 = \big\{\log(p/{{\eps_n}})/n\big\}^{3/4}, & m=1 \cr
(M_4/a_n)^4 = \log(p/{{\eps_n}})/n, & m = 2, 3, 4,
\end{cases}
\eel
Let $r_n = b_n(\log p)/n^{1/2}$.
By (\ref{pf-th-comp-empi-4}) and the condition $\log(p/{{\eps_n}})\le c_2n$.
\bes
&& a_nB_{n,1} = M_4(M_4/a_n)^2\le M_4c_2^{1/2},
\cr && a_n^mB_{n,m} = M_4^m(M_4/a_n)^{(4-m)}\le M_4^m c_2^{1-m/4},\quad 2\le m\le 4,
\cr && r_na_n = a_nb_n(\log p)/n^{1/2} \le {c_1},
\cr && r_nM_4 \le (M_4/a_n){c_1} \le {c_1}{c_2}^{1/4}.
\ees
It follows from (\ref{pf-th-comp-empi-1}), (\ref{pf-th-comp-empi-2}), (\ref{pf-th-comp-empi-3})
and (\ref{pf-th-comp-empi-4}) that
\bes
&& \sum_{m=2}^4 n^{1-m/2}b_n^m(\log p)^{m-1}\Big\|\nutil^{{(m)}} - \mutil^{{(m)}}\Big\|_{\max}
\cr &\le& C_{c_1,c_2}'\sum_{m=2}^4 \frac{n r_n^m}{\log p}
\bigg\{ a_n^mB_{n,m}
+ \sum_{k=1}^m
(a_nB_{n,1})^k\max\Big(M_4^{m-k},a_n^{m-k}B_{n,m-k}\Big)\bigg\}
\cr &\le& C_{c_1,c_2}''\sum_{m=2}^4\bigg\{\frac{(r_n a_n)^m\log(p/{{\eps_n}})}{\log p}
+  \frac{n(r_nM_4)^m}{\log p}\sum_{k=1}^m(M_4/a_n)^{2k}\bigg\}
\cr &\le & C_{c_1,c_2}\bigg\{\frac{(r_n a_n)^2\log(p/{{\eps_n}})}{\log p}
+  \frac{n(r_nM_4)^2}{\log p}(M_4/a_n)^{2}\bigg\}
\cr &=& 2C_{c_1,c_2}b_n^2(\log p)\{\log(p/{{\eps_n}})/n\}^{1/2}M_4^2
\ees
with at least probability $1-{{\eps_n}}$.
Thus, it follows from (\ref{th-comp-gene-2}) with $m=m^*=4$ that
\bes
\eta_n^*(1/b_n)
\le C_{c_1,c_2}\Big\{b_n^2(\log p)\{\log(p/{{\eps_n}})/n\}^{1/2}M_4^2 + \kappa_{n,4}\Big\}
 + \P\big\{\Omega_0\big\} + \P^*\big\{\Omega_0^*\big\}
\ees
with at least probability $1-{{\eps_n}}$,
where $\Omega_0$ and $\Omega^*_0$ are as in (\ref{Omega_0}).
Since $\eta_n^*(1/b_n)\le 1$,
\bes
\eta_n^*(1/b_n)
&\le& C_{c_1,c_2}b_n^2(\log p)\{\log(p/{{\eps_n}})/n\}^{1/2}M_4^2
 + \P\big\{\Omega_0\big\} + \P^*\big\{\Omega_0^*\big\}
\ees
with at least probability $1-{{\eps_n}}$. 
The conclusion then follows from Lemmas \ref{lm-truncate-new} and \ref{lm-truncate-empi} and a slight inflation of $\log p$ to $\log (p/\eps_n)$.
$\hfill\square$

\medskip
{\bf A2.5. Proof of Theorem \ref{th-comp-wild-a}}.
The following lemma is needed.

\begin{lemma}\label{lm-truncate-wild}
Let $\tbX$ be as in (\ref{Xtil-a_n}),
$X_i^*$ as in (\ref{wild}), $\Xtil_i^* = W_i\Xtil_i$
and $\Omega_0^*$ as in (\ref{Omega_0}).
Let $M_m$ be as in (\ref{M_m}) with $m>2$,
and with $0 < \eps_n \le \overline{\eps}_n$ and $c_1>0$ 
\bes
a_n = M_m\{n/\log(p/\overline{\eps}_n)\}^{1/m} \hbox{ and } \atil_n = c_1n^{1/2}/\big\{b_n (\log(p/\overline{\eps}_n))^{1/2} (\log(p/\eps_n))^{1/2} \big\}.
\ees
Suppose $\log(p/\eps_n) \le c_2n$ for $c_2 >0$ and $b_n^m (\log(p/ \overline{\eps}_n ))^{m/2-1} (\log(p/\eps_n))^{m/2} M_m^m/n^{m/2-1} \le 1/C_{m,\tau_0,c_1,c_2}$ for a sufficiently large $C_{m,\tau_0,c_1,c_2}$. Then,
\bel{lm-truncate-wild-1}
\P\Big\{\P^*\big\{\Omega_0^*\big\} >
 \overline{\eps}_n  \Big\}
&\le& \eps_{n} + \P\bigg\{ C_0\tau_0^2b_n^2\log(p/ \overline{\eps}_n )\max_{1\le j\le p}
\sum_{i=1}^n X_{i,j}^2I_{\{|X_{i,j}| > \atil_n\}} > n\bigg\}.
\eel
Moreover, with the $\M_{m,2}$ in \eqref{M_m-1}, 
\bel{lm-truncate-wild-2}
\P\bigg\{\P^*\big\{\Omega_0^*\big\} >
 \overline{\eps}_n  + C_{m, \tau_0, c_1} \frac{b_n^m(\log p)}{\eps_n} \Big(\frac{(\log(p/ \overline{\eps}_n ))(\log(p/\eps_n))}{n} \Big)^{m/2-1} \M_{m,2}^m \Big) \bigg\}
\le 2\eps_n.
\eel
\end{lemma}

{\sc Proof of Lemma \ref{lm-truncate-wild}.}
As $a_n / \atil_n \le C_{m, \tau_0, c_1, c_2}^{-1/m}/c_1 < 1$ for a sufficiently large $C_{m, \tau_0, c_1, c_2}$,
we write $X_{i,j}^*-\Xtil_{i,j}^* = \sum_{k=1}^4 \Xtil_{i,j,k}^*$ with
\bes
\Xtil_{i,j,1}^*&=&W_i\bigg(X_{i,j}I_{\{|X_{i,j}| > \atil_n\}}
- \frac{1}{n}\sum_{k=1}^n X_{k,j}I_{\{|X_{k,j}| > \atil_n\}}\bigg),
\cr \Xtil_{i,j,2}^*&=& W_i\bigg(X_{i,j}I_{\{a_n<|X_{i,j}|\le \atil_n\}}
- \frac{1}{n}\sum_{k=1}^n X_{k,j}I_{\{a_n<|X_{k,j}|\le \atil_n\}}\bigg),
\cr \Xtil_{i,j,3}^*&=& W_i 
\Big( \mu_{i,j} - \frac{1}{n}\sum_{k=1}^n \mu_{k,j} \Big)
\quad \Xtil_{i,j,4}^*= - W_i\bigg(\frac{1}{n}\sum_{k=1}^n \Xtil_{k,j}\bigg),
\ees
where $\mu_{i,j} = \E X_{i,j}I_{\{|X_{i,j}| \le a_n\}}$.
Recall that $\Xtil_{i,j} = X_{i,j}I_{\{|X_{i,j}| \le a_n\}}-\mu_{i,j}$.
Define
\bes
v^*_{n,j,1} = \frac{1}{n}\sum_{i=1}^n X_{i,j}^2I_{\{|X_{i,j}| > \atil_n\}},\quad
v^*_{n,j,2} = \frac{1}{n}\sum_{i=1}^n X_{i,j}^2I_{\{a_n<|X_{i,j}|\le \atil_n\}},\quad
v^*_{n,j,3} = \frac{1}{n}\sum_{i=1}^n \mu_{i,j}^2,
\ees
and $v^*_{n,j,4} = \big(n^{-1}\sum_{k=1}^n \Xtil_{k,j}\big)^2$.
Therefore, $\E^*\sum_{i=1}^n (\Xtil_{i,j,k}^*)^2/\sqrt{n} \le v^*_{n,j,k}$.
Because $\sum_i \Xtil_{i,j,4}^* = (\sum_i W_i)\sum_{k=1}^n \Xtil_{k,j}/n$,
it follows from (\ref{Omega_0}) and (\ref{subgaussian}) that
\bel{pf-lm-truncate-wild-1}
C_0' \tau_0^2b_n^2\bigg\{\log(p/{\overline{\eps}_n })\max_{1\le j\le p}\sum_{k=1}^3 v^*_{n,j,k}
+ \log(1/{\overline{\eps}_n }) \max_{1\le j\le p} v^*_{n,j,4}
\bigg\}\le 1
\ \Rightarrow\ \P^*\big\{\Omega_0^*\big\} \le {\overline{\eps}_n }.
\eel
Let $C_{m,\tau_0, c_1,c_2}$ be large enough to satisfy $1/C_{m,\tau_0, c_1,c_2}^{2/m} \le c_0 = 1/(4C_0')$. 
As $a_n = M_m(n/\log(p/{\overline{\eps}_n }))^{1/m}$ and $\eps_n \le \overline{\eps}_n$, 
\bes
&& v^*_{n,j,3} \le \E v^*_{n,j,2}
\le \frac{M_m^m}{a_n^{m-2}}
\le \frac{C_{m,\tau_0, c_1,c_2}^{-2/m}}{b_n^2 \log(p/\eps_n)} 
\le \frac{c_0}{b_n^2 \log(p/{\overline{\eps}_n })},
\cr && n\Var\Big(v^*_{n,j,2}\Big)
\le \frac{M_m^m}{a_n^{m-4}} = M_m^4\{\log(p/{\overline{\eps}_n })/n\}^{1-4/m}.
\ees
Thus, by the Bennett inequality, for any fixed $c_0>0$,
\bel{pf-lm-truncate-wild-2}
\P\bigg\{ v^*_{n,j,2} + v^*_{n,j,3}
> \frac{2c_0}{\tau_0^2b_n^2\log(p/{\overline{\eps}_n })} \bigg\}
\le \exp\bigg[ - \frac{(nc_0)\rho(u)/u}{\tau_0^2b_n^2\log(p/{\overline{\eps}_n })\atil_n^2}\bigg]
\eel
where $\rho(u) = (1+u)\log(1+u) -u$ 
and $u = \{c_0/(\tau_0^2b_n^2\log(p/{\overline{\eps}_n }))\}\atil_n^2/[M_m^4\{\log(p/{\overline{\eps}_n })/n\}^{1-4/m}]$.
This $u$ is large as
\bes
\frac{1}{u}
= \frac{\tau_0^2}{c_0c_1^2}\Big( 
\frac{ b_n^m (\log (p/{\overline{\eps}_n }))^{(3m)/4-1} (\log (p/\eps_n))^{m/4} M_m^m}
{n^{m/2-1}} \Big)^{4/m} 
\le \frac{\tau_0^2}{c_0c_1^2} C_{m,\tau_0, c_1,c_2}^{-4/m}.
\ees
Thus,
\bes
\frac{(nc_0)\rho(u)/u}{\tau_0^2b_n^2\log(p/{\overline{\eps}_n })\atil_n^2}
= \frac{c_0 \log (p/{{\eps_n}})\rho(u)/u}{\tau_0^2c_1^2}
\ge \log(2p/\eps_{n}).
\ees
This and (\ref{pf-lm-truncate-wild-2}) give
\bel{pf-lm-truncate-wild-3}
\P\bigg\{ \max_{1\le j \le p} (v^*_{n,j,2} + v^*_{n,j,3} )
> \frac{2c_0}{\tau_0^2b_n^2\log(p/{\overline{\eps}_n })} \bigg\}
\le \eps_{n}/2.
\eel
For $v^*_{n,j,4}$, the Bernstein inequality gives
\bes
\P\bigg\{\max_{1\le j \le p} v^*_{n,j,4} \ge \Big(M_m\sqrt{2\log(2p/\eps_{n})/n}
+ 4a_n \log(2p/\eps_{n})/(3n)\Big)^2 \bigg\} \le \eps_{n}/2
\ees
Because $\log p \le c_2 n$ and $\eps_n \le \overline{\eps}_n$,
\bes
&& b_n^2\log(1/{\overline{\eps}_n }) \Big(M_m\sqrt{2\log(2p/\eps_{n})/n} + 4a_n \log(2p/\eps_{n})/(3n)\Big)^2
\cr
&\le&
C_0 \Big( \frac{b_n^m(\log(p/{\overline{\eps}_n }))^{m/2-1}(\log(p/\eps_n))^{m/2-1}M_m^m}{n^{m/2}}\Big)^{2/m} \Big[ \Big(\frac{\log(p/\eps_n)}{n} \Big)^{2/m} + \frac{\log(p/\eps_n)}{n} \Big]
\cr
&\le& C_0C_{m, \tau_0, c_1, c_2}^{-2/m}(c_2^{2/m} + c_2)
\cr
&\le& 1/(2C_0'\tau_0^2)
\ees
for sufficiently large $C_{m, \tau_0, c_1, c_2}$. 
Thus,
$\P\big\{ C_0' \tau_0^2b_n^2\log(1/{\overline{\eps}_n }) \max_{1\le j \le p} v^*_{n,j,4} \ge 1/2 \big\} \le \eps_{n}/2$.
This and the inequalities in  (\ref{pf-lm-truncate-wild-1})
and (\ref{pf-lm-truncate-wild-3}) yield (\ref{lm-truncate-wild-1}).

Finally, as we observe 
\bes
\P^* \big\{ \Omega_0^* \big\} \le \P^* \Big\{ \max_{ 1\le j \le p} \Big| \sum_{i=1}^n \frac{\Xtil_{i,j,1}^*}{\sqrt{n}} \Big| \ge \frac{1}{8b_n} \Big\} + \P^* \Big\{ \max_{ 1\le j \le p} \Big| \sum_{i=1}^n \frac{\sum_{k=2}^4 \Xtil_{i,j,k}^*}{\sqrt{n}} \Big| \ge \frac{1}{8b_n} \Big\},
\ees
we bound the two quantities on the right-hand side. By the above analysis on $v_{n,j,2}, v_{n,j,3}$ and $v_{n,j,4}$, it yields that with at least probability $1 - \eps_n$ 
\bes
\P^* \Big\{ \max_{ 1\le j \le p} \Big| \sum_{i=1}^n \frac{\sum_{k=2}^4 \Xtil_{i,j,k}^*}{\sqrt{n}} \Big| \ge \frac{1}{8b_n} \Big\} \le {\overline{\eps}_n }.
\ees
On the other hand with $\Xtil_{i,j,1} = X_{i,j}I_{\{|X_{i,j}| > \atil_n\}}$,
\bes
&& \frac{8^2b_n^2} {\eps_n n } \E \max_{ 1\le j \le p} \Big| \sum_{i=1}^n \Xtil_{i,j,1}^* \Big|^2
\cr
&=& \frac{64b_n^2}{\eps_nn } \E \Bigg[ \E \bigg\{ \max_{1\le j \le p} \bigg(\frac{\sum_{i=1}^n \Xtil_{i,j,1}(W_i - \sum_{k=1}^n W_k/n)}{\sqrt{\sum_{i=1}^n \Xtil_{i,j,1}^2 }} \bigg)^2  \max_{1\le j \le p} \sum_{i=1}^n \Xtil_{i,j,1}^2 \bigg| \Xtil_{i,j,1} \bigg\} \Bigg] 
\cr
&\le& C_{\tau_0} \frac{b_n^2 \log p}{\eps_n} \E \max_{1\le j \le p} \frac{1}{n}\sum_{i=1}^n X_{i,j}^2I_{\{|X_{i,j}| \le \atil_n\}}
\cr
&\le& C_{\tau_0} \frac{b_n^2 \log p}{\eps_n} \frac{\M_{m,2}^m}{\atil_n^{m-2}}
\cr
&\le& C_{\tau_0, c_1} \frac{b_n^m (\log p)(\log (p/{\overline{\eps}_n } ))^{m/2-1} (\log (p/\eps_n))^{m/2-1} }{\eps_n \cdot n^{m/2-1}}\M_{m,2}^m,
\ees
where the first inequality comes from the sub-Gaussianity of $W_i$'s. The second conclusion \eqref{lm-truncate-wild-2} then follows from Markov's inequality as
\bes
&& \P \bigg\{ \P^* \Big\{ \max_{ 1\le j \le p} \Big| \sum_{i=1}^n \frac{\Xtil_{i,j,1}^*}{\sqrt{n}} \Big| \ge \frac{1}{8b_n} \Big\} > \frac{8^2b_n^2}{\eps_n n} \E \max_{ 1\le j \le p} \Big| \sum_{i=1}^n \Xtil_{i,j,1}^* \Big|^2\bigg\}
\cr
&\le& \P \Big\{ \max_{ 1\le j \le p} \Big| \sum_{i=1}^n \frac{\Xtil_{i,j,1}^*}{\sqrt{n}} \Big| \ge  \frac{1}{8b_n} \Big\} \Big/ \bigg( \frac{8^2b_n^2}{\eps_n n} \E \max_{ 1\le j \le p} \Big| \sum_{i=1}^n \Xtil_{i,j,1}^* \Big|^2 \bigg)
\cr
&\le& \eps_n.
\ees 
$\hfill\square$

\medskip
{\sc Proof of Theorem \ref{th-comp-wild-a}.} 
Let $\Xtil_{i,j}=X_{i,j}I_{\{|X_{i,j}|\le a_n\}}-\E X_{i,j}I_{\{|X_{i,j}|\le a_n\}}$ with
\bes
a_n = M_{m^*}\{n/\log(p/ {\overline{\eps}_n})\}^{1/m^*} \le M \{ n/\log(p/\overline{\eps}_n) \}^{1/4}.
\ees 
If $m^*=3$, $M = M_3 ( n/\log (p/\overline{\eps}_n ) )^{1/12}$. If $m^*=4$, $M=M_4$. Nevertheless, we assume that  
\bel{pf:th-comp-wild-a-assumption}
{b_n^{m^*}(\log(p/\overline{\eps}_n))^{m^*/2-1}(\log(p/\eps_n))^{m^*/2}}M_{m^*}^{m^*}  \big/n^{m^*/2-1} \le 1/C_{m^*,\tau_0, c_1, c_2}^{m^*/2}
\eel
as otherwise the bounds in \eqref{th-comp-wild-a-1} and \eqref{th-comp-wild-a-2} are trivial. It immediately implies that $a_n \le \atil_n$ for a sufficiently large $C_{m^*,\tau_0, c_1, c_2}$.}
Let $\Xtil_i^* = W_i\Xtil_i$. By (\ref{moment-match-2}) and the Bernstein inequality
\bes
&& \sum_{m=2}^{m^*}\frac{b_n^m(\log p)^{m-1}}{n^{m/2-1}}\Big\|\nutil^{{(m)}}  - \mutil^{{(m)}}\Big\|_{\max}
\cr &=& \sum_{m=2}^{m^*}\frac{b_n^m(\log p)^{m-1}}{n^{m/2-1}}
\bigg\|\frac{(\E W_i^m)}{n}\sum_{i=1}^n \Big(\Xtil_i^{\otimes m} - \E\Xtil_i^{\otimes m}\Big)\bigg\|_{\max}
\cr &\le& C_{m^*,\tau_0}
\sum_{m=2}^{m^*} b_n^2(\log p)\bigg\{\bigg(\frac{b_n \log p}{n^{1/2}}\bigg)^{m-2}
a_n^{m-2}\bigg\}\Big(\sqrt{a_n^{4-m^*} M_{m^*}^{m^*}\log(p/\eps_{n})/n}
+ a_n^2\log(p/\eps_{n})/n\bigg)
\cr &\le& C_{m^*,\tau_0,c_1} b_n^2 (\log (p/\overline{\eps}_n))^{1/2}(\log(p/\eps_{n}))M^2 /n^{1/2}.
\ees
with at least probability $1-\eps_n$.
Thus, by (\ref{th-comp-gene-2}), 
\bes
\eta_n^*(1/b_n)
&\le& C_{m^*,\tau_0,c_1}' b_n^2 (\log (p/\overline{\eps}_n))^{1/2}(\log(p/\eps_{n})) M^2 /n^{1/2} + \P\big\{\Omega_0\big\} + \P^*\big\{\Omega_0^*\big\}.
\ees
We apply Lemma \ref{lm-truncate-new} with its $\eps_n$ being $ {\overline{\eps}_n}$ and $m=m^*$, yielding 
\bel{apply-lemma-3}
\P\big\{\Omega_0\big\} \le {\overline{\eps}_n } + \P\big\{\Omegatil_0\big\}
\le {\overline{\eps}_n } + C_{c_1}\frac{b_n^{m^*} (\log (p/{\overline{\eps}_n }))^{m^*-1}}{n}\M_{m^*,2}^{m^*},
\eel
where
$\Omegatil_0 = \big\{\max_{1\le j\le p}\big|n^{-1/2}\sum_{i=1}^n X_{i,j}I_{\{|X_{i,j}|> \atil_n'\}}\big| >1/(8b_n)\big\}$
and $\atil_n' = c_1 \sqrt{n}/(b_n \log (p/{\overline{\eps}_n })) \ge \atil_n$ due to $\eps_n \le {\overline{\eps}_n }$, so that
\bes
\P\big\{\Omegatil_0\big\} \le \P\Big\{ \max_{1\le j\le p}\big|n^{-1/2}\sum_{i=1}^n X_{i,j}I_{\{|X_{i,j}|> \atil_n\}}\big| >1/(8b_n) \Big\}.
\ees
The required condition in Lemma \ref{lm-truncate-new}, a small enough $b_n^{m^*}(\log (p/{\overline{\eps}_n }))^{m^*} M_{m^*}^{m^*}/n^{m^*/2-1} $, is fulfilled with \eqref{pf:th-comp-wild-a-assumption}.
On the other hand, due to the assumption in \eqref{pf:th-comp-wild-a-assumption} and $\log p \le c_2 n$, Lemma \ref{lm-truncate-wild} is also applicable. It follows that $ \P^*\big\{\Omega_0^*\big\} \le {\overline{\eps}_n }$ holds with at least probability 
\bes
1 - \Big( \eps_{n} + \P\big\{ C_0\tau_0^2b_n^2\log(p/\overline{\eps}_n)\max_{1\le j\le p}\sum_{i=1}^n X_{i,j}^2I_{\{|X_{i,j}| > \atil_n\}} > n\big\} \Big).
\ees 
The conclusion in \eqref{th-comp-wild-a-1} therefore follows.

Alternatively, by the second inequality in \eqref{apply-lemma-3} and \eqref{lm-truncate-wild-2} in Lemma \ref{lm-truncate-wild},
\bes
\P\bigg\{ \eta_n^*(1/b_n)
> C_{m^*,\tau_0,c_1}' \Big( {\overline{\eps}_n } + \frac{b_n^{m^*}(\log(p/\overline{\eps}_n))^{m^*/2-1}(\log(p/\eps_n))^{m^*/2}}{\eps_n \cdot n^{m^*/2-1}}\M_{m^*,2}^{m^*} \Big) \bigg\} \le \eps_{n}.
\ees
The proof is complete. $\hfill\square$ 

\medskip
{\bf A2.6. Proof of Proposition \ref{prop-wild}.}  Define
\bes
&& D_{n,m,i} = \Big\langle \E^*{f}^{(m)}_i(\bU_{i}^0),
\big(X_{i}^{\otimes m} - \E\, X_{i}^{\otimes m}\big)\E\,W_i^m\Big\rangle,\quad 2\le m<m^*,
\cr && D_{n,m^*,i} = \Big\langle \E^* f_{\max}^{{(m)}}(\bU_{i}^0),
\E^*g_{m}(\|X_{i}^*\|/u_n)|X_{i}^*|^{\otimes m} - \E g_{m}(\|X_{i}^*\|/u_n)|X_{i}^*|^{\otimes m}\Big\rangle.
\ees
Because $D_{n,m,i}$ are martingale differences,
\bel{pf-prop-1-1}
\E \bigg\{\sum_{{i}=1}^n D_{n,m,i} \bigg\}^2
= \sum_{{i}=1}^n\E s_{n,m,i}^2,\quad 2\le m \le m^*.
\eel
It follows from (\ref{moment-match}) and (\ref{cond-f}) that in the expansion (\ref{Lindeberg-4}),
$\Delta_{n,m}^*
=  - (m!)^{-1}\sum_{{i}=1}^n D_{n,m,i}$ and
\bes
\big|\Rem\big| &=& \bigg|\E^*\bigg( \int_0^1 \frac{(1-t)^{m^*-1}}{(m^*-1)!}
\sum_{{i}=1}^n \Big\langle (\pa/\pa x_{i})^{\otimes m^*}f((1-t)\bU_{i}^0+t\bV_i),
(X_{i}^0)^{\otimes m^*}\Big\rangle dt
\cr && \qquad - \int_0^1 \frac{(1-t)^{m^*-1}}{(m^*-1)!}
\sum_{{i}=1}^n \Big\langle (\pa/\pa x_{i})^{\otimes m^*}f((1-t)\bU_{i}^0+t\bV_{i-1}),
(X_{i}^*)^{\otimes m^*}\Big\rangle dt\bigg) \bigg|
\cr &\le & \frac{1}{m^*!}
\sum_{{i}=1}^n \Big\langle \E^* f_{\max}^{{(m^*)}}(\bU_{i}^0),
\E g(\|X_{i}^0\|/u_n) |X_{i}^0|^{\otimes m^*} + \E^* g(\|X_{i}^*\|/u_n) |X_{i}^*|^{\otimes m^*} \Big\rangle
\cr & = & \frac{1}{m^*!}
\sum_{{i}=1}^n D_{n,m^*,i} + {\overline \Rem}.
\ees
Thus, a direct application of (\ref{pf-prop-1-1}) to the individual terms in (\ref{Lindeberg-4})
yields (\ref{prop-wild-1}).  $\hfill\square$

\medskip

{\bf A2.7. Proof of Theorem \ref{th-comp-wild-b}.}
Let $\Xtil_{i,j}=X_{i,j}I_{\{|X_{i,j}|\le a_n\}}-\E X_{i,j}I_{\{|X_{i,j}|\le a_n\}}$ with
\bes
a_n = \min\Big\{ M\{n/\log(p/\eps_{n})\}^{1/4}, c_1\sqrt{n/\log(p/\eps_{n})}/(b_n\sqrt{\log p})\Big\}.
\ees
Let $\{\Xtil_i^0\}$ be an independent copy of $\{\Xtil_i\}$ and $\Xtil_i^* = W_i\Xtil_i$.
Let 
$\Ttil_n^0 = \max_j\sum_{i=1}^n \Xtil_{i,j}^0/\sqrt{n}$ and
$\Ttil_n^* = \max_j\sum_{i=1}^n \Xtil_{i,j}^*/\sqrt{n}$.
Let $\etabar_n \in (0,1)$ and $t_0 = -\infty < t_1< \cdots<t_{k_n} < t_{k_n+1}=\infty$
such that
\bes
\P\big\{t_{k-1} < \Ttil_n^0 < t_k\big\} \le \etabar_n,\
\P\big\{t_{k-1} - \eps < \Ttil_n^0 < t_k - \eps \big\} \le \etabar_n,\ \eps = 1/(2b_n).
\ees
Such $t_k$ exists with $k_n \le 2/\etabar_n$.
By the definition of $\eta_n^{{(\P^*)}}(\eps,t;\Ttil_n^0,\Ttil_n^*)$, for $t\in (t_{k-1},t_k)$
\bes
& \P\big\{\Ttil_n^0\le t-\eps\big\} - \P^*\big\{\Ttil_n^*<t\big\} \le
\P\big\{\Ttil_n^0 < t_k-\eps\big\} - \P^*\big\{\Ttil_n^*<t_{k-1}\big\}
\le \eta_n^{{(\P^*)}}(\eps,t_{k-1};\Ttil_n^0,\Ttil_n^*)+\etabar_n,
\cr & \P^*\big\{\Ttil_n^*\le t-\eps\big\} - \P\big\{\Ttil_n^0<t\big\} \le
\P^*\big\{\Ttil_n^*\le t_k-\eps\big\} - \P\big\{\Ttil_n^0\le t_{k-1}\big\}
\le \eta_n^{{(\P^*)}}(\eps,t_k;\Ttil_n^0,\Ttil_n^*)+\etabar_n.
\ees
Thus, by (\ref{pf-th4-2}),
\bel{pf-th-comp-wild-b-1}
\eta_n^{{(\P^*)}}\big(1/b_n;T_n,T_n^*\big)
&\le& \P\big\{\Omega_0\big\} + \eta_n^{{(\P^*)}}\big(1/(2b_n);\Ttil_n^0,\Ttil_n^*\big)
+\P^*\big\{\Omega_0^*\big\}
\cr &\le& \max_{1\le k\le k_n} \eta_n^{{(\P^*)}}\big(1/(2b_n),t_k;\Ttil_n^0,\Ttil_n^*\big)
+\etabar_n + \P\big\{\Omega_0\big\} + \P^*\big\{\Omega_0^*\big\}.
\eel

Let $\beta_n = 4b_n\log p$.
By the argument leading to (\ref{pf-th-comp-gene-1}) in the proof of Theorem \ref{th-comp-gene},
\bes
I\{z_{\max} \le t_k-1/(2b_n)\}
\le h_{4b_nt_k}\big(4b_nF_{\beta_n}(z)\big) \le I\{z_{\max} < t_k\},
\ees
where $z_{\max}=\max(z_1,\ldots,z_p)$ and
$h_t(\cdot) = h_0(\cdot - t)$ is the location shift of a smooth function $h_0$.
Let $f_{(k)}(x_1,\ldots,x_n) = h_{4b_nt_k}\big(4b_nF_{\beta_n}(z)\big)$ with $z = (x_1+\ldots+x_n)/\sqrt{n}$. We have
\bel{pf-th-comp-wild-b-2}
\eta_n^{{(\P^*)}}(1/(2b_n),t_k;\Ttil_n^0,\Ttil_n^*) \le \big|\Delta_n^*(f_{(k)})\big|,\
\Delta_n^*(f_{(k)}) = \E^*\big\{f_{(k)}(\tbX^0) - f_{(k)}(\tbX^*)\big\}.
\eel

Let $f = f_{(k)}$ and $\tbU^0$ is the truncated version of $\bU^0$ corresponding to $\{\tbX^0,\tbX^*\}$.
As in the proof of Proposition \ref{prop-wild}, the expansion (\ref{Lindeberg-4}) can be further specified as
\bes
\Delta_n^*(f) &=& \sum_{m=2}^{m^*-1}\Delta_{n,m}^* + \Rem
= \sum_{m=2}^{m^*-1}\frac{-1}{m!}\sum_{{i}=1}^n D_{n,m,i} + \Rem
\cr \big|\Rem\big| &\le& \frac{1}{m^*!}
\sum_{{i}=1}^n D_{n,m^*,i} + {\overline \Rem}.
\ees
with martingale differences
\bes
&& D_{n,m,i} = \Big\langle \E^*{f}^{(m)}_i(\tbU_{i}^0),
\big(\Xtil_{i}^{\otimes m} - \E\, \Xtil_{i}^{\otimes m}\big)\E\,W_i^m\Big\rangle,\quad 2\le m<m^*,
\cr && D_{n,m^*,i} = \Big\langle \E^* f_{\max}^{{(m)}}(\tbU_{i}^0),
\E^*g_{m}(\|\Xtil_{i}^*\|/u_n)|\Xtil_{i}^*|^{\otimes m} - \E g_{m}(\|\Xtil_{i}^*\|/u_n)|\Xtil_{i}^*|^{\otimes m}\Big\rangle,
\ees
where $u_n=\sqrt{n}/(4b_n\log p)$, and
\bes
{\overline \Rem} = \frac{1}{m^*!}
\sum_{{i}=1}^n \Big\langle \E^* f_{\max}^{{(m^*)}}(\tbU_{i}^0),
\E\, g(\|\Xtil_{i}^0\|_\infty/u_n)|\Xtil_{i}^0|^{\otimes m^*} +
\E\, g(\|\Xtil_{i}^*\|_\infty/u_n)|\Xtil_{i}^*|^{\otimes m^*}\Big\rangle.
\ees
The quadratic variations corresponding to $\{D_{n,m,i}\}$ are given by
\bes
&& s_{n,m,i} = \Big\langle \big(\E^*{f}^{(m)}_i(\tbU_{i}^0)\big)^{\otimes 2},
\E \big(\Xtil_{i}^{\otimes m} - \E\, \Xtil_{i}^{\otimes m}\big)^{\otimes 2}\big(\E\,W_i^m\big)^2\Big\rangle^{1/2},
2\le m < m^*,
\cr && s_{n,m^*,i} = \Big\langle \big(\E^* f_{\max}^{{(m^*)}}(\tbU_{i}^0)\big)^{\otimes 2},
\E\big(\E^*g(\|\Xtil_{i}^*\|_\infty/u_n)|\Xtil_{i}^*|^{\otimes m^*}
- \E g(\|\Xtil_{i}^*\|_\infty/u_n)|\Xtil_{i}^*|^{\otimes m^*} \big)^{\otimes 2}\Big\rangle^{1/2},
\ees
Moreover, it follows from Lemmas \ref{f-derivatives} and \ref{H-stability}
that (\ref{cond-1a}) and (\ref{cond-1b}) 
hold for $2\le k\le m^*$, so that
\bes
\Big\|\E^* f_{\max}^{{(m)}}(\tbU_{i}^0)\Big\|_1
\le  C_m b_n^m(\log p)^{m-1}n^{-m/2},\quad 2\le m\le m^*.
\ees
By the definition of $a_n$, $\|\Xtil_{i}^*\|_\infty/u_n\le 2a_n/u_n\le 2c_1$,
$\|\Xtil_{i}^{\otimes m}\|_{\max}\le (2a_n)^m$,
\bes
\big|D_{n,m,i}\big| &\le& C_m b_n^m(\log p)^{m-1}n^{-m/2}\big|\E W_i^m\big| a_n^m
\cr &\le& C_{m, \tau_0, c_1} a_n^2 b_n^2(\log p)n^{-1}
\cr &\le& C_{m, \tau_0, c_1} b_n^2(\log p)n^{-1/2}M^2\sqrt{1/\log(1/\eps_{n})},
\cr \sum_{i=1}^n s_{n,m,i}^2
&\le& \sum_{i=1}^n \Big(C_m b_n^m(\log p)^{m-1}n^{-m/2}\Big)^2
\|\E(X_i^{\otimes m} - \E X_i^{\otimes m})^{\otimes 2}\|_{\max}(\E W_i^m)^2
\cr &\le& C_{m,\tau_0}\Big(n^m(\log p)^{m-1}n^{-m/2}\Big)^2 na_n^{2m-4}M^4
\cr &\le & C_{m,\tau_0,c_1}\Big(b_n^2(\log p)n^{-1/2}\Big)^2 M^4,
\cr \big|{\overline \Rem}\big|
&\le & C_{m^*,\tau_0,c_1}\kappa_{n,m^*}(\M_{m^*,1}/M_{m^*})^{m^*},
\ees
with the $\M_{m,1}$ in (\ref{M_m}).
By the martingale Bernstein inequality \citep{steiger1969best, freedman1975tail},
\bel{pf-th-comp-wild-b-3}
\big|\Delta_n^*(f)\big|
&\le& C_{m^*,\tau_0,c_1}b_n^2(\log p)n^{-1/2}M^2\sqrt{\log(1/\eps_{n})} + \big|{\overline \Rem}\big|
\cr &\le& C_{m^*,\tau_0,c_1}\Big\{b_n^2(\log p)n^{-1/2}M^2\sqrt{\log(1/\eps_{n})}
+\kappa_{n,m^*}(\M_{m^*,1}/M_{m^*})^{m^*}\Big\}
\eel
with at least probability $1 - \eps_n^2$.

Taking the maximum over $f_k$ in
(\ref{pf-th-comp-wild-b-1}) and (\ref{pf-th-comp-wild-b-2}), we find by (\ref{pf-th-comp-wild-b-3}) that
\bes
\eta_n^{{(\P^*)}}\big(1/b_n;T_n,T_n^*\big)
\le C_{m^*,\tau_0,c_1}\eps_{n} +\etabar_n + \P\big\{\Omega_0\big\} + \P^*\big\{\Omega_0^*\big\},
\ees
with at least probability $1 - k_n\eps_n^2$ and $k_n\le 2/\etabar_n$. We take $\etabar_n = \eps_{n}$.
We omit the rest of the proof as it involves applications of
Lemmas \ref{lm-truncate-new} and \ref{lm-truncate-wild} to bound $\P\big\{\Omega_0\big\}$ and
$\P^*\big\{\Omega_0^*\big\}$ in the same way as in the proof of Theorem \ref{th-comp-wild-a}.
$\hfill\square$

\medskip
{\bf A2.8. Proof of Theorem \ref{th-comp-wild}.}
Let $\beta_n = 2b_n\log p$.
Similar to (\ref{pf-th-comp-wild-b-2}), 
\bel{pf-th3-1}
\eta_n^{{(\P^*)}}(1/b_n,t_0;T_n^0,T_n^*) \le \big|\Delta_n^*(f)\big|,\
\Delta_n^*(f) = \E^*\big\{f(\bX^0) - f(\bX^*)\big\}
\eel
where $f(x_1,\ldots,x_n) = h\big(4b_nF_{\beta_n}(z)\big)$
with $z = (x_1+\ldots+x_n)/\sqrt{n}$ and $h$ is a smooth function.
It follows from Lemmas \ref{f-derivatives} and \ref{H-stability}
that (\ref{cond-1a}) and (\ref{cond-1b}) 
hold for $m^*=k$ with $\|x\|=\|x\|_\infty$,
\bes
{f}^{(k)}_{\max}(x_1,\ldots,x_n) = n^{-k/2} H^{(k)}_{b_n,\beta_n}(z),\
\big\|H^{(k)}_{b_n,\beta_n}(z)\big\|_1\le C_k b_n\beta_n^{k-1},\
g(t) = g_{k}(t) = \exp(2k t),
\ees
and $u_n = u_n = \sqrt{n}/\beta_n$.
As $\E|W_{i}|\le \E W_{i}^2=1$ and $e^{2k t|x|}t^{k}$ is convex in $t$ for $t>0$,
\bel{pf-th3-2}
& \E \exp(- 2k\|W_{i}X_{i}\|/u_{n}) \ge \E \exp(- 2k \|X_{i}\|/u_{n}),
\cr & \E\Big\{\exp\big(2k\|X_{i}\|/u_{n}\big)|X_{i}|^{\otimes k}\Big\}
\le \E\Big\{\exp\big(2k\|W_{i}X_{i}\|/u_{n}\big)|W_{i}X_{i}|^{\otimes k}\Big\}.
\eel

Let $\bU_{\sigma,{k}}^0 = (X_{\sigma_1}^0,\ldots,X_{\sigma_{{k}-1}}^0,
X_{\sigma_{{k}+1}}^*,\ldots,X_{\sigma_n}^*)$.
As in the proof of Theorem \ref{comparison}, (\ref{Lindeberg-2}) and (\ref{approx-Delta_{n,m}}) yield
\bel{th3-decomp}
\Delta_n^*(f) = \sum_{m=2}^{m^*-1}\A_\sigma\big(\Delta_{n,m,\sigma}^*\big) + \A_\sigma\big(\Rem_\sigma^*\big)
\eel
where $\Delta_{n,m,\sigma}^* =
(m!)^{-1}\sum_{k=1}^n \big\langle \E^* {f}^{(m,0)}(\bU_{\sigma,{k}}^0,0),
\E^*(X_{\sigma_k}^0)^{\otimes m} - \E^*(X_{\sigma_k}^*)^{\otimes m}\big\rangle$,
\bes
\A_\sigma\big(\Delta_{n,m,\sigma}^*\big)
&=& n\,\A_{\sigma,k}\Big((m!)^{-1}\Big\langle \E^* {f}^{(m,0)}(\bU_{\sigma,{k}}^0,0),
\E (X_{\sigma_k}^*)^{\otimes m} - \E^*(X_{\sigma_k}^*)^{\otimes m}\Big\rangle\Big),
\cr \A_\sigma\big(\Rem_\sigma^*\big) &=&
 n\, \A_{\sigma,k}\bigg(
\E^* \int_0^1 \left\langle {f}^{(m^*)}(\bU_{\sigma,{k}}^0,tX_{\sigma_k}^0),
\frac{(1-t)^{m^*-1}}{(m^*-1)!}(X_{\sigma_k}^0)^{\otimes 4}\right\rangle dt\bigg) \\
&& -  n\, \A_{\sigma,k}\bigg(
\E^* \int_0^1 \left\langle {f}^{(m^*)}(\bU_{\sigma,{k}}^0,tX_{\sigma_k}^*),
\frac{(1-t)^{m^*-1}}{(m^*-1)!}(X_{\sigma_k}^*)^{\otimes 4}\right\rangle dt\bigg).
\ees
Note that we applied (\ref{moment-match}) to replace $\E^*(X_{\sigma_k}^0)^{\otimes m}$
by $\E (X_{\sigma_k}^*)^{\otimes m}$ in the expression for $\A_\sigma\big(\Delta_{n,m,\sigma}^*\big)$.

Since $\big\langle \E^* {f}^{(m,0)}(\bU_{\sigma,{k}}^0,0),
\E(X_{\sigma_k}^*)^{\otimes m} - \E^*(X_{\sigma_k}^*)^{\otimes m}\big\rangle$
are martingale differences for fixed $\sigma$,
\bel{pf-th3-2b}
\|\A_\sigma\big(\Delta_{n,m,\sigma}^*\big)\|_{L_2(\P)}^2
&\le & \A_{\sigma}\,\E \bigg(\sum_{k=1}^n \Big\langle \E^* {f}^{(m,0)}(\bU_{\sigma,{k}}^0,0),
\E(X_{\sigma_k}^*)^{\otimes m} - \E^*(X_{\sigma_k}^*)^{\otimes m}\Big\rangle\bigg)^2
\cr &\le&
n\big(\E W_1^m\big)^2 \A_{\sigma,k}\,\Big\langle \E\{{f}^{(m,0)}(\bU_{\sigma,{k}}^0,0)\big\}^{\otimes 2},
\E(X_{\sigma_k}^0)^{\otimes (2m)}\Big\rangle
\cr &\le&
n\big(\E W_1^m\big)^2 \A_{\sigma,k}\,\bigg\langle
\E\{f_{\max}^{{(m)}}(\bU_{\sigma,{k}}^0,\zeta_{k,\sigma_k})\big\}^{\otimes 2},
\frac{\E(X_{\sigma_k}^0)^{\otimes (2m)}}{\E \exp( -4m\|X_{\sigma_k}\|_\infty/u_n)}\bigg\rangle
\cr &=&
n\big(\E W_1^m\big)^2 \bigg\langle
\A_{\sigma,k}\,\E\{f_{\max}^{{(m)}}(\bU_{\sigma,{k}}^0,\zeta_{k,\sigma_k})\big\}^{\otimes 2},
\A_{\sigma,k}\,\frac{\E(X_{\sigma_k}^0)^{\otimes (2m)}}{\E \exp( -4m\|X_{\sigma_k}\|_\infty/u_n)}\bigg\rangle
\cr &\le &
n\big(\E W_1^m\big)^2 \Big\|
\A_{\sigma,k}\,\E\{f_{\max}^{{(m)}}(\bU_{\sigma,{k}}^0,\zeta_{k,\sigma_k})\big\}^{\otimes 2}\Big\|_1
\M_{2m,{2}}^{2m}
\cr &\le& C_m \big(\E W_1^m\big)^2 \frac{b_n^{2m}(\log p)^{2m-2}}{n^{m-1}}\M_{2m,{2}}^{2m}
\eel

{\it Step 1: Proof of (\ref{th-comp-wild-1}).}
Here we need to bound the $L_1$ error $\eta_n^{(1)}(1/b_n)$. Let
\bes
\Rem' = \big(n/m^*!\big)\, \A_{\sigma,k}
\left\langle \E^*\,\fbar^{{(m^*)}}(\bU_{\sigma,{k}}^0,0),
\E |X_{\sigma_k}^0|^{\otimes m^*}e^{2m^*\|X_{\sigma_k}^0\|_\infty/u_n}
+ \E^* |X_{\sigma_k}^*|^{\otimes m^*}e^{2m^*\|X_{\sigma_k}^*\|_\infty/u_n} \right\rangle.
\ees
Let $\zeta_{{k},{i}}$ be as in Lemma 1.
Similar to the proofs of Theorems \ref{comparison} and \ref{th-comp-gene},
it follows from (\ref{cond-1a}), (\ref{cond-1b}), Lemma \ref{k-nonspecific},
the independence of $\bU_{\sigma,k}$ and $(X_{\sigma_k},X_{\sigma_k}^*)$
under both $\P^*$ and $\P$, and (\ref{pf-th3-2}) that
\bel{th3-L_1}
\E \big|\A_\sigma\big(\Rem_\sigma^*\big)\big|
&\le& \E\,\Rem'
\cr &\le & \frac{n}{m^*!} \A_{\sigma,k}
\left\langle \E\,f_{\max}^{{(m^*)}}(\bU_{\sigma,{k}}^0,\zeta_{k,\sigma_k}),
\frac{2 \E |X_{\sigma_k}^*|^{\otimes m^*}e^{2m^*\|X_{\sigma_k}^*\|_\infty/u_n}}
{\E \exp( - 2m^*\|X_{\sigma_k}\|/u_n)}\right\rangle
\cr & = &
2 \left\langle \E F_{\max}^{{(m^*)}},
\frac{1}{n}\sum_{i=1}^n\frac{ \E |W_iX_{i}|^{\otimes m^*}e^{2m^*\|W_iX_{i}\|_\infty/u_n}}
{\E \exp( - 2m^*\|X_{i}\|/u_n)}\right\rangle
\cr &\le& C_{m^*}\frac{b_n^{m^*}(\log p)^{m^*-1}}{n^{m^*/2-1}}\M_{m^*,{1}}^{m^*}.
\eel
Thus, by (\ref{pf-th3-1}), (\ref{th3-decomp}), (\ref{pf-th3-2b}), (\ref{th3-L_1}) and
the definition of $\eta_n^{(1)}(\eps)$ in (\ref{eta_n}),
\bes
\E\eta_n^{{(\P^*)}}(1/b_n,t_0;T_n^0,T_n^*)
\le C_{m^*} \sum_{m=2}^{m^*-1}
\big|\E W_1^m\big| \frac{b_n^{m}(\log p)^{m-1}}{n^{m/2-1/2}}\M_{2m,{2}}^{m}
+C_{m^*} \frac{b_n^{m^*}(\log p)^{m^*-1}}{n^{m^*/2-1}}\M_{m^*,{1}}^{m^*}.
\ees

{\it Step 2: Remainder term for bounded variables.}
Suppose in addition (\ref{subgaussian}) holds and $\|X_i\|_\infty\le 2a_n$
with $a_n \le {c_1}\sqrt{n}/(b_n\log p) = 2{c_1}u_n$.
Because $\|X_i^*\|_\infty/u_n\le 4{c_1}|W_i|$, (\ref{subgaussian}) implies that
\bes
\E^* |X_{\sigma_k}^*|^{\otimes m^*}e^{2m^*\|X_{\sigma_k}^*\|_\infty/u_n}
\le C_{m^*,\tau_0,{c_1}}|X_{\sigma_k}|^{\otimes m^*},\quad
\E \exp(- 2m^*\|X_i\|_\infty/u_n)\ge 1/C_{m^*,\tau_0,{c_1}}
\ees
for some $C_{m^*,\tau_0,{c_1}}$ depending on $(m^*,\tau_0,{c_1})$ only, so that
\bel{pf-th3-2c}
&& \E\Big(\Rem'\Big)^2
\cr &=& \E \Big((n/m^*!)\,\A_{\sigma,k}
\left\langle \E^*\,\fbar^{{(m^*)}}(\bU_{\sigma,{k}}^0,0),
\E |X_{\sigma_k}^0|^{\otimes m^*}e^{\|X_{\sigma_k}^0\|_\infty/u_n}
+ \E^* |X_{\sigma_k}^*|^{\otimes m^*}e^{\|X_{\sigma_k}^*\|_\infty/u_n} \right\rangle\Big)^2
\cr &\le & C_{m^*,\tau_0,{c_1}}'\, \E \bigg(n\,\A_{\sigma,k}
\left\langle \E^*\,\fbar^{{(m^*)}}(\bU_{\sigma,{k}}^0,0),\E |X_{\sigma_k}|^{\otimes m^*}\right\rangle\bigg)^2
\cr && +\, C_{m^*,\tau_0,{c_1}}'\, \E \bigg(n\,\A_{\sigma,k}
\left\langle \E^*\,\fbar^{{(m^*)}}(\bU_{\sigma,{k}}^0,0),
|X_{\sigma_k}|^{\otimes m^*} - \E |X_{\sigma_k}|^{\otimes m^*}
\right\rangle\bigg)^2
\cr &\le & C_{m^*,\tau_0,{c_1}}''\, \E \bigg(n\,\A_{\sigma,k}
\left\langle \E^*\,f_{\max}^{{(m^*)}}(\bU_{\sigma,{k}}^0,\zeta_{k,\sigma_k}),
\E |X_{\sigma_k}|^{\otimes m^*}\right\rangle\bigg)^2
\cr && +\, C_{m^*,\tau_0,{c_1}}'\, \A_{\sigma}\E \bigg( \sum_{k=1}^n
\left\langle \E^*\,\fbar^{{(m^*)}}(\bU_{\sigma,{k}}^0,0),
|X_{\sigma_k}|^{\otimes m^*} - \E |X_{\sigma_k}|^{\otimes m^*}
\right\rangle\bigg)^2
\cr &=& C_{m^*,\tau_0,{c_1}}''\, \E \bigg(
\bigg\langle n\,\A_{\sigma,k} \big(\E^*\,f_{\max}^{{(m^*)}}(\bU_{\sigma,{k}}^0,\zeta_{k,\sigma_k})\big),
n^{-1}\sum_{i=1}^n \E |X_{i}|^{\otimes m^*}\bigg\rangle\bigg)^2
\cr && +\, C_{m^*,\tau_0,{c_1}}'\, n\,\A_{\sigma,k}
\left\langle \E\big\{\E^*\,\fbar^{{(m^*)}}(\bU_{\sigma,{k}}^0,0)\big\}^{\otimes 2},
\E \big(|X_{\sigma_k}|^{\otimes m^*} - \E |X_{\sigma_k}|^{\otimes m^*}\big)^{\otimes 2}
\right\rangle
\cr &\le & C_{m^*,\tau_0,{c_1}}''' \bigg(\frac{b_n^{m^*}(\log p)^{m^*-1}}{n^{m^*/2-1}}M_{m^*}^{m^*}\bigg)^2
 + C_{m^*,\tau_0,{c_1}}'''\frac{b_n^{2m^*}(\log p)^{2m^*-2}}{n^{m^*-1}}M_{2m^*}^{2m^*},
\eel
for some $C_{m^*,\tau_0,{c_1}}'''$ depending on $(m^*,\tau_0,{c_1})$ only.
It follows from the condition $a_nb_n(\log p)/n^{1/2}\le {c_1}$ that
$b_n^{m-2}(\log p)^{m-2} n^{-(m-2)/2} M_{m}^{2m} \le c_1^{m-2} M_4^2$.
Thus, by (\ref{pf-th3-1}), (\ref{th3-decomp}), (\ref{pf-th3-2b}) and (\ref{pf-th3-2c}),
\bel{pf-th-wild-a}
&& \big\|\eta_n^{{(\P^*)}}(1/b_n,t_0;T_n^0,T_n^*)\big\|_{L_2(\P)}
\cr &\le& C_{m^*,\tau_0,{c_1}}'''\sum_{m=2}^{m^*}
\frac{b_n^{m}(\log p)^{m-1}}{n^{m/2-1/2}}M_{2m}^{m}
+ C_{m^*,\tau_0,{c_1}}''\frac{b_n^{m^*}(\log p)^{m^*-1}}{n^{m^*/2-1}}M_{m^*}^{m^*}
\cr &\le& C_{m^*,\tau_0,{c_1}}\Big\{b_n^{2}(\log p)n^{-1/2}M_4^2 + \kappa_{n,m^*}\Big\}.
\eel

{\it Step 3: Proof of (\ref{th-comp-wild-2}).}
 Let $\Xtil_i$ the centered truncation of $X_{i}$ in (\ref{Xtil-a_n}) and
$\Xtil_i^* = W_i\Xtil_i$ with the truncation level $a_n = c_1\sqrt{n}/(b_n\log p)$.
Let $\Ttil_n$ and $\Ttil^*_n$ be the maxima corresponding to $\{\Xtil_i\}$ and $\{\Xtil_i^*\}$.
Because (\ref{moment-match-2}) also holds for the truncated variables,
it follows from (\ref{pf-th4-2}) and (\ref{pf-th-wild-a}) that
\bes
\eta_n^{{(q)}}(1/b_n)
&\le& \etatil^{(q)}(1/(2b_n))
+ \P\big\{\Omega_0\big\} + \big\|\P^*\big\{\Omega_0^*\big\}\big\|_{L_q(\P)}
\\ \nonumber &\le& C_{m^*,\tau_0,{c_1}}\bigg(
\frac{b_n^2\log p}{n^{1/2}}M_{4}^2
+ \kappa_{n,m^*}\bigg) + \P\big\{\Omega_0\big\} + \big\|\P^*\big\{\Omega_0^*\big\}\big\|_{L_1(\P)}^{1/q}
\ees
where $\Omega_0$ and $\Omega_0^*$ are given in (\ref{Omega_0}).
Since the truncation level is $a_n = c_1\sqrt{n}/(b_n\log p)$, we have
\bes
\P\big\{\Omega_0\big\} + \big\|\P^*\big\{\Omega_0^*\big\}\big\|_{L_1(\P)}
\le C_{m^*,\tau_0,{c_1}}\kappa_{n,m^*}
+\E \min\Big\{2, C_{\tau_0}b_n^2(\log p)n^{-1}\max_{1\le j\le p}\sum_{i=1}^n X_{i,j}^2I_{\{|X_{i,j}|>a_n\}}\Big\}
\ees
as in the proof of Lemmas \ref{lm-truncate-new} and \ref{lm-truncate-wild}.
The conclusion follows.

{\it Step 4: Proof of (\ref{th-comp-wild-3}).}
Taking $T_n=\Ttil_n$ and $T_n^*=\Ttil_n^*$ in the proof of (\ref{pf-th-comp-wild-b-1}),
we find that for any positive number $\etabar_n$,
\bes
\eta_n^*(1/b_n) = \eta_n^{{(\P^*)}}\big(1/b_n;T_n,T_n^*\big)
\le \max_{1\le k\le k_n} \eta_n^{{(\P^*)}}\big(1/(2b_n),t_k;T_n^0,T_n^*\big)
+\etabar_n
\ees
with certain $t_1,\ldots,t_{k_n}$ and $k_n \le 2/\etabar_n$. Thus, for
$(-1/q)\etabar_n^{-1}(2/\etabar_n)^{1/q}\eta_n^{{(q)}}(1/b_n)+1=0$,
\bes
\Big(\E\Big| \eta_n^*(1/b_n)\Big|^q\Big)^{1/q}
\le (2/\etabar_n)^{1/q}\eta_n^{{(q)}}(1/b_n)+\etabar_n
= (1+q)\big\{q^{-1}2^{1/q}\eta_n^{{(q)}}(1/b_n)\big\}^{q/(q+1)}
\ees
The proof is complete.
$\hfill\square$

\medskip
{\bf A3. Proofs of the results in Section 4.} 
We first state some properties of the standard Gaussian hazard function in the following lemma. 

\begin{lemma}\label{lm-hazard}
Let $h(t) = \varphi(t)/\Phi(-t)$ be the $N(0,1)$ hazard function and $h^{-1}(t)$ its inverse. Then,
\begin{enumerate}[label=({\alph*})]
 	\item $t < h(t) < t+1/h(t)$ $\forall\, t\ge 0$;
 	\item $0<h'(t) < 1$ $\forall\, t \in \R$;
 	\item $t - 1/t < h^{-1}(t) < t$, $\forall\, t\ge h(0) = \sqrt{2/\pi}$.  
 \end{enumerate} 
\end{lemma}

{\sc Proof of Lemma \ref{lm-hazard}.} 
For $t>0$, it is well known that $t\Phi(t) < \varphi(t)<(t+1/t)\Phi(t)$ so that $t<h(t) < t+1/t$ and 
$h'(t) = - t h(t) + h^2(t) >0$. As $1/h(t) = \int_0^\infty e^{-x^2/2-tx}dx$, 
\bes
\lim_{t\to\infty} h'(t)
= \lim_{t\to\infty} \frac{t^2 h'(t)}{h^2(t)}
= \lim_{t\to\infty} t^2\int_0^\infty xe^{-x^2/2-tx}dx
= \lim_{t\to\infty} \int_0^\infty ye^{-y^2/(2t^2)-y}dy =1. 
\ees
Moreover, as $h''(t) = \{2h(t)- t\}h'(t)-h(t)$, 
$h'(t)\ge 1$ would imply $h''(t) \ge h(t)-t > 0$ for all $t\in\R$, so that $h'(t) < \lim_{t\to\infty} h'(t)=1$. 
This implies (b) as $h'(t)>0$ for $t<0$ is trivial. 
Therefore, (a) follows from $h(t) - t = h'(t)/h(t) < 1/h(t)$, and (c) follows from (a).
$\hfill\square$



\medskip
{\bf A3.1. Proof of Theorem \ref{anti-concentration}.}
Let $\xi'_j = \xi_j/\sigma_j$, $\mu_j' = \mu_j/\sigma_j$ and $\rho_{j,k}=\Cov(\xi'_j,\xi'_k)$. 
It suffices to consider fixed $\sigma_j$ and $\rho_{j,k}$. 
As the upper bound does not depend on $\rho_{j,k}$, we assume without loss of generality that 
the matrix $(\rho_{j,k})_{p\times p}$ is of full rank and $0< \sigma_1 < \cdots < \sigma_p$, 
taking limits if necessary. Let 
\bes
g^* = \sup_{\mu_1,\ldots,\mu_p}\sup_x\ (d/dx)\P\Big\{ \max_{1\le j\le p}\xi_j \le x\Big\}
\ees
As the location can be absorbed in the means, 
\bes
g^* &=& \sup_{\mu_1,\ldots,\mu_p} \bigg[(d/dx)\P\Big\{ \max_{1\le j\le p}\xi_j \le x\Big\}\bigg]_{x=0}
\cr &=& \sup_{\mu_1,\ldots,\mu_p} \sum_{j=1}^p 
\P\Big\{ \max_{1\le k\le p, k\neq j}\xi_k < 0\Big| \xi_j=0\Big\} 
\frac{\varphi(\mu_j/\sigma_j)}{\sigma_j}
\cr &=& \sup_{\mu_1',\ldots,\mu_p'} \sum_{j=1}^p 
\P\Big\{ \max_{1\le k\le p, k\neq j}\xi_k' < 0\Big| \xi_j'=0\Big\} 
\frac{\varphi(\mu_j')}{\sigma_j}. 
\ees
As $\xi_k'-\rho_{j,k}\xi_j'$ are independent of $\xi_j'$, 
\bes
&& \P\Big\{ \max_{1\le k\le p, k\neq j}\xi_k' < 0\Big| \xi_j'=0\Big\}\Phi(\mu'_j)
\cr &=& \P\Big\{\xi'_k-\rho_{j,k}\xi'_j < 0,\ \forall\, k\neq j\Big\}\P\Big\{\xi'_j > 0\Big\}
\cr &\le& \int_0^\infty \P\Big\{\xi'_k-\rho_{j,k}\xi'_j < (1- \rho_{j,k})x,\ \forall\, k\neq j\Big\} 
\varphi(x-\mu_j')dx
\cr & = & \int_0^\infty \P\Big\{\xi'_k < x,\ \forall\, k\neq j\Big| \xi_j'=x \Big\}\varphi(x-\mu_j')dx
\cr &=& \P\Big\{ \max_{k\neq j} \xi'_k < \xi'_j, \xi_j' \ge 0 \Big\}. 
\ees
Thus, as $\sum_{j=1}^p\P\big\{ \max_{k\neq j} \xi'_k < \xi'_j, \xi_j' \ge 0 \big\} 
= \P\big\{ \max_{1\le j\le p} \xi'_p \ge 0 \big\}\le 1$, 
we find that 
\bes
g^* &\le & \sup_{\mu_1',\ldots,\mu_p'}\sup
\bigg\{\sum_{j=1}^p \frac{\varphi(\mu_j')}{\sigma_j} 
\min\bigg(1, \frac{w_j}{\Phi(\mu_j')}\bigg): w_j\ge 0,\sum_j w_j=1\bigg\}
\cr & \le & \sup_{\mu_1' > 0,\ldots,\mu_p' > 0} \inf_{J\subseteq [p]}
\bigg\{\max_{j\in J}\frac{\varphi(\mu_j')}{\sigma_j\Phi(-\mu_j')} 
+ \sum_{j\in J^c} \frac{\varphi(\mu_j')}{\sigma_j} \bigg\}
\ees
Let $h(t)=\varphi(t)/\Phi(-t)$ be the standard normal hazard function, 
and $h^{-1}(t)$ its inverse for $t\ge 0$. 
Let $c_j = h(\mu_j')/\sigma_j$ so that $\mu_j' = h^{-1}(\sigma_jc_j)$. 
It follows that 
\bes
g^* &\le & \sup_{c_1> 0,\ldots,c_p > 0}\min_{1\le j\le p+1}
\bigg\{c_j + \sum_{c_k>c_j} \frac{\varphi(h^{-1}(\sigma_kc_k))}{\sigma_k} \bigg\}, 
\ees
with $c_{p+1}=0$ and $\sigma_{p+1} = \sigma_p$. 
As $\varphi(h^{-1}(\sigma_kc_k))/\sigma_k$ is decreasing in $\sigma_k$ for $c_k > 0$, 
the above supreme 
is attained when large $c_k$ is paired with small $\sigma_j$, $c_1\ge \cdots\ge c_p$. Thus, 
\bes
g^* &\le & \sup_{c_1\ge \cdots\ge c_p > 0} \min_{1\le j\le p+1}
\bigg\{c_j + \sum_{k=1}^{j-1} \frac{\varphi(h^{-1}(\sigma_kc_k))}{\sigma_k} \bigg\}. 
\ees
Let $m$ be the smallest $j \le p+1$ satisfying $c_j\sigma_j \le x_j$. We have 
\bes
g^* &\le & \max_{1\le m\le p+1}\bigg\{\frac{x_m}{\sigma_m}I\{m\le p\} + \sum_{k=1}^{m-1} 
\frac{\varphi(h^{-1}(x_k))}{\sigma_k}\bigg\}. 
\ees
This gives \eqref{lm-anti-2} as $h^{-1}(x_k)\ge x_k-1/x_k \ge 0$ 
for $x_k\ge 1$ by Lemma \ref{lm-hazard}. 
Taking $x_k=1+\sqrt{2\log k}$, \eqref{lm-anti-3} follows from 
$\sum_{k=1}^\infty\varphi(x_k-1/x_k)\le 1$. 

Finally, for \eqref{lm-anti-bd} we set $x_m = 1+\sqrt{2\log m}$ and consider integer $m$ satisfying 
$x_{{j}}/\sigma_{{j}} \le x_m/\sigma_m$ for all $1\le {{j}}\le p$. 
Let $\xi_j$, $k\le j\le m+1$, be independent $N(0,\sigma_j^2)$ variables.  
Set $\xi_j = \xi_k \sigma_j/\sigma_k$ for $j\le k$ and 
$\xi_j = \xi_{m+1} \sigma_j/\sigma_{m+1}$ for $m<j\le p+1$, with $\sigma_{p+1}=\sigma_p$. 
Let $J = \{k,\ldots,m,p\}$ and $\pi_x = \sum_{j=k}^m \Phi(-x/\sigma_j)$.  
For $\pi_x\in (0,1)$, 
\bes
\frac{d}{dx}\P\Big\{\max_{1\le j\le p} \xi_j \le x\Big\} 
= \sum_{j \in J} \frac{\varphi(x/\sigma_j)}{\sigma_j}\prod_{\ell\in J\setminus \{j\}}\Phi(x/\sigma_\ell)
\ge \frac{\pi_x(1-\pi_x)}{2} \frac{h(x/\sigma_m)}{\sigma_m}
\ge \frac{x\pi_x(1-\pi_x)}{2\sigma_m^2}. 
\ees
As $m\Phi(-x_m/2) \ge 1/4$ and $\pi_x \ge m\Phi(-x/\sigma_1)$ for $k=1$, 
for $\pi_x\le 1/4$ and $x_m\le 4$ we have 
\bel{lm-anti-bd-pf} 
\frac{d}{dx}\P\Big\{\max_{1\le j\le p} \xi_j \le x\Big\} 
\ge \frac{x_m\sigma_1\pi_x(1-\pi_x)}{4\sigma_m^2}
\ge \frac{x_m\pi_x(1-\pi_x)}{16\sigma_m} \ge \frac{\pi_x(1-\pi_x)(2+\sqrt{2\log p})}{16{\overline\sigma}}
\eel
For $x_m\ge 4$, $m\Phi(- x_m/2)\ge 2$ 
and $\pi_x\ge (m-k+1)\Phi(-x/\sigma_k)$, so that \eqref{lm-anti-bd-pf} is still valid 
for $\pi_x\le 1/4$ and $m-k+1\ge m(1/8)$, 
due to $\sigma_k/\sigma_m\ge x_k/x_m\ge1/4$ for such $(k,m)$. 
As $\P\{\max_{1\le j\le p} \xi_j \le x\}\ge 1 - \pi_x$, 
\eqref{lm-anti-bd} follows from \eqref{lm-anti-2} with $C_0\le 2^7/(1-1/4)$. 
The proof for independent $\xi_j\sim N(\mu_j,\sigma_j^2)$ follows from a nearly identical 
construction with the same $\xi_j$ for $j\in J$ and $\mu_j \approx - \infty$ for $j\not\in J$. 
$\hfill\square$

\medskip
{\bf A3.2. Proof of Theorem \ref{anti-mix}.} 
Assume without loss of generality $\sigma_1\le \ldots\le \sigma_p$.  
By (\ref{conditional}) the conditional variance of $Z_{n,j}^{**}$, 
\bes
\big(\sigma_j^{**}\big)^2 = \frac{1}{n}\sum_{i=1}^n a_0^2\delta_i X_{i,j}^2,
\ees
is an average of independent variables $a_0^2\delta_i X_{i,j}^2$ with
\bes
0\le a_0^2\delta_i X_{i,j}^2 \le a_n^2a_0^2,\quad \E \big(\sigma_j^{**}\big)^2 = a_0^2p_0\sigma^2_j,\quad
\Var\big(\big(\sigma_j^{**}\big)^2\big) \le a_n^2\sigma_j^2 a_0^4p_0/n.
\ees 
Let $\eps_0 =(\eps/{\overline\sigma})\sqrt{\log p}$. By the Bernstein inequality, 
\bes
\P\Big\{\Big|\big(\sigma_j^{**}\big)^2 - a_0^2p_0\sigma^2_j\Big|
> \sqrt{2 \log(j^2/\eps_0) a_n^2\sigma_j^2 a_0^4p_0/n}+ 2a_n^2a_0^2 \log(j^2/\eps_0) /(3n) \Big\} \le 
2e^{-\log(j^2/\eps_0)}\le 2\eps_0j^{-2}.
\ees
By \eqref{conc-anti-mix-0}, 
$\log\big(j^2/\eps_0\big) \le \sigma_j^2 p_0n/(8a_n^2)$ for all $j$, 
so that the above inequality yields 
\bes
\sqrt{2 \log(j^2/\eps_0) a_n^2\sigma_j^2 a_0^4p_0/n}+ 2a_n^2a_0^2 \log(j^2/\eps_0)/(3n) 
\le a_0^2p_0 \sigma_j^2/2+a_0^2p_0 \sigma_j^2/12 <
\,3a_0^2p_0 \sigma_j^2/4.
\ees
It follows that with at least probability 
$1-2\eps_0 \sum_{j=1}^p j^{-2} $ under $\P$,
\bel{pf-th-anti-mix-2}
\Big|\big(\sigma_j^{**}\big)^2 - a_0^2p_0\sigma^2_j\Big|
< 3a_0^2p_0 \sigma_j^2/4\quad \forall\,j,
\eel
which implies $\sigma_j^{**} >  \sigma_j\sqrt{p_0}a_0/2$.
Thus, by 
(\ref{lm-anti-3}), (\ref{un-mix}) and the fact that 
$0\le \omega_n^{{(\P^{**})}}\big(\eps;T_n^{**}\big)\le 1$,
\bes
\sup_{t\in \R}\P\Big\{ t \le T_n^{**} \le t+\eps\Big\}
\le
2\eps_0 \sum_{j=1}^p j^{-2}
+ \frac{\eps(1+\sqrt{2\log p})}{{\overline\sigma}\sqrt{p_0}a_0/2} \le
C_{a_0,b_0,p_0}\frac{\eps}{{\overline\sigma}}\sqrt{\log p},
\ees
as $\eps_0 =(\eps/{\overline\sigma}) \sqrt{\log p}$. 
${\hfill\square}$

\medskip
{\bf A3.3. Proof of Theorem \ref{anti-general}.} 
Let $\sigma_{(1)}=\min_j\sigma_j$ as in \eqref{sigmas}. Suppose
\bel{cond-a_n}
\max \bigg\{ \frac{2M_4^2}{\sigma_{(1)}}, M_4\Big(\frac{n}{\log (p/\kappa_{n,4})}\Big)^{1/4} \bigg\}
\le  a_n = \frac{c_1\sqrt{n}}{b_n\,\log p} \le
\min_{1\le j \le p} \frac{\sqrt{n}\sigma_{(j)}}{16
\sqrt{\log(j^2\, {\overline \sigma}b_n /\sqrt{\log p})}}.
\eel 
Let $\tbX =(\Xtil_{i,j})_{n\times p}=(\Xtil_1,\ldots,\Xtil_n)$ be as in (\ref{Xtil-a_n}) and
\bes
\Ttil_n = \max_{1\le j\le p}n^{-1/2}\sum_{i=1}^n \Xtil_{i,j},\quad
\Ttil_n^{**} = \max_{1\le j\le p}n^{-1/2}\sum_{i=1}^n W_i^{**}\Xtil_{i,j},\quad
{\widetilde \sigma}_j^2 = \frac{1}{n}\sum_{i=1}^n \E\Xtil_{i,j}^2,
\ees
where $W_i^{**}$ is as in (\ref{multiplier-mix}) with $p_0=1/2$.
It follows from (\ref{anti-conc-2}) that
\bel{pf-th-anti-general-1}
\omega_n^{{(\P)}}(1/b_n;T_n)\le 3\omega_n^{{(\P)}}(1/b_n;\Ttil_n^{**}) 
+ 2\eta_n^{{(\P)}}\big(1/b_n;T_n,\Ttil_n^{**}\big).
\eel

We shall use Theorem \ref{anti-mix} to bound the first term above.
The variance of $\Xtil_{i,j}$ are bounded by
\bes
\sigma_j^2 - {\widetilde \sigma}_j^2
= \frac{1}{n}\sum_{i=1}^n \E X_{i,j}^2I_{\{|X_{i,j}|>a_n\}}
+ \bigg(\frac{1}{n}\sum_{i=1}^n \E X_{i,j} I_{\{|X_{i,j}|>a_n\}}\bigg)^2
\le \frac{M_4^4}{a_n^2} + \bigg(\frac{M_2}{a_n}\bigg)^2\le \frac{2M_4^4}{a_n^2}.
\ees
By the first component of the first inequality in \eqref{cond-a_n}, 
$2M_4^4/a_n^2 \le \sigma_{(1)}^2/2$, 
so that $\sigma_j^2/2 \le {\widetilde \sigma}_j^2\le \sigma_j^2$. 
Thus, by the second inequality in \eqref{cond-a_n}, condition \eqref{conc-anti-mix-0} of Theorem \ref{anti-mix} holds for 
${\widetilde \sigma}_j$, $p_0=1/2$ and $\eps=1/b_n$.
Moreover, because $(a_0,b_0)$ is determined by the condition
$p_0=1/2$, Theorem \ref{anti-mix} yields
\bes
\omega_n^{{(\P)}}\big(1/b_n; \Ttil_n^{**}\big)
=\sup_{t\in \R}\P\Big\{ t \le \Ttil_n^{**} \le
{ t+1/b_n\Big\}
\le \frac{C_0}{b_n{\overline \sigma}}
\sqrt{\log p}}.
\ees

To bound $\eta_n^{{(\P)}}\big(1/b_n;T_n,\Ttil_n^{**}\big)$ 
in \eqref{pf-th-anti-general-1}, we 
apply Theorem \ref{th-comp-gene} (ii) to $\bX^* = \tbX^{*}= \tbX^{**}$.
We assume for a certain $C'_{c_1}$, $\kappa_{n,4} \le 1/C'_{c_1}$,
as the conclusion is otherwise trivial.
It follows from (\ref{th-comp-gene-2}) that 
\bes
\eta_n^{{(\P)}}\big(1/b_n;T_n,\Ttil_n^{**}\big) \le
C_{c_1}\kappa_{n,4} + \P\big\{\Omega_0 \big\}
\ees
because $\mutil^{{(m)}} = \nutil^{{(m)}}$ for $m=2, 3$ by (\ref{multiplier-mix-coef})
and $\P\big\{\Omega_0^*\big\}=0$ for $\tbX^* = \tbX^{**}$.
Moreover, due to the condition 
$M_4\big(n/\log (p/\kappa_{n,4})\big)^{1/4} \le a_n$ in \eqref{cond-a_n}, Lemma \ref{lm-truncate-new} with $(\atil_n,\eps_n)= (a_n,\kappa_{n,4})$ 
gives
\bes
\P\big\{\Omega_0 \big\} \le \kappa_{n,4}+
\P\bigg\{\max_{1\le j\le p}\bigg|\sum_{i=1}^n \frac{X_{i,j}I_{\{|X_{i,j}|> a_n
\}}}{\sqrt{n}}\bigg|
>\frac{1}{8b_n}\bigg\}. 
\ees
The conclusion follows by inserting the above three displayed inequalities into (\ref{pf-th-anti-general-1}).

It remains to verify \eqref{cond-a_n} for $C_0\sqrt{\log p}/(b_n{\overline \sigma})\le 1$. 
By the definition of ${\overline \sigma}$ in \eqref{sigmas},
	\bes
	\max\big(2/\sigma_{(1)}, \sqrt{\log j^2}/\sigma_{(j)}\big) 
	\le (2+\sqrt{2\log p})/{\overline \sigma} \le 5\sqrt{\log p}/{\overline \sigma}
	\ees 
for all $1\le j\le p$ and $p \ge 2$. 
	It follows that for sufficiently large $C_0$ 
\bes
\frac{c_1\sqrt{n}}{b_n\,\log p}\frac{16\sqrt{\log(j^2\, b_n{\overline \sigma} /\sqrt{\log p})}}
{\sqrt{n}\sigma_{(j)}}
\le \frac{16 c_1}{\log p}\bigg(\frac{5\sqrt{\log p}}
{b_n{\overline \sigma}} + \frac{5\sqrt{\log(b_n{\overline \sigma} /\sqrt{\log p})}}
{b_n{\overline \sigma}/\sqrt{\log p}}\bigg)
\le \frac{80 c_1(1+\sqrt{\log C_0})}{C_0\log p}, 
\ees
so that the second inequality in \eqref{cond-a_n} holds. For the first one, we have 
	\bes
	\frac{2M_4^2}{\sigma_{(1)}}\Big(\frac{b_n\,\log p}{c_1\sqrt{n}}\Big) 
	 = \frac{2\kappa_{n,4}^{1/2}}{\sigma_{(1)} c_1b_n \sqrt{\log p}}
	\le \frac{5\kappa_{n,4}^{1/2}}{c_1 b_n{\overline \sigma}} 
	\le \frac{5C_{c_1}^{-1/2}}{c_1C_0\sqrt{\log p}} \le 1
	\ees
for sufficiently large $C_0$ and $C_{c_1}$, and
\bes
M_4\Big(\frac{n}{\log (p/\kappa_{n,4})}\Big)^{1/4} \Big(\frac{b_n\,\log p}{c_1\sqrt{n}}\Big)
= \frac{1}{c_1} \Big( \frac{\kappa_{n,4}\log p}{\log (p/\kappa_{n,4})} \Big)^{1/4} 
\le c_1^{-1} C_{c_1}^{-1/4} \le 1, 
\ees
as $\kappa_{n,4} = b_n^4(\log p)^{3}n^{-1}M_4^4$. 
This completes the proof of \eqref{cond-a_n} and thus the entire theorem. 
${\hfill\square}$

\end{document}